\documentclass[
]{ceurart}

\usepackage{algorithm}
\usepackage{algpseudocode}

\usepackage{listings}

\usepackage{tabularx}
\usepackage{graphicx}
\usepackage{multirow}
\usepackage{booktabs}

\usepackage{float}

\usepackage{lscape}

\usepackage{todonotes}
\usepackage{enumitem}
\usepackage{numprint}

\usepackage{etoolbox}

\usepackage{caption}
\usepackage{xcolor}
\usepackage{amsmath}
\usepackage{pifont}
\usepackage{rotating}

\usepackage{pdflscape}
\usepackage{longtable}

\usepackage{placeins}

\usepackage{pbox}

\usepackage{cleveref}

\usepackage{tikz}
\usetikzlibrary{decorations.pathreplacing,calc}

\newcommand{\cmark}{\ensuremath{\checkmark}}
\newcommand{\xmark}{\text{\sffamily x}}
\newcommand{\pmark}{\textit{p}}

\newcommand{\BaseSurvey}{\citet{Zaveri2012}}

\newcommand{\BinaryMeasure}{\textit{binary measure}}
\newcommand{\RatioMeasure}{\textit{ratio measure}}
\newcommand{\CompositeMeasure}{\textit{composite measure}}

\newcommand{\shape}[1]{\textsf{SH#1}}

\newcommand{\dq}[1]{\textsf{#1}}

\newcommand{\rdftriple}[3]{$\langle${\small\texttt{#1,#2,#3}}$\rangle$}

\newcommand{\rdfproperty}[1]{{\small\texttt{#1}}}

\newcommand{\assumption}[1]{(A{#1})}

\newtoggle{appendixversion}
\toggletrue{appendixversion} 

\begin{document}

\copyrightyear{2022}
\copyrightclause{Copyright for this paper by its authors.
  Use permitted under Creative Commons License Attribution 4.0
  International (CC BY 4.0).}

\iftoggle{appendixversion}{
    \conference{}
}{
    \conference{WOP2025: 16th Workshop on Ontology Design and Patterns, November 2-3, 2025, Nara, Japan}
}
%


\title{Is SHACL Suitable for Data Quality Assessment?}



\author[1]{Carolina Cort\'es}[
email=carolina.cortes@hpi.de,
]
\address[1]{Hasso Plattner Institute, Potsdam, Germany}

\author[1]{Lisa Ehrlinger}[
email=lisa.ehrlinger@hpi.de
]

\author[2]{Lorena Etcheverry}[
email=lorenae@fing.edu.uy,
]
\address[2]{Universidad de la República, Montevideo, Uruguay}

\author[1]{Felix Naumann}[
email=felix.naumann@hpi.de,
]


\begin{abstract}
Knowledge graphs have been widely adopted in both enterprises, such as the Google Knowledge Graph, and open platforms like Wikidata, to represent domain knowledge and support artificial intelligence applications. 
They model real-world information as nodes and edges.
To embrace flexibility, knowledge graphs often lack enforced schemas (i.e., ontologies), leading to potential data quality issues, such as semantically overlapping nodes.
Yet ensuring their quality is essential, as issues in the data can affect applications relying on them. 
To assess the quality of knowledge graphs, existing works propose either high-level frameworks comprising various data quality dimensions without concrete implementations, define tools that measure data quality with ad-hoc SPARQL queries, or promote the usage of constraint languages, such as the Shapes Constraint Language (SHACL), to assess and improve the quality of the graph. Although the latter approaches claim to address data quality assessment, none of them comprehensively tries to cover all data quality dimensions.
In this paper, we explore this gap by investigating the extent to which SHACL core can be used to assess data quality in knowledge graphs. 
Specifically, we defined SHACL shapes for 69 data quality metrics proposed by~\citet{Zaveri2012} and implemented a prototype that automatically instantiates these shapes and computes the corresponding data quality measures from their validation results.
All resources are provided for repeatability.
\end{abstract}


\maketitle

\section{Introduction}

Knowledge graphs (KGs) have been increasingly used to represent domain knowledge and support artificial intelligence applications, such as information retrieval and question answering~\cite{Suchanek2024}. As a result, ensuring the quality of KGs is crucial for applications that rely on their input~\cite{Furber2018}.
Major technology companies, such as Google (Google Knowledge Graph~\cite{GoogleKG}) and Amazon (Alexa Knowledge Graph~\cite{AmazonAlexaKG}), have developed proprietary KGs~\cite{Suchanek2024}, while collaborative KGs, such as Wikidata~\cite{Wikidata}, DBpedia~\cite{DBpedia} and YAGO~\cite{YAGO}, have emerged as open-source alternatives. This growing adoption of KGs is reflected in the growth of the Linked Open Data (LOD) cloud, which has enabled the publication of numerous datasets across domains, with \numprint{1656} resources as of Nov.~2024~\cite{Tuozzo2025}.

Knowledge graphs represent information about the world as nodes and edges~\cite{Hogan2022}. 
They are typically constructed and enriched using diverse sources and (semi-)automated techniques, with some also supporting human curation~\cite{Furber2018, Hogan2022}. While several graph data models are available~\cite{Angles2018}, this paper focuses on the Resource Description Framework (RDF)~\cite{RDFconcepts}, which structures data as triples consisting of a subject, a predicate, and an object.
RDF-based graphs can include a semantic schema, such as a vocabulary or ontology, that defines their expected structure. However, to ensure flexibility and support the evolution of KGs over time, these schemata are generally not enforced~\cite{Polleres_2023}, which can introduce data quality issues. 
Furthermore, the quality of ontologies directly affects the quality of the data, as poorly defined classes or properties can lead to misclassified or ambiguous data in the graph, and inconsistent ontology axioms can propagate errors in reasoning over the graph~\cite{McGurk2017}. Consider the case of Wikidata, which contains several classes that are difficult to distinguish, such as ``geographical location'', ``location'', and ``geographic region'', and also classes and instances appear mixed, such as ``scientist'', which is both a subclass of ``researcher'' and an instance of ``profession''~\cite{Suchanek2024}. We focus on the quality assessment of the data graph, but the proposed methods can also be applied to ontologies.

Data quality (DQ) is generally defined as ``fitness for use''~\cite{WangStrong_1996}, which highlights the importance of considering the context in which the data is utilized when assessing its quality. DQ is typically evaluated across various \textit{dimensions}, such as \dq{Completeness}, \dq{Consistency}, and \dq{Understandability} \cite{Pipino2002}. These dimensions can be quantified using specific metrics, known as \textit{DQ metrics}, which are designed to measure different aspects of each dimension \cite{Pipino2002}. The process of data quality assessment (DQA) involves obtaining numerical values, referred to as \textit{DQ measures}, that characterize various aspects of DQ\@. While many classifications exist for both DQ dimensions and metrics, there is no universally accepted standard~\cite{Cichy_2019}. As a result, DQA remains a challenging task in practice.

Several works address the quality of KGs from a \emph{high-level} perspective, presenting definitions and metrics~\cite{Zaveri2012, Issa2021} or conceptual frameworks to assess the quality of KGs~\cite{Nayak2022, Chen2019} without a concrete implementation.
Other works approach the topic of KG quality from a \emph{bottom-up} approach, by developing tools to assess DQ (e.g., \cite{LuzzuDebattista2016, RDFUnitKontokostas2014, KGHeatBeatPellegrino2025, SWIQAFrber2011}). Most of these works define ad-hoc SPARQL (SPARQL Protocol and RDF Query Language) queries to obtain measures for DQ dimensions. 

In recent years, constraint languages, such as the Shapes Constraint Language (SHACL)~\cite{w3c_shacl} and Shapes Expression Language (ShEx)~\cite{Prudhommeaux2014}, have emerged to enable RDF graph validation by specifying constraints as shapes~\cite{Hogan2022}.
Despite syntactic differences, both languages enable the definition of constraints on nodes and their value nodes (i.e., values reachable via properties or paths), allowing for the detection of data violations~\cite {Rabbani2022}. These languages enable a \emph{low-level} perspective on DQ, focusing on concrete error detection through constraint checks. Unlike ad-hoc SPARQL-based approaches, constraint languages like SHACL offer a formal way to express validation rules.

\smallskip\noindent\textbf{Research gap and contributions.}
Since constraint languages for RDF were first introduced, different works have focused on the generation of shapes, either from the data or its metadata (e.g., ontologies)~\cite{SpahiuMP18, Rabbani2023_shactor, Boneva2019, Cimmino2020, Pandit2018, Duan2024, FernandezAlvarez2022, Yang2023, Luthfi2022}. Only some works attempt to bridge the gap between the generation of shapes and their potential to assess and improve the quality of the data \cite{SpahiuMP18,Rabbani2023_shactor,Yang2023,Luthfi2022}. In particular, \citet{Luthfi2022} approaches DQ from the dimensions perspective, defining shapes for \dq{Completeness}. 


This paper explores how SHACL core can be leveraged for DQA.
In particular, we want to investigate the extent to which we can connect the \textit{high-level} view on DQ dimensions with the \textit{low-level} view on DQ, which entails identifying constraint violations using SHACL shapes.
Thus, the contributions of this paper are as follows:
\begin{enumerate}
    \item Evaluation of the suitability of SHACL core for DQA by defining shapes for the set of 69 DQ metrics defined by \citet{Zaveri2012}.
    \item A prototype that (i) automatically instantiates the defined SHACL shapes, and (ii) computes DQ metrics based on the shape validation results.
\end{enumerate}

\smallskip\noindent\textbf{Outline.} 
\autoref{section:related_work} summarizes related work. \autoref{sec:approach} presents the definition of SHACL shapes for a single dimension and metric of each group defined by~\cite{Zaveri2012}. The complete list of all defined SHACL shapes for all dimensions is provided in
\iftoggle{appendixversion}{\autoref{section:Appendix_Shapes_Definition}}{the appendix of the extended version of this paper~\cite{AppendixVersion_2025}}. \autoref{sec:prototype} describes the implemented prototype for SHACL-based DQA, followed by a discussion of the suitability of SHACL for DQA in \autoref{sec:discussion}.
Finally, \autoref{sec:conclusion} concludes the paper with an outlook on future work.

\section{Related Work}\label{section:related_work}

\noindent\textbf{Data quality in knowledge graphs.} Several works address the quality of KGs from a high-level perspective. In particular, \citet{Zaveri2012} identified a set of DQ dimensions and metrics through a systematic literature review and compared different DQA tools. \citet{Issa2021} defined a set of DQ metrics for the \dq{Completeness} dimension, and also analyzed different tools capable of assessing KG \dq{Completeness}. 
\citet{Nayak2022} analyzed different tools for assessment, profiling, and improvement of linked data and proposed a DQ refinement lifecycle.
\citet{Chen2019} proposed a DQA framework by mapping KG application requirements to DQ dimensions from \cite{Zaveri2012}, and extending them with two new dimensions: \dq{Robustness} and \dq{Diversity}.
However, none of these works present a concrete implementation.

In addition, several tools have been proposed over the years to assess the quality of KGs and linked data (which we consider a type of KG, though not all KGs follow Linked Data principles). 
Some of these tools focus on specific DQ dimensions \cite{SPARQLESVandenbussche2017, LDSnifferMihindukulasooriya2017}, while others try to cover a wider range of them \cite{LuzzuDebattista2016, KGHeatBeatPellegrino2025, SemQuireLanger2018, TripleCheckMateKontokostas2013, SWIQAFrber2011, YummyDataYamamotoYS18, LiQuateRuckhaus2014, SecurityDashboardPizhukEDG25}. Moreover, some tools focus on assessing the quality of SPARQL endpoints \cite{YummyDataYamamotoYS18, LDSnifferMihindukulasooriya2017, SPARQLESVandenbussche2017}, while others focus on the quality of the data itself \cite{LuzzuDebattista2016, KGHeatBeatPellegrino2025, SemQuireLanger2018, TripleCheckMateKontokostas2013, SWIQAFrber2011, LiQuateRuckhaus2014, SecurityDashboardPizhukEDG25}. Assessment is primarily done via ad-hoc SPARQL queries \cite{KGHeatBeatPellegrino2025, SemQuireLanger2018, SPARQLESVandenbussche2017, SWIQAFrber2011, YummyDataYamamotoYS18, RDFUnitKontokostas2014, SecurityDashboardPizhukEDG25}, while some introduce more complex techniques \cite{LuzzuDebattista2016, LiQuateRuckhaus2014}. Only~\cite{SWIQAFrber2011} considers the usage of a constraint language. Additionally, some of these tools haven't been maintained for some time or aren't available\iftoggle{appendixversion}{. In \autoref{section:related_work_appendix} we present a table summarizing these tools.}{~(See Appendix A of~\cite{AppendixVersion_2025} for more details).}

\noindent\textbf{Constraint languages.} Constraint languages validate a set of conditions over RDF graphs. 
The Shapes Constraint Language (SHACL), a W3C recommendation \cite{w3c_shacl}, enables this via shapes that specify constraints on the graph. The \textit{shapes graph} holds these shapes, and the \textit{data graph} is the RDF graph being validated. When a shape is evaluated on a node (the \textit{focus node}), SHACL uses \textit{node shapes}, to constrain the node itself, and \textit{property shapes} to constrain \textit{value nodes} reached from the \textit{focus node} via a property or path in the graph.
\textit{Constraint components} define conditions to validate focus and value nodes. For example, \textit{MinCountConstraintComponent} defines the \textit{minCount} property, which can be used to specify a minimum number of values for a given property.
The validation process takes a \textit{data graph} and a \textit{shapes graph} as input and produces a validation report\iftoggle{appendixversion}{. This report, as defined by the \textit{RDF Validation Report Vocabulary}\cite{w3c_shacl}, provides insights on how to fix the errors causing the violation.}{, which provides insights on how to fix the errors causing the violations.}

While constraint languages provide a formal way to define and validate constraints, writing them is time-consuming and requires domain expertise~\cite{Rabbani2022}. To address this, several works aim to automatically~\cite{SpahiuMP18, Rabbani2023_shactor, Boneva2019, Cimmino2020, Pandit2018, Duan2024, FernandezAlvarez2022} or semi-automatically~\cite{Yang2023, Luthfi2022} generate shapes from existing data~\cite{Rabbani2023_shactor, Boneva2019, SpahiuMP18, FernandezAlvarez2022} or related artifacts~\cite{Pandit2018, Duan2024, Cimmino2020}, such as ontologies. Data-driven approaches usually cover basic constraints (e.g., required properties, ranges, cardinality)~\cite{Duan2024} but often produce many unreliable shapes, which are filtered using support/confidence~\cite{Rabbani2023_shactor} or trustworthiness scores~\cite{FernandezAlvarez2022}. Artifact-based methods, on the other hand, can generate richer constraints by leveraging formal restrictions like OWL axioms.

Few of these works try to bridge the gap between the generation of shapes and how these can help assess and improve DQ\@. \citet{SpahiuMP18} present an approach to generate SHACL shapes from semantic profiles created with a profiling tool, which are then used to assess the quality of different versions of a dataset over time. \citet{Rabbani2023_shactor} present the tool SHACTOR, which not only generates shapes from a KG, but also allows the user to generate SPARQL queries that retrieve the triples that produce low support and confidence shapes. These triples are considered to be erroneous and can be removed from the graph to improve its quality. \citet{Luthfi2022} consider the survey~\cite{Issa2021} and define shape patterns for different aspects of \dq{Completeness}, testing them on Wikidata and DBpedia. To instantiate the shape patterns, they consider specific information provided by Wikidata, such as property constraints. \citet{Yang2023} propose a SHACL-based DQ validation process for \dq{Completeness}, \dq{Accuracy}, and \dq{Consistency}, using shapes tailored to a health ontology. While constraint examples are given, full shape definitions are not publicly accessible, and the method is specific to a particular ontology and dataset.

Although the approaches discussed above claim to address DQ in some way, none of them comprehensively attempts to cover all DQ dimensions.

\section{SHACL shapes for DQ dimensions}\label{sec:approach}
This paper builds on the comprehensive DQ metrics survey by \citet{Zaveri2012}, which groups dimensions into four categories.
We follow their structure, defining SHACL core shapes for each metric or explaining SHACL core’s limitations when shapes cannot be defined.\footnote{
A discussion on SHACL extensions and the justification for the focus on SHACL core can be found in \autoref{sec:disc-shacl-extensions}. Across the paper we use SHACL interchangeably with SHACL core, unless stated otherwise.}
We present the results of this study in tables \ref{table:accessibility_metrics}--\ref{table:representational_metrics}, which illustrate the feasibility of implementing metrics from \cite{Zaveri2012} with SHACL core. Symbols \cmark, \pmark, and \xmark~denote the degree to which metric implementation was possible (full, partial, or not at all, respectively).
For the shape definition, we made some realistic assumptions (A1--3):
\begin{enumerate}[label=(A{\arabic*})]
    \item All entities are explicitly typed: for each entity \texttt{e} representing a real-world object, the triple \rdftriple{e}{rdf:type}{c} exists in the graph.

    \item The ontology used is sufficiently defined: each class \texttt{c} and property \texttt{p} has triples \rdftriple{c}{rdf:type}{rdfs:Class} and \rdftriple{p}{rdf:type} {rdf:Property}, respectively. Each property \texttt{p}, includes domain and range triples \rdftriple{p}{rdfs:domain}{d} and \rdftriple{p}{rdfs:range}{r}. Other property characteristics (e.g., irreflexivity, asymmetry) are also defined.
    For every instance \texttt{i} of a class \textit{c} defined in the ontology, \rdftriple{i}{rdf:type}{owl:NamedIndividual} exists.
    
    \item Relevant domain knowledge, typically provided by domain experts, is available for shape instantiation, e.g., expected number of values for certain properties, gold standard value sets, or definitions of when data is considered to be up to date.
\end{enumerate}



We acknowledge that these assumptions may not always hold in real-world KGs. If~\assumption{1} is violated, untyped instances are excluded from validation. Likewise, when~\assumption{2} is not satisfied, schema characteristics such as range, domain, or property constraints (e.g., symmetry, irreflexivity) cannot be checked. One possible direction to relax these requirements is to mine such characteristics automatically or leverage profiling statistics to approximate range and domain information, thereby supporting the construction of an ``initial'' ontology. Regarding~\assumption{3}, this is inherent to DQA, as some quality dimensions rely on domain knowledge, limiting their assessment.
Tables \ref{table:accessibility_metrics}--\ref{table:representational_metrics}, mark the use of assumptions with $*_{X}$, where \textit{X} is the assumption number. 
In the following subsections, we summarize SHACL core coverage per group, discuss a representative DQ metric for a single dimension of each group, and, if applicable, provide the corresponding shape in Turtle syntax. For the implemented shapes, we also indicate the DQ measure type: \BinaryMeasure~(1 if no violations, 0 otherwise), \RatioMeasure~(violation-based ratios), or \CompositeMeasure~(aggregated score across shape instances for specific properties or classes).

\subsection{Accessibility}\label{section:accessibility}

The \textit{Accessibility} group includes dimensions related to accessing, retrieving, and verifying the authenticity of data: \dq{Availability}, \dq{Licensing}, \dq{Interlinking}, \dq{Security}, and \dq{Performance}~\cite{Zaveri2012}. In the following, we exemplarily present the shape definition for a metric of the dimension \dq{Performance}. \autoref{table:accessibility_metrics} summarizes which metrics from the entire \textit{Accessibility} group can be implemented using SHACL core.
The remaining shapes' definitions can be found in \iftoggle{appendixversion}{Appendix \ref{section:accessibility_appendix}}{Appendix B of~\cite{AppendixVersion_2025}}.

\dq{Performance} refers to how efficiently a system that hosts a large dataset can process data. For this dimension, \BaseSurvey~defined four metrics. 
We showcase the assessment of \dq{Performance} with metric P1 in \shape{\ref{shape:performance_hash_uris_entities}}. P1 was defined by~\cite{Flemming2011} and checks for slash-URIs in datasets with over \numprint{500000} triples. 
According to W3C recommendations, slash-URIs are preferred to identify entities in large graphs because hash (``\#'') URIs can cause performance issues~\cite{CoolURIs_2008}. 
We follow the W3C Best Practices~\cite{BestPractices_2014} and apply this rule to entities' URIs.
Here, \shape{\ref{shape:performance_hash_uris_entities}} targets subjects of triples with predicate \texttt{rdf:type}, applying a regex pattern that does not allow \# to appear in URIs. 
Note that the regex pattern identifies any hash occurrence (not just at the end of the URI) to cover cases such as \url{https://www.example.org\#Example1/}, which still loads all entities with \url{https://www.example.org\#} as base namespace. Therefore, the validation result will output nodes whose URI contains a hash in any part of the URI, not just at the end. 

\noindent
\begin{minipage}{\linewidth}
\begin{lstlisting}[basicstyle=\ttfamily\scriptsize, caption={Performance - Use of Hash URIs in Entities}, label={shape:performance_hash_uris_entities}]
ex:UsageHashURIsShape a sh:NodeShape ;
  sh:targetSubjectsOf rdf:type; 
  sh:or (
    [ sh:path rdf:type; sh:hasValue rdfs:Class; ] [ sh:path rdf:type; sh:hasValue rdf:Property; ] 
    [ sh:path rdf:type; sh:hasValue owl:NamedIndividual; ] [ sh:pattern "^[^#]*$"; ]
  ).
\end{lstlisting}
\end{minipage}

For this metric, assumption \assumption{1} is required; without it, URIs of untyped entities cannot be checked. Assumption \assumption{2} is also needed to restrict the constraint to entities, excluding classes, properties, and named individuals. 
Instances of \texttt{owl:NamedIndividual} are excluded to avoid mixing ontology-level individuals with graph entities, since validation runs over a graph containing both instance data and schema definitions. This ``filtering'' approach is used across the whole study when the metric verifies a constraint across all entities in the graph. 
The DQ measure derived from the validation result is a \RatioMeasure, calculated with \(\frac{\text{\# violations}}{\text{\# entities}}\).

\begin{table}[h!]
\scriptsize
    \centering
    \begin{tabularx}{\textwidth} { 
  | >{\centering\arraybackslash}c 
  | >{\centering\arraybackslash}c 
  | >{\arraybackslash}X 
  | >{\centering\arraybackslash}c |  }
        \hline
        Dimension & Metric Id & Metric & Implemented with SHACL core \\
        \hline
        \multirow{5}{*}{Availability} 
        & A1 & Accessibility of the SPARQL endpoint and the server & \xmark \\ \cline{2-4}
        & A2 & Accessibility of the RDF dumps & \pmark \\ \cline{2-4}
        & A3 & Dereferenceability of the URI & \xmark \\ \cline{2-4}
        & A4 & No misreported content types & \xmark \\ \cline{2-4}
        & A5 & Dereferenced forward-links & \xmark \\ \hline

        \multirow{3}{*}{Licensing}
        & L1 & Machine-readable indication of a license in the VoID description & \cmark \\ \cline{2-4}
        & L2 & Human-readable indication of a license in the documentation & \xmark \\ \cline{2-4}
        & L3 & Specifying the correct license & \xmark \\ \hline

        \multirow{3}{*}{Interlinking}
        & I1 & Detection of good quality interlinks & \xmark \\ \cline{2-4}
        & I2 & Existence of links to external data providers & \cmark \\ \cline{2-4}
        & I3 & Dereferenced back-links & \xmark \\ \hline

        \multirow{2}{*}{Security}
        & S1 & Usage of digital signatures & \cmark \\ \cline{2-4}
        & S2 & Authenticity of the dataset & \cmark \\ \hline
  
        \multirow{4}{*}{Performance}
        & P1 & Usage of slash-URIs & \cmark ~ $*_{1}*_{2}$ \\ \cline{2-4}
        & P2 & Low latency & \xmark \\ \cline{2-4}
        & P3 & High throughput & \xmark \\ \cline{2-4}
        & P4 & Scalability of a data source & \xmark \\ \hline
        \end{tabularx}
    \caption{SHACL core coverage of the \textit{Accessibility} group metrics.}
    \label{table:accessibility_metrics}
\end{table}

\subsection{Intrinsic}\label{section:instrinsic}

The \textit{Intrinsic} category groups dimensions that are independent of the user’s context, and assess whether data accurately (syntactically and semantically), compactly, and completely represents the real world, and whether it is logically consistent. The dimensions in this category are  \dq{Syntactic Validity}, \dq{Semantic Accuracy}, \dq{Consistency}, \dq{Conciseness}, and \dq{Completeness} \cite{Zaveri2012}. In the following, we
exemplarily present the shape definition for a metric of the dimension \dq{Consistency}. \autoref{table:intrinsic_metrics} summarizes which metrics from the \textit{Intrinsic} group can be implemented using SHACL core.
The remaining shapes' definitions can be found in \iftoggle{appendixversion}{Appendix \ref{section:intrinsic_appendix}}{Appendix B of~\cite{AppendixVersion_2025}}.

\dq{Consistency} means that a knowledge base contains no (logical or formal) contradictions according to its knowledge representation and inference mechanisms \cite{Zaveri2012}.
For this dimension, \BaseSurvey~identified 10 metrics. We showcase the assessment of \dq{Consistency} with metric CN5, which checks the correct use of inverse-functional properties. For this metric, \BaseSurvey~discuss two ways to check inverse-functional properties: (i)~verifying the uniqueness of their values and (ii)~defining a SPARQL constraint for such properties. While SHACL core cannot cover~(ii), we defined \shape{\ref{shape:consistency_uniqueness_inverse_functional_property}} to address~(i), which ensures that no two different subjects share the same value, by verifying that each object has only one incoming link.

\noindent
\begin{minipage}{\linewidth}
\begin{lstlisting}[basicstyle=\ttfamily\scriptsize, caption={Consistency - Uniqueness of inverse functional properties}, label={shape:consistency_uniqueness_inverse_functional_property}]
ex:InverseFunctionalPropertyShape a sh:NodeShape ;
  sh:targetObjectsOf PROPERTY_URI; 
  sh:property [ sh:path [ sh:inversePath PROPERTY_URI ]; sh:maxCount 1; ].
\end{lstlisting}
\end{minipage}

\shape{\ref{shape:consistency_uniqueness_inverse_functional_property}} shape is meant to be instantiated by replacing the placeholder \rdfproperty{PROPERTY\_URI} with specific inverse-functional properties defined in the ontology/vocabulary. Shape instantiation is needed in cases where metrics apply to particular classes or properties: instead of defining a separate shape for each one, we define a generic shape with a placeholder and instantiate it with the relevant URI(s) before validation. Moreover, in some cases, shapes may need to be instantiated with domain knowledge (e.g. \shape{\ref{shape:timeliness_entities}} in \autoref{section:contextual}).

The validation report outputs a violation for each property value that is used more than once. Additionally, the DQ measure is a \CompositeMeasure, so we compute an individual score for each property as: 1 if no violations are found, 0 otherwise. The final metric score is then aggregated with the formula: 
\(\frac{\texttt{\# inverse-functional properties correctly used}}{\texttt{\# inverse-functional properties used to instantiate the shape}}\). We consider an inverse-functional property correctly used if the value of its individual score is~1.

\begin{table}[h!]
\scriptsize
    \centering
        \begin{tabularx}{\textwidth} { 
  | >{\centering\arraybackslash}c 
  | >{\centering\arraybackslash}c 
  | >{\raggedright\arraybackslash}X 
  | >{\centering\arraybackslash}c |  }
        \hline
        Dimension & Metric Id & Metric & Implemented with SHACL core \\
        \hline

        \multirow{3}{*}{\shortstack{Syntactic\\validity}}
        & SV1 & No syntax errors of the documents & \xmark \\ \cline{2-4}
        & SV2 & Syntactically accurate values & \pmark ~ $*_{3}$ \\ \cline{2-4}
        & SV3 & No malformed datatype literals & \cmark \\ \hline

        \multirow{5}{*}{\shortstack{Semantic\\accuracy}}
        & SA1 & No outliers & \xmark \\ \cline{2-4}
        & SA2 & No inaccurate values & \pmark ~ $*_{3}$\\ \cline{2-4}
        & SA3 & No inaccurate annotations, labellings or classifications & \pmark ~ $*_{3}$ \\ \cline{2-4}
        & SA4 & No misuse of properties & \xmark \\ \cline{2-4}
        & SA5 & Detection of valid rules & \xmark \\ \hline

        \multirow{10}{*}{Consistency}
        & CN1 & No use of entities as members of disjoint classes & \cmark ~ $*_{2}$ \\ \cline{2-4}
        & CN2 & No misplaced classes or properties & \cmark ~ $*_{1}*_{2}$ \\ \cline{2-4}
        & CN3 & No misuse of \textit{owl:DatatypeProperty} or \textit{owl:ObjectProperty} & \cmark ~ $*_{2}$ \\ \cline{2-4}
        & CN4 & Members of \textit{owl:DeprecatedClass} or \textit{owl:DeprecatedProperty} not used & \cmark ~ $*_{1}*_{2}$ \\ \cline{2-4}
        & CN5 & Valid usage of inverse-functional properties & \pmark ~ $*_{2}$ \\ \cline{2-4}
        & CN6 & Absence of ontology hijacking & \xmark \\ \cline{2-4}
        & CN7 & No negative dependencies/correlation among properties & \pmark~$*_{3}$ \\ \cline{2-4}
        & CN8 & No inconsistencies in spatial data & \xmark \\ \cline{2-4}
        & CN9 & Correct domain and range definition & \cmark ~ $*_{2}$ \\ \cline{2-4}
        & CN10 & No inconsistent values & \pmark ~ $*_{2}$ \\ \hline

        \multirow{3}{*}{Conciseness}
        & CS1 & High intensional conciseness & \xmark \\ \cline{2-4}
        & CS2 & High extensional conciseness & \pmark ~ $*_{3}$ \\ \cline{2-4}
        & CS3 & Usage of unambiguous annotations/labels & \xmark \\ \hline

        \multirow{4}{*}{Completeness}
        & CP1 & Schema completeness & \pmark ~ $*_{1}*_{2}$ \\ \cline{2-4}
        & CP2 & Property completeness & \pmark~$*_{3}$ \\ \cline{2-4}
        & CP3 & Population completeness & \pmark~$*_{3}$ \\ \cline{2-4}
        & CP4 & Interlinking completeness & \pmark ~ $*_{1}*_{2}$ \\ \hline

   \end{tabularx}
    \caption{SHACL core coverage of the \textit{Intrinsic} group metrics.}
    \label{table:intrinsic_metrics}
\end{table}

\subsection{Contextual}\label{section:contextual}

\textit{Contextual} dimensions are those that depend on the specific task at hand or on the context. This group includes four dimensions: \dq{Relevancy}, \dq{Trustworthiness}, \dq{Understandability}, and \dq{Timeliness}.
We exemplarily present the shape definition for a metric of the dimension \dq{Timeliness}. \autoref{table:contextual_metrics} summarizes which metrics from the entire \textit{Contextual} group can be implemented using SHACL core.
The remaining shapes' definitions can be found in \iftoggle{appendixversion}{Appendix \ref{section:contextual_appendix}}{Appendix B of~\cite{AppendixVersion_2025}}. 

\dq{Timeliness} measures how current (or up-to-date) data is in relation to a specific task. For this dimension, \BaseSurvey present two metrics. We illustrate the assessment of \dq{Timeliness} with T1, which verifies the freshness of the dataset based on currency and volatility. This metric uses the formula \(max\{0, 1 - \frac{\text{currency}}{\text{volatility}}\}\), where volatility refers to the length of time the data remains valid, and currency describes the age of the data at the time it is delivered to the user.\iftoggle{appendixversion}{ In this formula, a value of 1 means that the data is up-to-date, while a value of 0 means the data is outdated.}{}
In this case, we are not able to calculate the formula, but we can use SHACL core to identify outdated nodes. Therefore, for the definition of \shape{\ref{shape:timeliness_entities}}, we assume there's some temporal annotation in the data, for example, using properties like \rdfproperty{dcterms:date} or \rdfproperty{dcterms:temporal} from the Dublin Core vocabulary. The defined shape identifies outdated entities by constraining the value of the temporal property to be after a certain point in time, indicating that the node is up-to-date. For this shape, we need to consider all assumptions, given that we are targeting entities. Additionally, we require a vocabulary or ontology to determine which properties are used to annotate entities with temporal facts, and domain knowledge indicating when entities are considered up-to-date. The validation report for this shape outputs a violation for each of the entities whose \texttt{\small dcterms:date} value is older than \texttt{\small DATE\_RANGE\_MIN\_BOUND}. The DQ measure in this case is a \RatioMeasure, calculated as  
\(\frac{\texttt{\#violations}}{\texttt{\#entities}}\).

\noindent
\begin{minipage}{0.8\linewidth}
\begin{lstlisting}[basicstyle=\ttfamily\scriptsize, caption={Timeliness - Outdated entities}, label={shape:timeliness_entities}]
ex:TimelinessEntitiesShape a sh:NodeShape ;
  sh:targetSubjectsOf rdf:type; 
  sh:or ( [ sh:path rdf:type; sh:hasValue rdfs:Class; ]
          [ sh:path rdf:type; sh:hasValue rdf:Property; ]
          [ sh:path rdf:type; sh:hasValue owl:NamedIndividual; ]
          [ sh:path dcterms:date; sh:minInclusive "DATE_RANGE_MIN_BOUND"; ]  ).
\end{lstlisting}
\end{minipage}

\begin{table}[h!]
\scriptsize
    \centering
    \begin{tabularx}{\textwidth} { 
      | >{\centering\arraybackslash}c 
      | >{\centering\arraybackslash}c 
      | >{\arraybackslash}X 
      | >{\centering\arraybackslash}c |  }
        \hline
        Dimension & Metric Id & Metric & Implemented with SHACL core \\ \hline

        \multirow{2}{*}{Relevancy}
        & R1 & Relevant terms within meta-information attributes & \xmark \\ \cline{2-4}
        & R2 & Coverage & \pmark~$*_{3}$ \\ \hline
        
        \multirow{6}{*}{\parbox{1.7cm}{Understan\-dability}}
        & U1 & Human-readable labelling of classes, properties and entities as well as presence of metadata & \pmark ~ $*_{1}*_{2}$ \\ \cline{2-4}
        & U2 & Indication of one or more exemplary URIs & \cmark \\ \cline{2-4}
        & U3 & Indication of a regular expression that matches the URIs of a dataset & \cmark ~ $*_{1}*_{2}$ \\ \cline{2-4}
        & U4 & Indication of an exemplary SPARQL query & \xmark \\ \cline{2-4}
        & U5 & Indication of the vocabularies used in the dataset & \cmark \\ \cline{2-4}
        & U6 & Provision of message boards and mailing lists & \xmark \\ \hline

        \multirow{7}{*}{\parbox{1.7cm}{Trust\-worthiness}}
        & TW1 & Trustworthiness of statements & \xmark \\ \cline{2-4}
        & TW2 & Trustworthiness through reasoning & \pmark ~ $*_{1}*_{2}$ \\ \cline{2-4}
        & TW3 & Trustworthiness of statements, datasets and rules & \pmark ~ $*_{1}*_{2}$ \\ \cline{2-4}
        & TW4 & Trustworthiness of a resource & \xmark \\ \cline{2-4}
        & TW5 & Trustworthiness of the information provider & \pmark ~ $*_{2}*_{3}$ \\ \cline{2-4}
        & TW6 & Trustworthiness of information provided (content trust) & \cmark ~ $*_{1}*_{2}*_{3}$ \\ \cline{2-4}
        & TW7 & Reputation of the dataset & \xmark \\ \hline

        \multirow{2}{*}{Timeliness}
        & T1 & Freshness of datasets based on currency and volatility & \pmark ~ $*_{1}*_{2}*_{3}$ \\ \cline{2-4}
        & T2 & Freshness of datasets based on their data source & \pmark~$*_{3}$\\ \hline

    \end{tabularx}
    \caption{SHACL core coverage of the \textit{Contextual} group metrics.}
    \label{table:contextual_metrics}
\end{table}

\subsection{Representational}\label{section:representational}
The \textit{Representational} group addresses design aspects of the data. The dimensions in this category are \dq{Representational Conciseness}, \dq{Interoperability}, \dq{Versatility}, and \dq{Interpretability} \cite{Zaveri2012}.
We exemplarily present the shape definition for a metric of the dimension \dq{Versatility}. \autoref{table:representational_metrics} summarizes which metrics from the entire \textit{Representational} group can be implemented using SHACL core. The remaining shapes' definitions can be found in \iftoggle{appendixversion}{Appendix \ref{section:representational_appendix}}{Appendix B of~\cite{AppendixVersion_2025}}.

\dq{Versatility} refers to the availability of data in multiple representations and its support for internationalization \cite{Zaveri2012}.
For this dimension, \BaseSurvey~present two metrics. We showcase the assessment of \dq{Versatility} with V2, defined in \cite{Flemming2011}, which checks whether data is available in multiple languages by verifying the use of language tags in literals used for entity labels and descriptions. This metric can be partially covered by SHACL core, as it allows us to check for the presence of language tags on labels and descriptions. However, it does not verify whether the literal values are actually written in the specified language, which would require semantic analysis beyond SHACL’s capabilities. Therefore, for this metric, we defined shapes \shape{\ref{shape:versatility_languages_labels}} and \shape{\ref{shape:versatility_languages_descriptions}}, which check that labels and descriptions in entities have language tags. \iftoggle{appendixversion}{For both shapes, we consider assumptions \assumption{1} and \assumption{2}, as classes, properties, and individuals may have labels or descriptions; however, our focus is on verifying the use of language tags in labels and descriptions of entities. Moreover, we only verify these constraints over entities that have a label/description.}{For both shapes, we consider assumptions \assumption{1} and \assumption{2}, as we verify the use of language tags in entity labels and descriptions. Moreover, constraints are only checked for entities that have a label or description.}

\noindent
\begin{minipage}[t]{0.48\linewidth}
\begin{lstlisting}[basicstyle=\ttfamily\scriptsize, caption={Versatility - Languages in entities labels}, label={shape:versatility_languages_labels}]
ex:DifferentLanguagesLabelsShape a sh:NodeShape ;
  sh:targetSubjectsOf rdfs:label; 
  sh:or (
    [sh:path rdf:type; sh:hasValue rdfs:Class;]
    [sh:path rdf:type; sh:hasValue rdf:Property;]
    [sh:path rdf:type; sh:hasValue owl:NamedIndividual;]
    [sh:path rdfs:label; sh:datatype rdf:langString;]
  ).
  
\end{lstlisting}
\end{minipage}
\hfill
\begin{minipage}[t]{0.49\linewidth}
\begin{lstlisting}[basicstyle=\ttfamily\scriptsize, caption={Versatility - Languages in entities descriptions}, label={shape:versatility_languages_descriptions}]
ex:DifferentLanguagesDescriptionsShape a sh:NodeShape;
  sh:targetSubjectsOf rdfs:comment; 
  sh:or (
    [sh:path rdf:type; sh:hasValue rdfs:Class;]
    [sh:path rdf:type; sh:hasValue rdf:Property;]
    [sh:path rdf:type; sh:hasValue owl:NamedIndividual;]
    [sh:path rdfs:comment; sh:datatype rdf:langString;]
  ).
\end{lstlisting}
\end{minipage}

The validation report for \shape{\ref{shape:versatility_languages_labels}} and \shape{\ref{shape:versatility_languages_descriptions}} outputs a violation for each entity that has a label or description without a language tag, respectively. In both cases, the DQ measure is a \RatioMeasure, calculated using the formulas {\scriptsize\(\frac{\texttt{\# violations}}{{\texttt{\# entities with labels}}}\)} and {\scriptsize\(\frac{\texttt{\# violations}}{{\texttt{\# entities with descriptions}}}\)}, respectively.

\begin{table}[h!]
\scriptsize
    \centering
    \begin{tabularx}{\textwidth} { 
      | >{\centering\arraybackslash}c 
      | >{\centering\arraybackslash}c 
      | >{\arraybackslash}X 
      | >{\centering\arraybackslash}c |  }
        \hline
        Dimension & Metric ID & Metric & Implemented with SHACL core \\ \hline

        \multirow{2}{*}{\shortstack{Representational\\conciseness}}
        & RC1 & Keeping URIs short & \cmark ~ $*_{1}*_{2}$ \\ \cline{2-4}
        & RC2 & No use of prolix RDF features & \cmark ~ $*_{1}*_{2}$ \\ \hline

        \multirow{2}{*}{Interoperability}
        & ITO1 & Re-use of existing terms & \cmark ~ $*_{3}$ \\ \cline{2-4}
        & ITO2 & Re-use of existing vocabularies & \pmark ~ $*_{3}$ \\ \hline

        \multirow{2}{*}{Versatility}
        & V1 & Provision of the data in different serialization formats & \cmark \\ \cline{2-4}
        & V2 & Checking whether data is available in different languages & \pmark ~$*_{1}*_{2}$ \\ \hline

        \multirow{4}{*}{Interpretability}
        & ITP1 & Use of self-descriptive formats & \cmark ~ $*_{1}*_{2}$ \\ \cline{2-4}
        & ITP2 & Detecting the interpretability of data & \pmark ~ $*_{3}$ \\ \cline{2-4}
        & ITP3 & Invalid usage of undefined classes and properties & \cmark ~ $*_{2}$ \\ \cline{2-4}
        & ITP4 & No misinterpretation of missing values & \cmark ~ $*_{1}*_{2}$ \\ \hline

    \end{tabularx}
    \caption{SHACL core coverage of the \textit{Representational} group metrics.}
    \label{table:representational_metrics}
\end{table}

\section{Prototype}\label{sec:prototype}
This section presents our prototype for DQA using SHACL core. It is built with the Python libraries \textit{rdflib} and \textit{PySHACL} and provides a Streamlit dashboard to visualize results. The code is available on GitHub at \cite{shacl-dqa-prototype}.

\smallskip\noindent\textbf{Overview of the DQ assessment process.}
Our DQA process takes as input the data graph to be evaluated, optionally a metadata file (either the VoID or DCAT description of the graph), and an ontology, along with a set of vocabularies used in the data graph. 
Without an ontology, vocabularies, or a metadata file, fewer shapes are instantiated, and some, such as those targeting \texttt{void:Dataset} or those checking for class/property labels, are not validated, as they rely on these artifacts. 
A configuration file allows users to customize preferred properties needed for instantiating SHACL shapes.
\iftoggle{appendixversion}{ We use, for example,  \texttt{rdf:type}, \texttt{rdfs:label}, \texttt{rdfs:comment}, and \texttt{owl:sameAs} as default properties for defining shapes, but users may change this.}{}

\begin{figure}[ht]
    \centering
    \includegraphics[width=\linewidth]{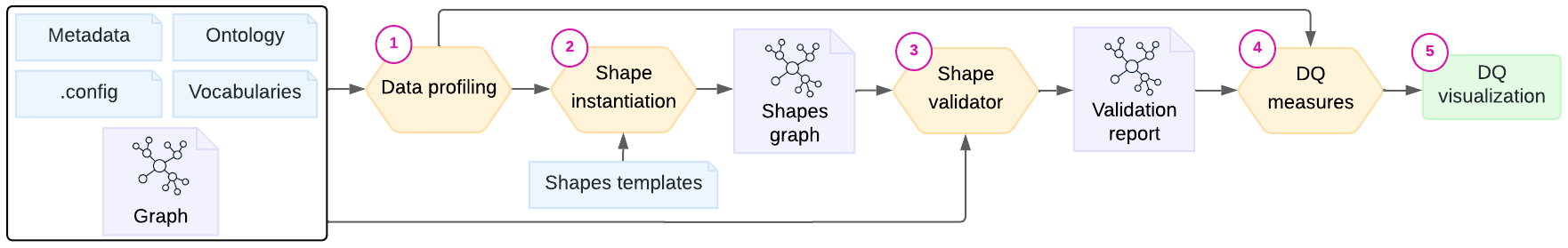}
    \caption{SHACL-based data quality assessment architecture.}
    \label{fig:shape_instantiation_architecture}
\end{figure}

\autoref{fig:shape_instantiation_architecture} shows the architecture of the prototype.
The DQA process begins by Step~(1) profiling the graph, ontology, and vocabularies to obtain the necessary information for shape instantiation and metric calculation. For example,  \shape{\ref{shape:consistency_uniqueness_inverse_functional_property}} is instantiated with properties marked as \texttt{owl:InverseFunctionalProperty} in the ontology; while to compute the measure associated with \shape{\ref{shape:performance_hash_uris_entities}}, we retrieve the total number of entities.

After obtaining this information, we (2) instantiate the shapes' templates and generate the shapes graph. Our prototype stores SHACL shapes as reusable templates with variables 
that are replaced with actual values during instantiation -- a process where the template is populated with specific properties from either the graph profiling results or the configuration file of the dataset. For example, shape \shape{\ref{shape:performance_hash_uris_entities}} uses the type property specified in the configuration file to create a concrete shape from its template form. Moreover, shape \shape{\ref{shape:consistency_uniqueness_inverse_functional_property}} is instantiated with inverse-functional properties used in the dataset, obtained from the graph profiling results. Before shape validation, we perform a pre-processing step, based on assumption \assumption{2}, to enrich the data graph with necessary triples\iftoggle{appendixversion}{. Specifically, we add type declarations for OWL-defined properties and classes defined in ontology/vocabularies, marking each property as an \texttt{rdf:Property} and each class as an \texttt{rdfs:Class}. We also explicitly type instances in vocabularies/ontology as \texttt{owl:NamedIndividual} to distinguish data entities from schema elements. For terms defined as \texttt{rdfs:Datatype}, we type them as \texttt{rdfs:Class}, in line with the RDFS entailment rule \textit{rdfs11}. Additionally, when multiple vocabularies are reused, we include all relevant axioms in the data graph, as stated by the SHACL specification~\cite{w3c_shacl}.
}{ (described in detail in Section 4 of~\cite{AppendixVersion_2025}).} Then, we use the validator~(3) provided by the PySHACL library to validate the shapes against the data graph, ontology, vocabularies, or metadata file, depending on the shape. Once the validation is completed, we (4)~calculate the DQ measures from the validation result, which are later stored in a CSV file. Finally, the results can be visualized in a dashboard~(5).

\smallskip\noindent\textbf{Shape instantiation details.} For the 69 metrics defined in \BaseSurvey, we defined 64~\emph{shapes}. The mapping is not one-to-one: some metrics have no shape, others (e.g., V2 in \autoref{section:representational}) have multiple. We identified 38 shapes that could be instantiated without domain expert input, and excluded five more: one requiring merged graphs (SH11) and 4 using non-standard vocabularies (S13, S14, S53, S55) (see~\iftoggle{appendixversion}{ \autoref{section:Appendix_Shapes_Definition}}{Appendix B of~\cite{AppendixVersion_2025}}). Of the 38, 11 rely on vocabulary or ontology terms, and are instantiated only with those found in the data graph (e.g., \shape{\ref{shape:consistency_uniqueness_inverse_functional_property}} uses only inverse-functional properties present in the data). This avoids generating unnecessary shapes, as many data graphs may partially reuse vocabularies. 
The exceptions are the shapes for metrics CN2 and CP1: CN2 checks for property/class misuse by instantiating with all available classes and properties since misused ones do not appear in profiling results, while CP1 uses all defined classes in the vocabulary to check if they are used. See \iftoggle{appendixversion}{Appendix \ref{section:evaluation_appendix}}{Appendix C of~\cite{AppendixVersion_2025}} for additional aspects of shape instantiation aimed at improving runtime efficiency.

\smallskip\noindent\textbf{Evaluation.} We evaluated the prototype with three datasets from the LOD Cloud\iftoggle{appendixversion}{\footnote{All datasets are available at \url{https://zenodo.org/records/16644385}}}{~\cite{cortes2025shacl_datasets}}:  \textit{Temples of the Classical World}
(\numprint{15326} triples and \numprint{1363} entities), \textit{DBTunes - John Peel Sessions}
(\numprint{271369} triples and \numprint{76056} entities), and \textit{Drugbank}
(\numprint{3646181} triples and \numprint{316555} entities). Validation time ranged from 40~seconds to 3~hours for 500–740 instantiated shapes, depending on the dataset. Longer times are likely due to both more triples and more shapes. As the prototype aimed to test the SHACL-based DQA approach, performance aspects were left outside its scope, and no runtime experiments were conducted since validation relied on an external library.
\iftoggle{appendixversion}{
\autoref{table:dqa_results_temples} presents the assessment results for the dataset \textit{Temples of the Classical World} in detail. The results for the other datasets are provided on GitHub at \cite{shacl-dqa-prototype}.
While \textit{Temples of the Classical World} demonstrates strengths in aspects such as URI design and labeling, it exhibits several quality issues, particularly in metadata aspects, and correct vocabulary usage.

\begin{table}[h!]
\scriptsize
    \centering
    \begin{tabularx}{\textwidth} { 
  | >{\centering\arraybackslash}p{1.9cm} 
  | >{\centering\arraybackslash}X 
  | >{\centering\arraybackslash}p{1.5cm} |
  }
        \hline
        Dimension & DQA results & Metrics \\
        \hline

        Availability & No machine-readable indication of an RDF dump, via \texttt{void:dataDump} & A2 in \autoref{table:accessibility_metrics} \\ \hline

        Performance & All entities use slash uris & P1 \autoref{table:accessibility_metrics} \\ \hline

        Understandability & Includes title, description, homepage, and an example URI. Missing: \texttt{void:uriRegexPattern}, \texttt{void:uriSpace}, and \texttt{void:vocabulary}. Label completeness: entities (99\%), FOAF (classes: 93\%, properties: 100\%), GN (classes: 100\%, properties: 90\%), LAWD (classes: 6.7\%, properties: 0\%). SKOS missing 20\% of class labels & U1, U2, U3 and U5 in \autoref{table:contextual_metrics} \\ \hline

        Security & Includes contributor/publisher/creator, but lacks provenance (dcterms:source, dcterms:provenance) & S2 in \autoref{table:accessibility_metrics}. \\ \hline

        Licensing & No indication of a machine-readable license (via \texttt{dcterms:license}) & L1 in \autoref{table:accessibility_metrics}. \\ \hline

        Interlinking & 100\% of interlinked instances refer to external providers & I2 in \autoref{table:accessibility_metrics}. \\ \hline

        Versatility & No indication of serialization formats (via \texttt{void:feature}), and no language tags in labeled entities & V1 and V2 in \autoref{table:representational_metrics} \\ \hline

        Representational Conciseness & No entities are instances of prolix constructs (e.g., \texttt{rdf:List}, \texttt{rdf:Seq}). All URIs are short ($\leq$ 80 characters (\url{https://www.w3.org/TR/chips/})) and do not use parameters & RC1 and RC2 in \autoref{table:representational_metrics} \\ \hline

        Interpretability & All entities are identified with URIs, and only 40\% of properties use IRIs as values. A misuse of \texttt{void:dataDump} as a class triggers a violation for the metric ITP3. No blank nodes are used & ITP1, ITP3 and ITP4 in \autoref{table:representational_metrics} \\ \hline

        Consistency & \texttt{void:dataDump} is misused as a class; all object properties and inverse functional properties are correctly used. Only 20\% of properties with defined ranges and 33\% with defined domains are used consistently & CN2, CN3, CN5, CN9 in \autoref{table:intrinsic_metrics} \\ \hline

        Completeness & 59.8\% interlinking completeness and 1.2\% schema completeness & CP4 and CP1 in \autoref{table:intrinsic_metrics} \\ \hline
        
    \end{tabularx}
    \caption{DQA results for the dataset \textit{Temples of the Classical World}.}
    \label{table:dqa_results_temples}
\end{table}
}{Section 4 of~\cite{AppendixVersion_2025} presents detailed results for the \textit{Temples of the Classical World} dataset, while results for other datasets are available on GitHub at \cite{shacl-dqa-prototype}}

    
    


\section{Suitability of SHACL core to assess RDF data quality}
\label{sec:discussion}
We now discuss the suitability of SHACL core to assess RDF DQ. \autoref{sec:disc-coverage} highlights where SHACL core falls short in covering certain DQ dimensions -- metrics not mentioned are considered covered (see \autoref{sec:approach} and \iftoggle{appendixversion}{\autoref{section:Appendix_Shapes_Definition}}{Appendix B of~\cite{AppendixVersion_2025}}).
We then examine SHACL core’s strengths and limitations (\autoref{sec:disc-limitations}), and discuss SHACL core's extensions (\autoref{sec:disc-shacl-extensions}).

\begin{figure*}
    \centering
    \includegraphics[width=\linewidth]{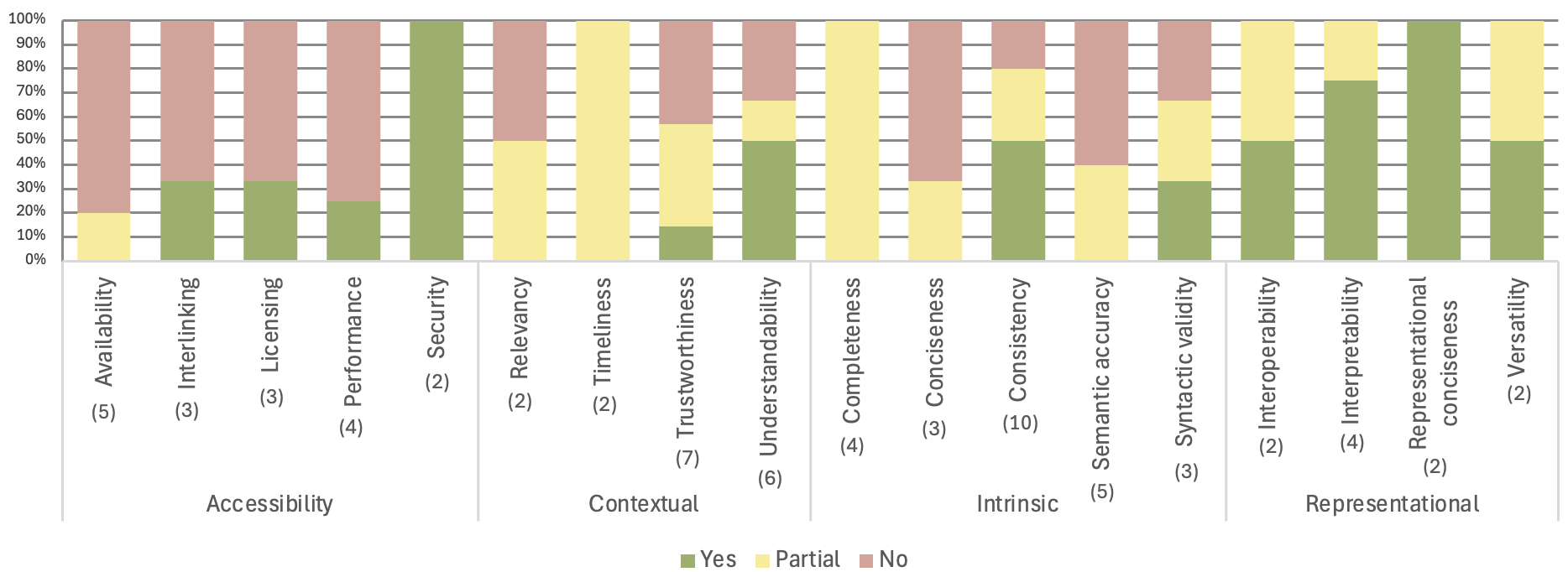}
    \caption{Coverage of DQ dimensions by SHACL core (including number of metrics for each dimension).}
    \label{fig:shacl_coverage_dimensions}
\end{figure*}

\subsection{Coverage of DQ dimensions by SHACL core}
\label{sec:disc-coverage}

For each of the 18 DQ dimensions discussed in~\BaseSurvey, we present SHACL core coverage as the normalized sum of coverage values assigned to each DQ metric: 1 for full coverage, 0.5 for partial coverage, and 0 for no coverage. The resulting sum is divided by the total number of metrics to obtain an overall coverage percentage.
Figure~\ref{fig:shacl_coverage_dimensions} shows these results, where two dimensions with 100\% coverage stand out: \dq{Representational Conciseness} and \dq{Security}. In these cases, the implemented metrics check for the presence or use of specific properties or classes. 

Seven dimensions present a coverage between 50-90\% : \dq{Timeliness}, \dq{Understandability}, \dq{Consistency}, \dq{Completeness}, \dq{Syntactic Validity}, \dq{Interoperability}, \dq{Interpretability}, and \dq{Versatility}. 
Regarding \dq{Timeliness}, SHACL core partially covers both of the proposed metrics. While it cannot directly compute the formulas these metrics rely on, like T1’s formula or T2’s time-distance calculation, it can still check related aspects. For example, for T1, we defined a shape that detects outdated entities via temporal annotations, and for T2, the shape verifies if the dataset is up to date, even though we cannot compute distances with SHACL core.

In the context of \dq{Understandability}, \dq{Versatility}, and \dq{Interpretability}, SHACL only partially covers metrics U1, V2, and ITP2 due to its limitations to do ``semantic'' checks. It can confirm the presence of labels (U1) and language tags (V2, ITP2), but it cannot assess the readability of labels or whether literals align with their specified language.  Additionally, for \dq{Understandability}, SHACL core cannot cover metrics U4 and U6. U4 is not covered because there’s no standard method to declare example SPARQL queries. While datasets might use properties like \rdfproperty{rdfs:comment} to include such queries, this is not standard and would require language processing to verify. U6 involves checking external resources (e.g., mailing lists), which SHACL core cannot handle since it only works on RDF graphs.

In terms of \dq{Consistency}, several metrics rely on SPARQL queries or reasoning, which are beyond the capabilities of SHACL core (specifically metrics CN5, CN6, and CN10). Additionally, SHACL core does not cover CN8, as it requires specialized knowledge related to the representation of spatial data.

In the case of \dq{Completeness}, all metrics are partially covered. CP1 measures schema completeness, but SHACL core can only check class usage, not property usage, since SHACL core cannot target triples in the graph to check whether a property is used or not. CP2 defines two aspects for measuring property completeness, where the second one uses property distribution statistics to assess completeness, which SHACL core does not support. CP3 measures population completeness; while SHACL core can check for cardinality and allowed values, it cannot leverage semantic constructs explicitly stated in the graph (e.g., recognizing equivalent entities defined using \rdfproperty{owl:sameAs}). Finally, CP4 (interlinking completeness) is partially covered: SHACL core cannot verify linkset completeness because it involves checking the existence of links between instances of equivalent classes. However, SHACL node shapes require a specific target, while this check verifies the existence of certain triples in the graph and is not associated with any particular node.

When it comes to \dq{Syntactic Validity}, SHACL core cannot cover SV1 since it requires checking syntax errors of RDF documents, while SHACL works after parsing the RDF document. SV2 is only partially covered because some aspects of this metric involve complex techniques like clustering.

For \dq{Interoperability}, SHACL core does not cover all metrics due to its limited target capabilities. In particular, ITO2 requires checking vocabulary usage. While SHACL can check the usage of classes, it cannot check property usage, as this would entail checking if a property is used in any triple, but SHACL cannot target the predicate position in triples.

Finally, we turn to dimensions that are covered below 50\% (\dq{Availability}, \dq{Interlinking}, \dq{Licensing} \dq{Performance}, \dq{Relevancy}, \dq{Trustworthiness}, \dq{Conciseness}, and \dq{Semantic accuracy}).

For \dq{Availability}, most of the metrics cannot be covered with SHACL core, as most require accessing resources on the web, such as checking the dereferenceability of URIs or detecting broken links. 

In the case of \dq{Interlinking}, SHACL core cannot cover I1, as it requires computing graph-based measures (e.g., interlinking degree), which cannot be expressed as constraints over nodes and their properties. I3 also cannot be covered: SHACL cannot dereference URIs (task 1), nor verify whether there is any triple with the resource as the object (in-links) (task 2). For task 2, SHACL core also falls short because, while it can check for the existence of values for specific properties using \texttt{sh:path}, it cannot verify the existence of any triple where the resource is the object, regardless of the predicate.

As for \dq{Licensing}, the metric L2 cannot be covered with SHACL core since it entails verifying the existence of a human-readable license in the documentation of the dataset, which usually is an HTML document.
For metric L3, it also cannot be covered with SHACL core, as it requires checking license clauses and comparing licenses between datasets, which involves natural language processing.


Regarding \dq{Performance}, SHACL core cannot cover metrics (P2 - P4), as they describe system-level behaviors (e.g., low-latency or high throughput), rather than constraints on nodes and properties.

For \dq{Relevancy}, R1 is not covered, as it requires identifying relevant data via ranking or crowd-sourcing, which SHACL, as a constraint language, does not support. R2 is only partially covered, since it entails measuring the coverage (i.e., number of entities) and level of detail of entities (i.e., number of properties) in the dataset, to ensure that the data is appropriate for the task at hand. While SHACL core does not provide a way of counting entities and properties, we were able to define a shape that states the expected properties (level of detail) for instances of a certain class. In the validation results, we obtain entities lacking the expected level of detail.

Regarding \dq{Trustworthiness}, SHACL core fully covers one metric, partially covers three, and cannot cover the remaining three. Metrics TW2 and TW3 require annotating the data with trust values, for example, using a trust ontology. While SHACL core cannot annotate the data, it can check the presence of these annotations, so these are partially covered. TW5 can only be partially covered since one of the aspects of this metric states checking the trustworthiness of the information provider using decision networks, which SHACL core cannot handle. To conclude, the metrics that cannot be covered for this dimension require trust value computations (TW1, TW4, and TW7) or human input (TW7), both beyond SHACL's capabilities.

For \dq{Conciseness}, SHACL core is not able to cover the metric CS1, as it requires identifying semantically equivalent properties or classes. CS2 is only partially covered, as one of its approaches requires identifying duplicated entities (i.e., entities with different URIs but similar or identical property values), which involves cross-entity comparisons and similarity measures, both unsupported by SHACL core. Finally, CS3 cannot be covered by SHACL core, as detecting ambiguous labels and annotations requires semantic interpretation beyond SHACL’s capabilities.

Finally, for \dq{Semantic Accuracy}, three metrics (SA1, SA4, and SA5) are not covered, since they rely on outlier detection (SA1), profiling (SA4), and association rule generation via induction and analogy methods (SA5) - all of which are unsupported by SHACL core. The other two (SA2 and SA3) are only partially covered. SA2 specifies three checks to detect inaccurate values; the third (i.e., validating functional dependencies) is not supported by SHACL core, as it requires comparing property values for triples that do not share the same subject. Metric SA3 checks for inaccurate labels and classifications. We defined a SHACL shape to verify labels and types against a list of allowed values, but we cannot compute the original metric defined in \BaseSurvey~since it requires similarity measures, and SHACL core supports only exact value matching.

\subsection{Strengths and Limitations of SHACL core}
\label{sec:disc-limitations}

SHACL core has many strengths for specific aspects of DQA\@.
For example, it can perform syntactic validation, including pattern validation, datatype checks, and verifying whether values fall within specified ranges or lists. It can validate consistency between the data and vocabulary or ontology definitions, such as ensuring correct use of a property’s domain or range. It can also support best practices for publishing Linked Data, such as the use of hash URIs for entities, labels in resources (i.e., entities, classes, and properties), and some characteristics of URI designs (i.e., short URIs and no parameters). Moreover, SHACL can verify the correct application of property characteristics such as irreflexive, functional, asymmetric, and inverse-functional. Finally, it is well-suited for defining the expected structure of class instances, specifying required properties, their types, and other basic constraints~\cite{Polleres_2023}.

However, its applicability to DQA is still limited in several ways:

\noindent\textbf{Lack of External Access.} SHACL core operates solely within the RDF graph and cannot access external web resources or perform system-level checks. This restricts its applicability to metrics related to \dq{Availability} and \dq{Performance}, which often require testing endpoints, dereferencing URIs, or measuring response times.

\noindent\textbf{Node-Centric Scope.} Constraints in SHACL core are evaluated for individual \textit{focus nodes} and their immediate property values. This node-centric approach limits the ability to compute network-level measures, which are essential for evaluating the \dq{Interlinking} dimension (covered by SHACL-SPARQL).

\noindent\textbf{No Cross-Entity Comparison.} SHACL core lacks mechanisms to compare property values across different entities in the graph. As a result, it cannot assess functional dependencies or detect duplicated entities, limiting its applicability for \dq{Semantic Accuracy} and \dq{Conciseness} (covered by SHACL-SPARQL).

\noindent\textbf{No Arithmetic or Dynamic Expressions.} SHACL core lacks support for arithmetic operations and dynamic expressions. For instance, it cannot evaluate conditions like {\small\texttt{age = now() - birthDate}}. This limits the applicability of SHACL core for metrics that depend on computed values, such as \dq{Timeliness} and \dq{Trustworthiness} (covered by SHACL-SPARQL).

\noindent\textbf{Limited Semantic Awareness.} SHACL core lacks mechanisms to verify semantic aspects of the data, which are crucial for dimensions like \dq{Semantic Accuracy}, \dq{Understandability}, and \dq{Versatility}. Additionally, it cannot take advantage of semantic declarations present in the graph, such as entity alignments defined through properties like {\small\texttt{owl:sameAs}}.

\noindent\textbf{Restricted Targeting Mechanism.} SHACL provides only a small set of predefined target types (e.g., {\small \texttt{sh:targetClass}, \texttt{sh:targetNode}}). This makes it difficult to define metrics that need to assess patterns over arbitrary triples or general predicate-based constraints (covered by SHACL-SPARQL).

\subsection{SHACL extensions}
\label{sec:disc-shacl-extensions}

The SHACL language consists of two main components: SHACL core and SHACL-SPARQL. Additionally, there exist non-standard SHACL extensions such as the \textit{DASH Data Shapes Vocabulary}~\cite{dash_shapes} and \textit{SHACL Advanced Features}~\cite{shacl_af}, all of which rely heavily on SPARQL. This paper focuses on SHACL core for several reasons.
First, studies that extract constraints from RDF graphs primarily use SHACL core components. Notably, research by \citet{SpahiuMP18} and \citet{Rabbani2023_shactor} indicate that SHACL can enhance DQ\@. Thus, we aim to determine if SHACL core alone is sufficient for this purpose.
Second, while SHACL can be implemented in various programming languages (e.g., \textit{jena-shacl} in Java and \textit{PySHACL} in Python), each with their own optimizations, once SPARQL-based constraints are used, performance depends on the underlying query engine. Furthermore, while the SHACL recommendation details its extension via SPARQL, other languages could be used.
Lastly, we choose to focus on SHACL core because it is typically easier to understand and use than writing complete SPARQL queries, allowing for more intuitive formulation of constraints accessible to domain experts~\cite{Hartmann2016}. \iftoggle{appendixversion}{Appendix \ref{section:extensions_appendix}}{Appendix D of~\cite{AppendixVersion_2025}}~provides more details on SHACL extensions.


\section{Conclusion and Future Work}
\label{sec:conclusion}
In this paper, we assess the suitability of SHACL for DQA by defining shapes for the 69 data quality (DQ) metrics identified by \BaseSurvey, whenever possible. We also developed a prototype to automatically instantiate and validate the defined shapes, and compute DQ measures from the validation results.

Our findings indicate that SHACL is well-suited for syntactic validation (pattern validation, datatype checks, and value range enforcement), structural validation of class instances, ensuring correct use of properties and classes, and enforcing linked data best practices. This makes SHACL particularly useful for assessing dimensions such as \dq{Syntactic Validity}, \dq{Interpretability}, \dq{Security}, \dq{Representational Conciseness}, and specific aspects of \dq{Consistency}, \dq{Versatility}, and \dq{Understandability}. However, SHACL core has several limitations: it cannot access external resources, perform cross-entity comparisons, support network-based measures, or assess data semantics. This limits its use for \dq{Availability}, \dq{Conciseness}, \dq{Interlinking}, and \dq{Semantic Accuracy}. Its lack of dynamic calculations and limited targeting also limits its applicability for \dq{Interoperability}, \dq{Trustworthiness}, and \dq{Timeliness}.

The DQ metrics defined by \BaseSurvey~may benefit from refinement, as some are overly specific (e.g., those that require spatial data representations) and difficult to generalize, limiting their practical use in KG DQA. Moreover, new metrics (e.g., bias~\cite{KraftU22}) are not covered in the survey; we suggest extending this work to test whether SHACL core can cover them.

Assessing DQ goes beyond checking constraints, encompassing the data source, the system processing the data, downstream tasks, and human factors~\cite{Mohammed_2025}. SHACL core focuses on data graphs, limiting its applicability for comprehensive DQA. Future work should explore hybrid approaches combining SHACL’s constraint checking with tools that access external resources and LLMs to overcome its limited semantic awareness. It would also be useful to identify which limitations are inherent to SHACL core design that can be addressed through extensions.



\bibliography{bibliography}

\begin{thebibliography}{65}
\expandafter\ifx\csname natexlab\endcsname\relax\def\natexlab#1{#1}\fi
\providecommand{\url}[1]{\texttt{#1}}
\providecommand{\href}[2]{#2}
\providecommand{\path}[1]{#1}
\providecommand{\DOIprefix}{doi:}
\providecommand{\ArXivprefix}{arXiv:}
\providecommand{\URLprefix}{URL: }
\providecommand{\Pubmedprefix}{pmid:}
\providecommand{\doi}[1]{\href{http://dx.doi.org/#1}{\path{#1}}}
\providecommand{\Pubmed}[1]{\href{pmid:#1}{\path{#1}}}
\providecommand{\bibinfo}[2]{#2}
\ifx\xfnm\relax \def\xfnm[#1]{\unskip,\space#1}\fi
\bibitem[{Zaveri et~al.(2012)Zaveri, Rula, Maurino, Pietrobon, Lehmann, and Auer}]{Zaveri2012}
\bibinfo{author}{A.~Zaveri}, \bibinfo{author}{A.~Rula}, \bibinfo{author}{A.~Maurino}, \bibinfo{author}{R.~Pietrobon}, \bibinfo{author}{J.~Lehmann}, \bibinfo{author}{S.~Auer},
\newblock \bibinfo{title}{{Quality assessment for Linked Data: A Survey}},
\newblock \bibinfo{journal}{Semantic Web} \bibinfo{volume}{7} (\bibinfo{year}{2012}) \bibinfo{pages}{63--93}. \DOIprefix\doi{10.3233/SW-150175}.
\bibitem[{Suchanek et~al.(2024)Suchanek, Alam, Bonald, Chen, Paris, and Soria}]{Suchanek2024}
\bibinfo{author}{F.~M. Suchanek}, \bibinfo{author}{M.~Alam}, \bibinfo{author}{T.~Bonald}, \bibinfo{author}{L.~Chen}, \bibinfo{author}{P.~H. Paris}, \bibinfo{author}{J.~Soria},
\newblock \bibinfo{title}{Yago 4.5: A large and clean knowledge base with a rich taxonomy},
\newblock \bibinfo{journal}{Proceedings of the International Conference on Information retrieval (SIGIR)}  (\bibinfo{year}{2024}) \bibinfo{pages}{131--140}. \DOIprefix\doi{10.1145/3626772.3657876}.
\bibitem[{Fur(2018)}]{Furber2018}
\bibinfo{title}{{Linked data quality of DBpedia, Freebase, OpenCyc, Wikidata, and YAGO}},
\newblock \bibinfo{journal}{Semantic Web} \bibinfo{volume}{9} (\bibinfo{year}{2018}) \bibinfo{pages}{77--129}. \DOIprefix\doi{10.3233/SW-170275}.
\bibitem[{Singhal(2012)}]{GoogleKG}
\bibinfo{author}{A.~Singhal}, \bibinfo{title}{Introducing the knowledge graph: things, not strings}, \bibinfo{howpublished}{\url{https://blog.google/products/search/introducing-knowledge-graph-things-not/}}, \bibinfo{year}{2012}. \bibinfo{note}{Accessed: 2025-07-29}.
\bibitem[{Developer(2023)}]{AmazonAlexaKG}
\bibinfo{author}{A.~Developer}, \bibinfo{title}{Alexa entities reference}, \bibinfo{howpublished}{\url{https://developer.amazon.com/en-US/docs/alexa/custom-skills/alexa-entities-reference.html}}, \bibinfo{year}{2023}. \bibinfo{note}{Accessed: 2025-07-29}.
\bibitem[{Vrande{\v{c}}i{\'c} and Kr{\"o}tzsch(2014)}]{Wikidata}
\bibinfo{author}{D.~Vrande{\v{c}}i{\'c}}, \bibinfo{author}{M.~Kr{\"o}tzsch},
\newblock \bibinfo{title}{Wikidata: A free collaborative knowledgebase},
\newblock \bibinfo{journal}{Communications of the ACM} \bibinfo{volume}{57} (\bibinfo{year}{2014}) \bibinfo{pages}{78--85}.
\bibitem[{Auer et~al.(2007)Auer, Bizer, Kobilarov, Lehmann, Cyganiak, and Ives}]{DBpedia}
\bibinfo{author}{S.~Auer}, \bibinfo{author}{C.~Bizer}, \bibinfo{author}{G.~Kobilarov}, \bibinfo{author}{J.~Lehmann}, \bibinfo{author}{R.~Cyganiak}, \bibinfo{author}{Z.~Ives},
\newblock \bibinfo{title}{Dbpedia: A nucleus for a web of open data},
\newblock in: \bibinfo{booktitle}{The Semantic Web}, \bibinfo{publisher}{Springer}, \bibinfo{year}{2007}, pp. \bibinfo{pages}{722--735}.
\bibitem[{Suchanek et~al.(2007)Suchanek, Kasneci, and Weikum}]{YAGO}
\bibinfo{author}{F.~M. Suchanek}, \bibinfo{author}{G.~Kasneci}, \bibinfo{author}{G.~Weikum},
\newblock \bibinfo{title}{Yago: A core of semantic knowledge},
\newblock in: \bibinfo{booktitle}{Proceedings of the 16th international conference on World Wide Web}, \bibinfo{organization}{ACM}, \bibinfo{year}{2007}, pp. \bibinfo{pages}{697--706}.
\bibitem[{Tuozzo(2025)}]{Tuozzo2025}
\bibinfo{author}{G.~Tuozzo},
\newblock \bibinfo{title}{Navigating the lod subclouds: Assessing linked open data quality by domain},
\newblock in: \bibinfo{booktitle}{Companion Proceedings of the {ACM} on Web Conference}, \bibinfo{publisher}{ACM}, \bibinfo{year}{2025}, pp. \bibinfo{pages}{2141--2148}. \DOIprefix\doi{10.1145/3701716.3717569}.
\bibitem[{Hogan et~al.(2022)Hogan, Gutierrez, Cochez, de~Melo, Kirrane, Polleres, Navigli, Ngomo, Rashid, Schmelzeisen, Staab, Blomqvist, d’Amato, Gayo, Neumaier, Rula, Sequeda, and Zimmermann}]{Hogan2022}
\bibinfo{author}{A.~Hogan}, \bibinfo{author}{C.~Gutierrez}, \bibinfo{author}{M.~Cochez}, \bibinfo{author}{G.~de~Melo}, \bibinfo{author}{S.~Kirrane}, \bibinfo{author}{A.~Polleres}, \bibinfo{author}{R.~Navigli}, \bibinfo{author}{A.-C.~N. Ngomo}, \bibinfo{author}{S.~M. Rashid}, \bibinfo{author}{L.~Schmelzeisen}, \bibinfo{author}{S.~Staab}, \bibinfo{author}{E.~Blomqvist}, \bibinfo{author}{C.~d’Amato}, \bibinfo{author}{J.~E.~L. Gayo}, \bibinfo{author}{S.~Neumaier}, \bibinfo{author}{A.~Rula}, \bibinfo{author}{J.~Sequeda}, \bibinfo{author}{A.~Zimmermann},
\newblock \bibinfo{title}{{Knowledge Graphs}}  (\bibinfo{year}{2022}). \DOIprefix\doi{10.1007/978-3-031-01918-0}.
\bibitem[{Angles(2018)}]{Angles2018}
\bibinfo{author}{R.~Angles},
\newblock \bibinfo{title}{{The Property Graph Database Model}},
\newblock in: \bibinfo{booktitle}{Proceedings of the 12th Alberto Mendelzon International Workshop on Foundations of Data Management}, \bibinfo{year}{2018}.
\bibitem[{Wood et~al.(2014)Wood, Lanthaler, and Cyganiak}]{RDFconcepts}
\bibinfo{author}{D.~Wood}, \bibinfo{author}{M.~Lanthaler}, \bibinfo{author}{R.~Cyganiak}, \bibinfo{title}{{RDF 1.1 Concepts and Abstract Syntax}}, \bibinfo{year}{2014}.
\bibitem[{Polleres et~al.(2023)Polleres, Pernisch, Bonifati, Dell'Aglio, Dobriy, Dumbrava, Etcheverry, Ferranti, Hose, Jim{\'e}nez-Ruiz et~al.}]{Polleres_2023}
\bibinfo{author}{A.~Polleres}, \bibinfo{author}{R.~Pernisch}, \bibinfo{author}{A.~Bonifati}, \bibinfo{author}{D.~Dell'Aglio}, \bibinfo{author}{D.~Dobriy}, \bibinfo{author}{S.~Dumbrava}, \bibinfo{author}{L.~Etcheverry}, \bibinfo{author}{N.~Ferranti}, \bibinfo{author}{K.~Hose}, \bibinfo{author}{E.~Jim{\'e}nez-Ruiz}, et~al.,
\newblock \bibinfo{title}{{How does knowledge evolve in open knowledge graphs?}},
\newblock \bibinfo{journal}{{Transactions on Graph Data and Knowledge}} \bibinfo{volume}{1} (\bibinfo{year}{2023}) \bibinfo{pages}{11--1}.
\bibitem[{Gurk et~al.(2017)Gurk, Abela, and Debattista}]{McGurk2017}
\bibinfo{author}{S.~M. Gurk}, \bibinfo{author}{C.~Abela}, \bibinfo{author}{J.~Debattista},
\newblock \bibinfo{title}{Towards ontology quality assessment},
\newblock in: \bibinfo{booktitle}{Joint proceedings of the Workshop on Managing the Evolution and Preservation of the Data Web (MEPDaW) and the Workshop on Linked Data Quality {(LDQ}) co-located with European Semantic Web Conference {(ESWC})}, volume \bibinfo{volume}{1824} of \textit{\bibinfo{series}{{CEUR} Workshop Proceedings}}, \bibinfo{publisher}{CEUR-WS.org}, \bibinfo{year}{2017}, pp. \bibinfo{pages}{94--106}.
\bibitem[{Wang and Strong(1996)}]{WangStrong_1996}
\bibinfo{author}{R.~Y. Wang}, \bibinfo{author}{D.~M. Strong},
\newblock \bibinfo{title}{{Beyond Accuracy: What Data Quality Means to Data Consumers}},
\newblock \bibinfo{journal}{Journal of Management Information Systems} \bibinfo{volume}{12} (\bibinfo{year}{1996}) \bibinfo{pages}{5--33}.
\bibitem[{Pipino et~al.(2002)Pipino, Lee, and Wang}]{Pipino2002}
\bibinfo{author}{L.~L. Pipino}, \bibinfo{author}{Y.~W. Lee}, \bibinfo{author}{R.~Y. Wang},
\newblock \bibinfo{title}{{Data quality assessment}},
\newblock \bibinfo{journal}{Communications of the ACM} \bibinfo{volume}{45} (\bibinfo{year}{2002}) \bibinfo{pages}{211--218}. \DOIprefix\doi{10.1145/505248.506010}.
\bibitem[{Cichy and Rass(2019)}]{Cichy_2019}
\bibinfo{author}{C.~Cichy}, \bibinfo{author}{S.~Rass},
\newblock \bibinfo{title}{{An Overview of Data Quality Frameworks}},
\newblock \bibinfo{journal}{IEEE Access} \bibinfo{volume}{7} (\bibinfo{year}{2019}) \bibinfo{pages}{24634--24648}.
\bibitem[{Issa et~al.(2021)Issa, Adekunle, Hamdi, Cherfi, Dumontier, and Zaveri}]{Issa2021}
\bibinfo{author}{S.~Issa}, \bibinfo{author}{O.~Adekunle}, \bibinfo{author}{F.~Hamdi}, \bibinfo{author}{S.~S. Cherfi}, \bibinfo{author}{M.~Dumontier}, \bibinfo{author}{A.~Zaveri},
\newblock \bibinfo{title}{{Knowledge Graph Completeness: A Systematic Literature Review}},
\newblock \bibinfo{journal}{{IEEE} Access} \bibinfo{volume}{9} (\bibinfo{year}{2021}) \bibinfo{pages}{31322--31339}. \DOIprefix\doi{10.1109/ACCESS.2021.3056622}.
\bibitem[{Nayak et~al.(2021)Nayak, Bozic, and Longo}]{Nayak2022}
\bibinfo{author}{A.~Nayak}, \bibinfo{author}{B.~Bozic}, \bibinfo{author}{L.~Longo},
\newblock \bibinfo{title}{{Linked Data Quality Assessment: A Survey}},
\newblock in: \bibinfo{booktitle}{Web Services - {ICWS} - International Conference, Held as Part of the Services Conference Federation, {SCF}}, volume \bibinfo{volume}{12994} of \textit{\bibinfo{series}{Lecture Notes in Computer Science}}, \bibinfo{publisher}{Springer}, \bibinfo{year}{2021}, pp. \bibinfo{pages}{63--76}. \DOIprefix\doi{10.1007/978-3-030-96140-4\_5}.
\bibitem[{Chen et~al.(2019)Chen, Cao, Chen, and Ding}]{Chen2019}
\bibinfo{author}{H.~Chen}, \bibinfo{author}{G.~Cao}, \bibinfo{author}{J.~Chen}, \bibinfo{author}{J.~Ding},
\newblock \bibinfo{title}{A practical framework for evaluating the quality of knowledge graph},
\newblock \bibinfo{journal}{Communications in Computer and Information Science} \bibinfo{volume}{1134 CCIS} (\bibinfo{year}{2019}) \bibinfo{pages}{111--122}. \DOIprefix\doi{10.1007/978-981-15-1956-7_10}.
\bibitem[{Debattista et~al.(2016)Debattista, Auer, and Lange}]{LuzzuDebattista2016}
\bibinfo{author}{J.~Debattista}, \bibinfo{author}{S.~Auer}, \bibinfo{author}{C.~Lange},
\newblock \bibinfo{title}{{Luzzu—A Methodology and Framework for Linked Data Quality Assessment}},
\newblock \bibinfo{journal}{Journal on Data and Information Quality (JDIQ)} \bibinfo{volume}{8} (\bibinfo{year}{2016}). \DOIprefix\doi{10.1145/2992786}.
\bibitem[{Kontokostas et~al.(2014)Kontokostas, Westphal, Auer, Hellmann, Lehmann, and Cornelissen}]{RDFUnitKontokostas2014}
\bibinfo{author}{D.~Kontokostas}, \bibinfo{author}{P.~Westphal}, \bibinfo{author}{S.~Auer}, \bibinfo{author}{S.~Hellmann}, \bibinfo{author}{J.~Lehmann}, \bibinfo{author}{R.~Cornelissen},
\newblock \bibinfo{title}{{Databugger: a test-driven framework for debugging the web of data}},
\newblock in: \bibinfo{booktitle}{International World Wide Web Conference}, \bibinfo{publisher}{{ACM}}, \bibinfo{year}{2014}, pp. \bibinfo{pages}{115--118}. \DOIprefix\doi{10.1145/2567948.2577017}.
\bibitem[{Pellegrino et~al.(2024)Pellegrino, Rula, and Tuozzo}]{KGHeatBeatPellegrino2025}
\bibinfo{author}{M.~A. Pellegrino}, \bibinfo{author}{A.~Rula}, \bibinfo{author}{G.~Tuozzo},
\newblock \bibinfo{title}{Kgheartbeat: An open source tool for periodically evaluating the quality of knowledge graphs},
\newblock in: \bibinfo{booktitle}{The Semantic Web - International Semantic Web Conference}, volume \bibinfo{volume}{15233} of \textit{\bibinfo{series}{Lecture Notes in Computer Science}}, \bibinfo{publisher}{Springer}, \bibinfo{year}{2024}, pp. \bibinfo{pages}{40--58}. \DOIprefix\doi{10.1007/978-3-031-77847-6\_3}.
\bibitem[{F{\"{u}}rber and Hepp(2011)}]{SWIQAFrber2011}
\bibinfo{author}{C.~F{\"{u}}rber}, \bibinfo{author}{M.~Hepp},
\newblock \bibinfo{title}{{SWIQA - A Semantic Web Information Quality Assessment Framework}},
\newblock in: \bibinfo{booktitle}{European Conference on Information Systems}, \bibinfo{year}{2011}, p.~\bibinfo{pages}{76}.
\bibitem[{W3C(2017)}]{w3c_shacl}
\bibinfo{author}{W3C}, \bibinfo{title}{{Shapes Constraint Language (SHACL)}}, \bibinfo{howpublished}{https://www.w3.org/TR/shacl/}, \bibinfo{year}{2017}. \bibinfo{note}{W3C Recommendation}.
\bibitem[{Prud'hommeaux et~al.(2014)Prud'hommeaux, Gayo, and Solbrig}]{Prudhommeaux2014}
\bibinfo{author}{E.~Prud'hommeaux}, \bibinfo{author}{J.~E.~L. Gayo}, \bibinfo{author}{H.~R. Solbrig},
\newblock \bibinfo{title}{{Shape expressions: an {RDF} validation and transformation language}},
\newblock in: \bibinfo{booktitle}{Proceedings of the International Conference on Semantic Systems, SEMANTiCS}, \bibinfo{publisher}{{ACM}}, \bibinfo{year}{2014}, pp. \bibinfo{pages}{32--40}. \DOIprefix\doi{10.1145/2660517.2660523}.
\bibitem[{Rabbani et~al.(2022)Rabbani, Lissandrini, and Hose}]{Rabbani2022}
\bibinfo{author}{K.~Rabbani}, \bibinfo{author}{M.~Lissandrini}, \bibinfo{author}{K.~Hose},
\newblock \bibinfo{title}{{SHACL} and shex in the wild: {A} community survey on validating shapes generation and adoption},
\newblock in: \bibinfo{booktitle}{Companion of The Web Conference 2022, Virtual Event / Lyon, France, April 25 - 29, 2022}, \bibinfo{publisher}{{ACM}}, \bibinfo{year}{2022}, pp. \bibinfo{pages}{260--263}. \DOIprefix\doi{10.1145/3487553.3524253}.
\bibitem[{Spahiu et~al.(2018)Spahiu, Maurino, and Palmonari}]{SpahiuMP18}
\bibinfo{author}{B.~Spahiu}, \bibinfo{author}{A.~Maurino}, \bibinfo{author}{M.~Palmonari},
\newblock \bibinfo{title}{{Towards Improving the Quality of Knowledge Graphs with Data-driven Ontology Patterns and SHACL}},
\newblock in: \bibinfo{booktitle}{{Proceedings of the Workshop on Ontology Design and Patterns (WOP) co-located with International Semantic Web Conference (ISWC)}}, volume \bibinfo{volume}{2195} of \textit{\bibinfo{series}{{CEUR Workshop Proceedings}}}, \bibinfo{publisher}{CEUR-WS.org}, \bibinfo{year}{2018}, pp. \bibinfo{pages}{52--66}.
\bibitem[{Rabbani et~al.(2023)Rabbani, Lissandrini, and Hose}]{Rabbani2023_shactor}
\bibinfo{author}{K.~Rabbani}, \bibinfo{author}{M.~Lissandrini}, \bibinfo{author}{K.~Hose},
\newblock \bibinfo{title}{{SHACTOR: Improving the Quality of Large-Scale Knowledge Graphs with Validating Shapes}},
\newblock in: \bibinfo{booktitle}{{Companion of the International Conference on Management of Data, SIGMOD/PODS}}, \bibinfo{publisher}{{ACM}}, \bibinfo{year}{2023}, pp. \bibinfo{pages}{151--154}. \DOIprefix\doi{10.1145/3555041.3589723}.
\bibitem[{Boneva et~al.(2019)Boneva, Dusart, Fern{\'{a}}ndez{-}{\'{A}}lvarez, and Gayo}]{Boneva2019}
\bibinfo{author}{I.~Boneva}, \bibinfo{author}{J.~Dusart}, \bibinfo{author}{D.~Fern{\'{a}}ndez{-}{\'{A}}lvarez}, \bibinfo{author}{J.~E.~L. Gayo},
\newblock \bibinfo{title}{{Shape Designer for ShEx and SHACL constraints}},
\newblock in: \bibinfo{booktitle}{{Proceedings of the ISWC Satellite Tracks (Posters {\&} Demonstrations, Industry, and Outrageous Ideas) co-located with International Semantic Web Conference (ISWC)}}, volume \bibinfo{volume}{2456} of \textit{\bibinfo{series}{{CEUR} Workshop Proceedings}}, \bibinfo{publisher}{CEUR-WS.org}, \bibinfo{year}{2019}, pp. \bibinfo{pages}{269--272}.
\bibitem[{Cimmino et~al.(2020)Cimmino, Fern{\'{a}}ndez{-}Izquierdo, and Garc{\'{\i}}a{-}Castro}]{Cimmino2020}
\bibinfo{author}{A.~Cimmino}, \bibinfo{author}{A.~Fern{\'{a}}ndez{-}Izquierdo}, \bibinfo{author}{R.~Garc{\'{\i}}a{-}Castro},
\newblock \bibinfo{title}{Astrea: Automatic generation of {SHACL} shapes from ontologies},
\newblock in: \bibinfo{booktitle}{The Semantic Web - International Conference, {ESWC}, Proceedings}, volume \bibinfo{volume}{12123} of \textit{\bibinfo{series}{Lecture Notes in Computer Science}}, \bibinfo{publisher}{Springer}, \bibinfo{year}{2020}, pp. \bibinfo{pages}{497--513}. \DOIprefix\doi{10.1007/978-3-030-49461-2\_29}.
\bibitem[{Pandit et~al.(2018)Pandit, O'Sullivan, and Lewis}]{Pandit2018}
\bibinfo{author}{H.~J. Pandit}, \bibinfo{author}{D.~O'Sullivan}, \bibinfo{author}{D.~Lewis},
\newblock \bibinfo{title}{{Using Ontology Design Patterns to Define SHACL Shapes}},
\newblock in: \bibinfo{booktitle}{Proceedings of the Workshop on Ontology Design Patterns (WOP) co-located with the International Semantic Web Conference (ISWC)}, \bibinfo{year}{2018}.
\bibitem[{Duan et~al.(2024)Duan, Chaves-Fraga, Derom, and Dimou}]{Duan2024}
\bibinfo{author}{X.~Duan}, \bibinfo{author}{D.~Chaves-Fraga}, \bibinfo{author}{O.~Derom}, \bibinfo{author}{A.~Dimou},
\newblock \bibinfo{title}{{SCOOP All the Constraints’ Flavours for Your Knowledge Graph}},
\newblock in: \bibinfo{booktitle}{Proceedings of the Extended Semantic Web Conference (ESWC)}, volume \bibinfo{volume}{14665 LNCS}, \bibinfo{publisher}{Springer}, \bibinfo{year}{2024}, pp. \bibinfo{pages}{217--234}. \DOIprefix\doi{10.1007/978-3-031-60635-9_13}.
\bibitem[{Fernandez-Álvarez et~al.(2022)Fernandez-Álvarez, Labra-Gayo, and Gayo-Avello}]{FernandezAlvarez2022}
\bibinfo{author}{D.~Fernandez-Álvarez}, \bibinfo{author}{J.~E. Labra-Gayo}, \bibinfo{author}{D.~Gayo-Avello},
\newblock \bibinfo{title}{{Automatic extraction of shapes using sheXer}},
\newblock \bibinfo{journal}{Knowledge-Based Systems} \bibinfo{volume}{238} (\bibinfo{year}{2022}) \bibinfo{pages}{107975}. \DOIprefix\doi{10.1016/J.KNOSYS.2021.107975}.
\bibitem[{Yang et~al.(2023)Yang, Zhang, Li, and Li}]{Yang2023}
\bibinfo{author}{X.~Yang}, \bibinfo{author}{H.~Zhang}, \bibinfo{author}{J.~Li}, \bibinfo{author}{S.~Li},
\newblock \bibinfo{title}{{A Method for Data Quality Validation Based on Shapes Constraint Language}},
\newblock \bibinfo{journal}{Proceedings - International Conference on Big Data, Information and Computer Network, BDICN}  (\bibinfo{year}{2023}) \bibinfo{pages}{83--87}. \DOIprefix\doi{10.1109/BDICN58493.2023.00024}.
\bibitem[{Luthfi et~al.(2022)Luthfi, Darari, and Ashardian}]{Luthfi2022}
\bibinfo{author}{M.~J. Luthfi}, \bibinfo{author}{F.~Darari}, \bibinfo{author}{A.~C. Ashardian},
\newblock \bibinfo{title}{{SoCK: SHACL on Completeness Knowledge}},
\newblock in: \bibinfo{booktitle}{Proceedings of the Workshop on Ontology Design and Patterns (WOP) co-located with the International Semantic Web Conference (ISWC)}, \bibinfo{publisher}{{CEUR-WS.org}}, \bibinfo{year}{2022}.
\bibitem[{Vandenbussche et~al.(2017)Vandenbussche, Umbrich, Matteis, Hogan, and Buil-Aranda}]{SPARQLESVandenbussche2017}
\bibinfo{author}{P.~Y. Vandenbussche}, \bibinfo{author}{J.~Umbrich}, \bibinfo{author}{L.~Matteis}, \bibinfo{author}{A.~Hogan}, \bibinfo{author}{C.~Buil-Aranda},
\newblock \bibinfo{title}{{SPARQLES: Monitoring Public SPARQL Endpoints}},
\newblock \bibinfo{journal}{Semantic Web} \bibinfo{volume}{8} (\bibinfo{year}{2017}) \bibinfo{pages}{1049--1065}. \DOIprefix\doi{10.3233/SW-170254}.
\bibitem[{Mihindukulasooriya et~al.(2016)Mihindukulasooriya, Garc{\'{\i}}a{-}Castro, and G{\'{o}}mez{-}P{\'{e}}rez}]{LDSnifferMihindukulasooriya2017}
\bibinfo{author}{N.~Mihindukulasooriya}, \bibinfo{author}{R.~Garc{\'{\i}}a{-}Castro}, \bibinfo{author}{A.~G{\'{o}}mez{-}P{\'{e}}rez},
\newblock \bibinfo{title}{{LD Sniffer: A Quality Assessment Tool for Measuring the Accessibility of Linked Data}},
\newblock in: \bibinfo{booktitle}{Knowledge Engineering and Knowledge Management - {EKAW} Satellite Events, {EKM} and Drift-an-LOD}, volume \bibinfo{volume}{10180} of \textit{\bibinfo{series}{Lecture Notes in Computer Science}}, \bibinfo{publisher}{Springer}, \bibinfo{year}{2016}, pp. \bibinfo{pages}{149--152}. \DOIprefix\doi{10.1007/978-3-319-58694-6\_20}.
\bibitem[{Langer et~al.(2018)Langer, Siegert, G{\"{o}}pfert, and Gaedke}]{SemQuireLanger2018}
\bibinfo{author}{A.~Langer}, \bibinfo{author}{V.~Siegert}, \bibinfo{author}{C.~G{\"{o}}pfert}, \bibinfo{author}{M.~Gaedke},
\newblock \bibinfo{title}{Semquire - assessing the data quality of linked open data sources based on {DQV}},
\newblock in: \bibinfo{booktitle}{Current Trends in Web Engineering - {ICWE} International Workshops, MATWEP, EnWot, KD-WEB, WEOD, TourismKG}, volume \bibinfo{volume}{11153} of \textit{\bibinfo{series}{Lecture Notes in Computer Science}}, \bibinfo{publisher}{Springer}, \bibinfo{year}{2018}, pp. \bibinfo{pages}{163--175}. \DOIprefix\doi{10.1007/978-3-030-03056-8\_14}.
\bibitem[{Kontokostas et~al.(2013)Kontokostas, Zaveri, Auer, and Lehmann}]{TripleCheckMateKontokostas2013}
\bibinfo{author}{D.~Kontokostas}, \bibinfo{author}{A.~Zaveri}, \bibinfo{author}{S.~Auer}, \bibinfo{author}{J.~Lehmann},
\newblock \bibinfo{title}{{TripleCheckMate: A Tool for Crowdsourcing the Quality Assessment of Linked Data}},
\newblock in: \bibinfo{booktitle}{Knowledge Engineering and the Semantic Web - International Conference}, volume \bibinfo{volume}{394} of \textit{\bibinfo{series}{Communications in Computer and Information Science}}, \bibinfo{publisher}{Springer}, \bibinfo{year}{2013}, pp. \bibinfo{pages}{265--272}. \DOIprefix\doi{10.1007/978-3-642-41360-5\_22}.
\bibitem[{Yamamoto et~al.(2018)Yamamoto, Yamaguchi, and Splendiani}]{YummyDataYamamotoYS18}
\bibinfo{author}{Y.~Yamamoto}, \bibinfo{author}{A.~Yamaguchi}, \bibinfo{author}{A.~Splendiani},
\newblock \bibinfo{title}{{YummyData: providing high-quality open life science data}},
\newblock \bibinfo{journal}{Database J. Biol. Databases Curation} \bibinfo{volume}{2018} (\bibinfo{year}{2018}). \DOIprefix\doi{10.1093/DATABASE/BAY022}.
\bibitem[{Ruckhaus et~al.(2014)Ruckhaus, Vidal, Castillo, Burguillos, and Baldizan}]{LiQuateRuckhaus2014}
\bibinfo{author}{E.~Ruckhaus}, \bibinfo{author}{M.~Vidal}, \bibinfo{author}{S.~Castillo}, \bibinfo{author}{O.~Burguillos}, \bibinfo{author}{O.~Baldizan},
\newblock \bibinfo{title}{{Analyzing Linked Data Quality with LiQuate}},
\newblock in: \bibinfo{booktitle}{The Semantic Web: {ESWC} Satellite Events}, volume \bibinfo{volume}{8798} of \textit{\bibinfo{series}{Lecture Notes in Computer Science}}, \bibinfo{publisher}{Springer}, \bibinfo{year}{2014}, pp. \bibinfo{pages}{488--493}. \DOIprefix\doi{10.1007/978-3-319-11955-7\_72}.
\bibitem[{Pizhuk et~al.(2025)Pizhuk, Ehrlinger, Denk, and Geist}]{SecurityDashboardPizhukEDG25}
\bibinfo{author}{D.~Pizhuk}, \bibinfo{author}{L.~Ehrlinger}, \bibinfo{author}{G.~Denk}, \bibinfo{author}{V.~Geist},
\newblock \bibinfo{title}{A data quality dashboard for (security) knowledge graphs},
\newblock in: \bibinfo{booktitle}{{Datenbanksysteme f{\"{u}}r Business, Technologie und Web {(BTW}), Fachtagung des GI-Fachbereichs, Datenbanken und Informationssysteme" (DBIS), Proceedings}}, volume \bibinfo{volume}{{P-361}} of \textit{\bibinfo{series}{{LNI}}}, \bibinfo{publisher}{Gesellschaft f{\"{u}}r Informatik e.V.}, \bibinfo{year}{2025}, pp. \bibinfo{pages}{803--810}. \DOIprefix\doi{10.18420/BTW2025-45}.
\bibitem[{Flemming(2011)}]{Flemming2011}
\bibinfo{author}{A.~Flemming}, \bibinfo{title}{{Qualitätsmerkmale von Linked Data-veröffentlichenden Datenquellen}}, Master's thesis, Universität Leipzig, \bibinfo{year}{2011}. \bibinfo{note}{Diplomarbeit (Quality Criteria for Linked Data Sources)}.
\bibitem[{Berrueta and Phipps(2008)}]{CoolURIs_2008}
\bibinfo{author}{D.~Berrueta}, \bibinfo{author}{J.~Phipps}, \bibinfo{title}{Cool URIs for the Semantic Web}, \bibinfo{type}{W3C Recommendation}, W3C, \bibinfo{year}{2008}. \bibinfo{note}{Accessed: 2025-07-24}.
\bibitem[{Bernadette~Hyland and Villazón-Terrazas(2014)}]{BestPractices_2014}
\bibinfo{author}{G.~A. Bernadette~Hyland}, \bibinfo{author}{B.~Villazón-Terrazas}, \bibinfo{title}{{Best Practices for Publishing Linked Data}}, \bibinfo{type}{W3C Working Group Note}, W3C, \bibinfo{year}{2014}. \bibinfo{note}{Accessed: 2025-07-24}.
\bibitem[{Cortés(2025)}]{shacl-dqa-prototype}
\bibinfo{author}{C.~Cortés}, \bibinfo{title}{{SHACL-DQA-prototype}}, \bibinfo{howpublished}{\url{https://github.com/HPI-Information-Systems/SHACL-DQA}}, \bibinfo{year}{2025}. \bibinfo{note}{GitHub repository}.
\bibitem[{{TopQuadrant}(2024)}]{dash_shapes}
\bibinfo{author}{{TopQuadrant}}, \bibinfo{title}{{DASH - Data Shapes}}, \bibinfo{howpublished}{\url{https://datashapes.org/dash}}, \bibinfo{year}{2024}. \bibinfo{note}{Accessed: 2025-07-29}.
\bibitem[{{W3C SHACL Community Group}(2023)}]{shacl_af}
\bibinfo{author}{{W3C SHACL Community Group}}, \bibinfo{title}{{SHACL Advanced Features}}, \bibinfo{howpublished}{\url{https://www.w3.org/TR/shacl-af/}}, \bibinfo{year}{2023}. \bibinfo{note}{Accessed: 2025-07-29}.
\bibitem[{Hartmann et~al.(2016)Hartmann, Zapilko, Wackerow, and Eckert}]{Hartmann2016}
\bibinfo{author}{T.~Hartmann}, \bibinfo{author}{B.~Zapilko}, \bibinfo{author}{J.~Wackerow}, \bibinfo{author}{K.~Eckert},
\newblock \bibinfo{title}{{Validating RDF Data Quality Using Constraints to Direct the Development of Constraint Languages}},
\newblock \bibinfo{journal}{{Proceedings - IEEE International Conference on Semantic Computing (ICSC)}}  (\bibinfo{year}{2016}) \bibinfo{pages}{116--123}. \DOIprefix\doi{10.1109/ICSC.2016.43}.
\bibitem[{Kraft and Usbeck(2022)}]{KraftU22}
\bibinfo{author}{A.~Kraft}, \bibinfo{author}{R.~Usbeck},
\newblock \bibinfo{title}{The lifecycle of "facts": {A} survey of social bias in knowledge graphs},
\newblock in: \bibinfo{booktitle}{Proceedings of the Conference of the Asia-Pacific Chapter of the Association for Computational Linguistics and the International Joint Conference on Natural Language Processing}, \bibinfo{year}{2022}, pp. \bibinfo{pages}{639--652}. \DOIprefix\doi{10.18653/V1/2022.AACL-MAIN.49}.
\bibitem[{Mohammed et~al.(2025)Mohammed, Ehrlinger, Harmouch, Naumann, and Srivastava}]{Mohammed_2025}
\bibinfo{author}{S.~Mohammed}, \bibinfo{author}{L.~Ehrlinger}, \bibinfo{author}{H.~Harmouch}, \bibinfo{author}{F.~Naumann}, \bibinfo{author}{D.~Srivastava},
\newblock \bibinfo{title}{{The Five Facets of Data Quality Assessment}},
\newblock \bibinfo{journal}{SIGMOD Record} \bibinfo{volume}{54} (\bibinfo{year}{2025}) \bibinfo{pages}{18--27}.
\bibitem[{Mendes et~al.(2012)Mendes, M{\"{u}}hleisen, and Bizer}]{SieveMendes2012}
\bibinfo{author}{P.~N. Mendes}, \bibinfo{author}{H.~M{\"{u}}hleisen}, \bibinfo{author}{C.~Bizer},
\newblock \bibinfo{title}{{Sieve: linked data quality assessment and fusion}},
\newblock in: \bibinfo{booktitle}{Proceedings of the Joint {EDBT/ICDT} Workshops}, \bibinfo{publisher}{{ACM}}, \bibinfo{year}{2012}, pp. \bibinfo{pages}{116--123}. \DOIprefix\doi{10.1145/2320765.2320803}.
\bibitem[{Hogan et~al.(2010)Hogan, Harth, Passant, Decker, and Polleres}]{HoganHPDP10}
\bibinfo{author}{A.~Hogan}, \bibinfo{author}{A.~Harth}, \bibinfo{author}{A.~Passant}, \bibinfo{author}{S.~Decker}, \bibinfo{author}{A.~Polleres},
\newblock \bibinfo{title}{{Weaving the Pedantic Web}},
\newblock in: \bibinfo{booktitle}{{Proceedings of the WWW Workshop on Linked Data on the Web, LDOW}}, volume \bibinfo{volume}{628} of \textit{\bibinfo{series}{{CEUR} Workshop Proceedings}}, \bibinfo{publisher}{CEUR-WS.org}, \bibinfo{year}{2010}.
\bibitem[{Hogan et~al.(2012)Hogan, Umbrich, Harth, Cyganiak, Polleres, and Decker}]{Hogan2012}
\bibinfo{author}{A.~Hogan}, \bibinfo{author}{J.~Umbrich}, \bibinfo{author}{A.~Harth}, \bibinfo{author}{R.~Cyganiak}, \bibinfo{author}{A.~Polleres}, \bibinfo{author}{S.~Decker},
\newblock \bibinfo{title}{{An empirical survey of Linked Data conformance}},
\newblock \bibinfo{journal}{{Journal of Web Semantics}} \bibinfo{volume}{14} (\bibinfo{year}{2012}) \bibinfo{pages}{14--44}. \DOIprefix\doi{10.1016/J.WEBSEM.2012.02.001}.
\bibitem[{Lei et~al.(2007)Lei, Uren, and Motta}]{Lei2007}
\bibinfo{author}{Y.~Lei}, \bibinfo{author}{V.~S. Uren}, \bibinfo{author}{E.~Motta},
\newblock \bibinfo{title}{A framework for evaluating semantic metadata},
\newblock in: \bibinfo{booktitle}{Proceedings of the International Conference on Knowledge Capture {(K-CAP})}, \bibinfo{publisher}{{ACM}}, \bibinfo{year}{2007}, pp. \bibinfo{pages}{135--142}. \DOIprefix\doi{10.1145/1298406.1298431}.
\bibitem[{Chen and Garcia(2010)}]{Chen2010}
\bibinfo{author}{P.~Chen}, \bibinfo{author}{W.~Garcia},
\newblock \bibinfo{title}{{Hypothesis generation and data quality assessment through association mining}},
\newblock \bibinfo{journal}{{Proceedings of the IEEE International Conference on Cognitive Informatics, ICCI}}  (\bibinfo{year}{2010}) \bibinfo{pages}{659--666}. \DOIprefix\doi{10.1109/COGINF.2010.5599828}.
\bibitem[{Böhm et~al.(2010)Böhm, Naumann, Abedjan, Fenz, Grütze, Hefenbrock, Pohl, and Sonnabend}]{Bohm2010}
\bibinfo{author}{C.~Böhm}, \bibinfo{author}{F.~Naumann}, \bibinfo{author}{Z.~Abedjan}, \bibinfo{author}{D.~Fenz}, \bibinfo{author}{T.~Grütze}, \bibinfo{author}{D.~Hefenbrock}, \bibinfo{author}{M.~Pohl}, \bibinfo{author}{D.~Sonnabend},
\newblock \bibinfo{title}{{Profiling linked open data with ProLOD}},
\newblock \bibinfo{journal}{{Proceedings - International Conference on Data Engineering}}  (\bibinfo{year}{2010}) \bibinfo{pages}{175--178}. \DOIprefix\doi{10.1109/ICDEW.2010.5452762}.
\bibitem[{Antonie and Za{\"{\i}}ane(2004)}]{Antonie2004}
\bibinfo{author}{M.~Antonie}, \bibinfo{author}{O.~R. Za{\"{\i}}ane},
\newblock \bibinfo{title}{Mining positive and negative association rules: An approach for confined rules},
\newblock in: \bibinfo{booktitle}{Knowledge Discovery in Databases: {PKDD}, European Conference on Principles and Practice of Knowledge Discovery in Databases, Proceedings}, volume \bibinfo{volume}{3202} of \textit{\bibinfo{series}{Lecture Notes in Computer Science}}, \bibinfo{publisher}{Springer}, \bibinfo{year}{2004}, pp. \bibinfo{pages}{27--38}. \DOIprefix\doi{10.1007/978-3-540-30116-5\_6}.
\bibitem[{Mostafavi et~al.(2004)Mostafavi, Edwards, and Jeansoulin}]{Mostafavi2004}
\bibinfo{author}{M.-A. Mostafavi}, \bibinfo{author}{G.~Edwards}, \bibinfo{author}{R.~Jeansoulin},
\newblock \bibinfo{title}{An ontology-based method for quality assessment of spatial data bases},
\newblock in: \bibinfo{editor}{A.~Frank}, \bibinfo{editor}{E.~Grum} (Eds.), \bibinfo{booktitle}{Proceedings of the 3rd International Symposium on Spatial Data Quality}, GeoInfo, \bibinfo{address}{Bruck an der Leitha, Austria}, \bibinfo{year}{2004}, pp. \bibinfo{pages}{49--66}.
\bibitem[{Zaveri et~al.(2013)Zaveri, Kontokostas, Sherif, Bühmann, Morsey, Auer, and Lehmann}]{Zaveri2013}
\bibinfo{author}{A.~Zaveri}, \bibinfo{author}{D.~Kontokostas}, \bibinfo{author}{M.~A. Sherif}, \bibinfo{author}{L.~Bühmann}, \bibinfo{author}{M.~Morsey}, \bibinfo{author}{S.~Auer}, \bibinfo{author}{J.~Lehmann},
\newblock \bibinfo{title}{{User-driven quality evaluation of DBpedia}},
\newblock \bibinfo{journal}{{ACM International Conference Proceeding Series}}  (\bibinfo{year}{2013}) \bibinfo{pages}{97--104}. \DOIprefix\doi{10.1145/2506182.2506195}.
\bibitem[{Albertoni and Pérez(2013)}]{Albertoni2013}
\bibinfo{author}{R.~Albertoni}, \bibinfo{author}{A.~G. Pérez},
\newblock \bibinfo{title}{{Assessing linkset quality for complementing third-party datasets}},
\newblock \bibinfo{journal}{{ACM International Conference Proceeding Series}}  (\bibinfo{year}{2013}) \bibinfo{pages}{52--59}. \DOIprefix\doi{10.1145/2457317.2457327}.
\bibitem[{Alexander et~al.(2011)Alexander, Cyganiak, Hausenblas, and Zhao}]{void2011}
\bibinfo{author}{K.~Alexander}, \bibinfo{author}{R.~Cyganiak}, \bibinfo{author}{M.~Hausenblas}, \bibinfo{author}{J.~Zhao}, \bibinfo{title}{{Describing Linked Datasets with the VoID Vocabulary}}, \bibinfo{howpublished}{\url{https://www.w3.org/TR/void/}}, \bibinfo{year}{2011}. \bibinfo{note}{W3C Interest Group Note}.
\bibitem[{Jacobi et~al.(2011)Jacobi, Kagal, and Khandelwal}]{Jacobi2011}
\bibinfo{author}{I.~Jacobi}, \bibinfo{author}{L.~Kagal}, \bibinfo{author}{A.~Khandelwal},
\newblock \bibinfo{title}{Rule-based trust assessment on the semantic web},
\newblock in: \bibinfo{booktitle}{Rule-Based Reasoning, Programming, and Applications - 5th International Symposium, RuleML. Proceedings}, volume \bibinfo{volume}{6826} of \textit{\bibinfo{series}{Lecture Notes in Computer Science}}, \bibinfo{publisher}{Springer}, \bibinfo{year}{2011}, pp. \bibinfo{pages}{227--241}. \DOIprefix\doi{10.1007/978-3-642-22546-8\_18}.
\bibitem[{Golbeck et~al.(2003)Golbeck, Parsia, and Hendler}]{Golbeck2003}
\bibinfo{author}{J.~Golbeck}, \bibinfo{author}{B.~Parsia}, \bibinfo{author}{J.~A. Hendler},
\newblock \bibinfo{title}{Trust networks on the semantic web},
\newblock in: \bibinfo{booktitle}{Cooperative Information Agents VII, International Workshop, {CIA}, Proceedings}, volume \bibinfo{volume}{2782} of \textit{\bibinfo{series}{Lecture Notes in Computer Science}}, \bibinfo{publisher}{Springer}, \bibinfo{year}{2003}, pp. \bibinfo{pages}{238--249}. \URLprefix \url{https://doi.org/10.1007/978-3-540-45217-1\_18}. \DOIprefix\doi{10.1007/978-3-540-45217-1\_18}.

\end{thebibliography}

\iftoggle{appendixversion}{
    \appendix

\section{Related Work}\label{section:related_work_appendix}

\autoref{table:dqa_tools_kgs} presents a summary of DQA tools proposed in the literature, including their name, input type, the DQ dimensions they cover, and whether they are currently maintained.

\begin{table*}[!ht]
\scriptsize
\centering
\begin{tabular}{p{2.5cm}|p{2cm}|p{7cm}|p{1.8cm}}
\toprule
\textbf{Tool} & \textbf{Input} & \textbf{Dimensions} & \textbf{Maintained} \\ \midrule

KGHeartbeat \cite{KGHeatBeatPellegrino2025} & Dump/SPARQL & Availability, Believability, Completeness, Conciseness, Consistency, Currency, Interlinking, Interoperability, Interpretability, Licensing, Performance, Representational Conciseness, Reputation, Security, Semantic Accuracy, Timeliness, Understandability, Verifiability, Versatility \cite{Zaveri2012} & Yes \\ 

DQ dashboard~\cite{SecurityDashboardPizhukEDG25} & Graph DB & Completeness, Consistency, Readability & Yes \\ 

SemQuire \cite{SemQuireLanger2018} & Dump/SPARQL & Multiple dimensions from \cite{Zaveri2012} & N/A \\ 

YummyData \cite{YummyDataYamamotoYS18} & SPARQL & Availability, Freshness, Usefulness, Validity, Performance & Yes \\ 

SPARQLES \cite{SPARQLESVandenbussche2017} & SPARQL & Accessibility, Discoverability, Performance, Interoperability & No \\

LDSniffer \cite{LDSnifferMihindukulasooriya2017} & SPARQL & Accessibility & No \\ 

Luzzu~\cite{LuzzuDebattista2016} & Dump/SPARQL & Representational Conciseness, Interoperability, Interpretability, Versatility, Understandability, Timeliness, Semantic Accuracy, Consistency, Conciseness, Availability, Licensing, Interlinking, Security, Performance \cite{Zaveri2012} & No \\ 


LiQuate~\cite{LiQuateRuckhaus2014} & Dump & Completeness, Consistency & N/A \\ 

RDFUnit \cite{RDFUnitKontokostas2014} & Dump/SPARQL & Not specified & No \\ 

TripleCheckMate \cite{TripleCheckMateKontokostas2013} & SPARQL & Accuracy, Relevancy, Interlinking, Conciseness & No \\ 

Sieve \cite{SieveMendes2012} & Dump/SPARQL & Recency, Reputation & No \\ 

SWIQA \cite{SWIQAFrber2011} & Dump & Accuracy, Completeness, Uniqueness, Timeliness & N/A \\ 

RDF alerts \cite{HoganHPDP10} & URI & Not specified & No \\ 
\bottomrule
\end{tabular}
\caption{Summary of DQA tools for linked data.}
\label{table:dqa_tools_kgs}
\end{table*}

\section{Shapes Definitions}\label{section:Appendix_Shapes_Definition}
This appendix presents additional shape definitions that were omitted from \autoref{sec:approach} due to space constraints. It includes the remaining shapes for the dimensions of each group and, for the implemented shapes, we specify the type of DQ measure derived from each shape's validation result. A summary of these DQ measures can be found in \autoref{section:evaluation_appendix}.

\subsection{Accessibility}\label{section:accessibility_appendix}

This section presents the shape definitions for the dimensions \dq{Availability}, \dq{Licensing}, \dq{Security}, and \dq{Interlinking}.
It also includes the remaining shape definitions for the dimension \dq{Performance}, which were omitted in \autoref{section:accessibility}.

\subsubsection{Availability}

According to \BaseSurvey~, the \dq{Availability} dimension refers to the degree to which a dataset, or part of it, is accessible, retrievable, and ready for use. Five metrics are proposed for this dimension, most of which require accessing resources on the web (e.g., checking the dereferenceability of URIs or detecting dead or broken links).
From the five metrics, only A2 can be partially checked, and it refers to the availability of an RDF dump, and whether it can be downloaded. For this metric, we defined shape \shape{\ref{shape:availability_void_shape}}, which checks if there's an RDF dump provided, by verifying the existence of the property \rdfproperty{void:dataDump} in the VoID description of the dataset.
This can also be checked for the property \rdfproperty{dcat:downloadURL} from the DCAT vocabulary. However, in both cases, SHACL cannot verify if the referenced dump can be downloaded, as it cannot perform network requests. Therefore, we state that this metric can be partially covered with SHACL core.

The validation report for \shape{\ref{shape:availability_void_shape}} outputs a result for each instance of the class \rdfproperty{void:Dataset} that doesn't have a value for either of the properties \rdfproperty{void:dataDump} and \rdfproperty{dcat:downloadURL} (after following the path \rdfproperty{dcat:distribution dcat:downloadURL}). The DQ measure calculated from the validation result is a \BinaryMeasure.

\begin{minipage}{\linewidth}
\begin{lstlisting}[basicstyle=\ttfamily\scriptsize, caption={Availability - RDF Dump}, label={shape:availability_void_shape}]
ex:AvailabilityDumpShape a sh:NodeShape ;
    sh:targetClass void:Dataset ; 
    sh:or ( 
        [ sh:path void:dataDump; sh:minCount 1; ] 
        [ sh:path (dcat:distribution dcat:downloadURL); sh:minCount 1; ]
    ).
\end{lstlisting}
\end{minipage}

\subsubsection{Licensing}

The dimension \dq{Licensing} refers to the granting of permission to users to re-use a dataset under specified conditions \cite{Zaveri2012}.

\BaseSurvey~identified three metrics for this dimension, where only one of them can be covered with SHACL core. The metric L1 checks if there's a machine-readable indication of a license in the VoID description of the dataset or in the dataset itself. A SHACL shape can be defined for this metric, that targets instances of the class \rdfproperty{void:Dataset} and enforces the existence (\rdfproperty{sh:minCount 1}) of a property, such as {\rdfproperty{dcterms:license} (See \shape{\ref{shape:licensing_void_shape}}).

\begin{minipage}[t]{\linewidth}
\begin{lstlisting}[basicstyle=\ttfamily\scriptsize, caption={Licensing - Machine-readable license}, label={shape:licensing_void_shape}]
ex:MachineReadableLicenseShape a sh:NodeShape ;
    sh:targetClass void:Dataset ; 
    sh:property [
        sh:path dcterms:license;
        sh:class dcterms:LicenseDocument;
        sh:minCount 1;
    ].
\end{lstlisting}
\end{minipage}

The validation report for \shape{\ref{shape:licensing_void_shape}} outputs a result for each instance of the class \rdfproperty{void:Dataset} that doesn't have a value for the property \rdfproperty{dcterms:license}, or that it has a value but isn't of the correct type (\rdfproperty{dcterms:LicenseDocument}). The DQ measure calculated from the validation result is a \BinaryMeasure.

With regards to the metric L2, it can't be covered with SHACL core, since it checks for a human-readable specification of a license in the documentation of the dataset, typically an HTML document. 

The last metric (L3) checks for the usage of the correct license. This metric is defined in \cite{Flemming2011}, where they mention the importance of taking into account the requirements specified inside licenses. One example of these requirements is licenses that require that the derivative dataset be published under the same license as the original dataset (e.g., ShareAlike \footnote{\url{https://creativecommons.org/licenses/by-sa/4.0/deed.en}} licenses). However, to cover this metric, we would need to access the original dataset, review the clauses in its license, and verify that both datasets have the same license if the license requires it. However, SHACL core cannot handle these tasks, as it operates over RDF graphs and cannot parse natural language (e.g., to check whether a license contains a specific clause).\\

\subsubsection{Interlinking}\label{section:interlinking}

\dq{Interlinking} describes the extent to which entities representing the same concept are connected, either within a single data source or across multiple sources, for which \BaseSurvey~identified three metrics (I1-I3). 

Metric I1 evaluates the quality of interlinks, with six sub-aspects identified in \cite{Zaveri2012} (I1M1–I1M6). The first three (I1M1–I1M3) focus on graph topology characteristics, such as interlinking degree (e.g., detecting hubs in a network). These aspects can't be expressed with SHACL core because they involve extracting characteristics of the graph or a part of it (subgraph), rather than expressing constraints over nodes or properties. 

The aspect I1M4 checks for open sameAs chains, which implies checking that a node is reachable from itself via a full \rdfproperty{owl:sameAs} path. In this case, SHACL core can partially cover this metric, as it allows us to check only if a node has outgoing and incoming \rdfproperty{owl:sameAs} links locally. 

Initially, we defined shape \shape{\ref{shape:interlinking_open_sameas_chains}}, which targets a specific node and constrains that the end of one or more \rdfproperty{owl:sameAs} paths must point back to the target node. However, this shape fails when a node has multiple sameAs chains: if at least one chain is properly closed, the shape considers the constraint satisfied, even if other chains remain open. In other words, the shape only checks for the existence of a single closed sameAs path, not whether all such paths are closed. Therefore, we defined \shape{\ref{shape:interlinking_open_sameas_pairs}}, which verifies that for every node with an outgoing \rdfproperty{owl:sameAs} link, a corresponding inverse triple exists. In other words, for every triple like \rdftriple{ex:A}{owl:sameAs}{ex:B}, \rdftriple{ex:B}{owl:sameAs}{ex:A} must also exist.

\noindent
\begin{minipage}[t]{0.48\linewidth}
\begin{lstlisting}[basicstyle=\ttfamily\scriptsize, caption={Interlinking - Open sameAs chains}, label={shape:interlinking_open_sameas_chains}]
ex:OpenSameAsChainsShapes a sh:NodeShape ;
    sh:targetNode ENTITY_URI; 
    sh:property [
        sh:path [sh:oneOrMorePaths owl:sameAs];
        sh:hasValue ENTITY_URI;
    ].
\end{lstlisting}
\end{minipage}
\hfill
\begin{minipage}[t]{0.48\linewidth}
\begin{lstlisting}[basicstyle=\ttfamily\scriptsize, caption={Interlinking - Open sameAs pairs}, label={shape:interlinking_open_sameas_pairs}]
ex:OpenSameAsPairsShape a sh:NodeShape ;
    sh:targetSubjectsOf owl:sameAs; 
    sh:property [
        sh:path [sh:inversePath owl:sameAs];
        sh:equals owl:sameAs;
    ].
\end{lstlisting}
\end{minipage}

\shape{\ref{shape:interlinking_open_sameas_pairs}} would return a violation for each node that is missing one of its inverse \rdfproperty{owl:sameAs} links. Additionally, the property \rdfproperty{owl:sameAs} is generally used to link equivalent entities across different datasets. As a result, for the shape to function correctly, the datasets containing such links should be ``merged'' into a single data graph, which would then be used for validation.

The aspect I1M5 checks how much value is added through the use of sameAs edges. This metric involves analyzing multiple graphs to compare properties from linked entities and determine whether one entity contributes new information to the other. However, this comparison is non-trivial: it requires a semantic understanding of the properties involved, as similar or even repeated properties may exist under different IRIs. Even if the linked graphs were merged into a single one, SHACL core cannot express constraints that compare property values across different entities within the same graph. This is because SHACL core restricts property value comparisons between triples that share the same subject. 

Finally, the aspect I1M6 checks for good quality interlinks via crowd-sourcing. However, SHACL cannot cover this metric, since it lacks support for human input.

The second metric (I2) can be expressed with SHACL core, by checking that the values of \rdfproperty{owl:sameAs} use a different URI than the one used in the dataset (See shape \shape{\ref{shape:interlinking_external_uris}}).
We target the subjects of \rdfproperty{rdf:type} to ensure that the validation report outputs the entities without links to external resources. The DQ measure calculated from the validation result is a \RatioMeasure.

\begin{minipage}{\linewidth}
\begin{lstlisting}[basicstyle=\ttfamily\scriptsize, caption={Interlinking - External URIs}, label={shape:interlinking_external_uris}]
ex:UsageExternalURIShape
    a sh:NodeShape ;
    sh:targetSubjectsOf owl:sameAs ; 
    sh:property [
        sh:path owl:sameAs ;
        sh:pattern "^(?!DATASET_URI)";
    ].
\end{lstlisting}
\end{minipage}

The metric I3 aims to detect local in-links \footnote{in-links: all triples from a dataset that have the resource’s URI as the object \cite{Hogan2012}} when dereferencing a URI. This requires accessing resources on the web to obtain the description document of the URI, something SHACL core cannot do. Moreover, even if the document was obtained externally (e.g., using an external tool), SHACL core is still unable to check for local in-links. This is because checking for local in-links requires identifying triples where the resource appears as the object. However, to verify the existence of values for a property, SHACL core requires specifying a certain property, using \rdfproperty{sh:path} in a property shape. In contrast, in-links are any triple that has the resource as the object, regardless of the property.

\subsubsection{Security}

\dq{Security} refers to the degree to which data is protected against changes and misuse.
For this dimension, \BaseSurvey~identified two metrics, S1 and S2. Metric S1 checks for the usage of digital signatures. The Verifiable Credential Data Integrity 1.0\footnote{\url{https://www.w3.org/TR/vc-data-integrity/}} was published by the W3C as a recommendation. It outlines cryptographic mechanisms for verifying the authenticity and integrity of credentials and digital documents, primarily through the application of digital signatures and associated mathematical proofs. In this specification, they define \textit{data integrity proofs} and state that a digital signature is a type of data integrity proof. Therefore, to check for the usage of a digital signature, we defined shape  \shape{\ref{shape:security_digital_signature_usage}} that checks for the usage of the class \rdfproperty{DataIntegrityProof}. We also defined shape \shape{\ref{shape:security_digital_signature_properties}}, which checks for specific properties that an instance of this class should have. According to the specification, every instance of \texttt{DataIntegrityProof} must have:

\begin{enumerate}
\item \rdfproperty{sec:cryptoSuite}, with a value of type \rdfproperty{sec:cryptoSuiteString}.
\item \rdfproperty{sec:proofValue}, a base-encoded binary data used for verifying the digital proof. This value starts with ``u'' or ``z'', indicating the encoding used \footnote{\url{https://www.w3.org/TR/cid-1.0/\#multibase-0}}.
\item \rdfproperty{sec:proofOfPurpose} with one of the values: \rdfproperty{sec:authentication}, \rdfproperty{sec:assertionMethod}, \rdfproperty{sec:keyAgreement}, \rdfproperty{sec:capabilityDelegation} or \rdfproperty{sec:capabilityInvocation}
\end{enumerate}

\noindent
\begin{minipage}[t]{0.48\linewidth}
\begin{lstlisting}[basicstyle=\ttfamily\scriptsize, caption={Security - Digital Signatures}, label={shape:security_digital_signature_usage}]
ex:DigitalSignatureShape a sh:NodeShape ;
    sh:targetNode sec:DataIntegrityProof; 
    sh:property [
        sh:path [ sh:inversePath rdf:type ] ;
        sh:minCount 1 ;
    ].
\end{lstlisting}
\end{minipage}
\hfill
\begin{minipage}[t]{0.48\linewidth}
\begin{lstlisting}[basicstyle=\ttfamily\scriptsize, caption={Security - Digital Signature properties}, label={shape:security_digital_signature_properties}]
ex:DigitalSignaturePropertiesShape a sh:NodeShape ;
    sh:targetClass sec:DataIntegrityProof; 
    sh:property [
        sh:path sec:proofPurpose;
        sh:minCount 1;
        sh:in (
            sec:assertionMethod
            sec:authentication
            sec:keyAgreement
            sec:capabilityInvocation
            sec:capabilityDelegation
        );
    ];
    sh:property [
        sh:path sec:cryptosuite;
        sh:datatype sec:cryptosuiteString;
        sh:minCount 1 ;
    ];
    sh:property [
        sh:path sec:proofValue;
        sh:datatype xsd:string;
        sh:minCount 1 ;
    ].
\end{lstlisting}
\end{minipage}

Note that shapes \shape{\ref{shape:security_digital_signature_usage}} and \shape{\ref{shape:security_digital_signature_properties}} use properties from the Security Vocabulary\footnote{\url{https://www.w3.org/2025/credentials/vcdi/vocab/v2/vocabulary.html}} (namespace: \texttt{sec}). Although published by a W3C Working Group in March 2025, this vocabulary is not an official W3C recommendation. These shapes should be updated to use the recommended vocabulary in the future.

For shape \shape{\ref{shape:security_digital_signature_usage}}, the validation report would indicate a violation if there is no instance of the class \rdfproperty{sec:DataIntegrityProof}. For \shape{\ref{shape:security_digital_signature_properties}}, the validation report would flag a violation if at least one of the properties \rdfproperty{sec:proofPurpose}, \rdfproperty{sec:cryptosuite}, or \rdfproperty{sec:proofValue} is either missing or does not conform to its specified range. 

The metric S2 is defined in \cite{Flemming2011}, where they state that this metric should check for the specification of someone responsible for the data, by checking for properties such as \rdfproperty{dcterms:contributor}, \rdfproperty{dcterms:creator}, \rdfproperty{dcterms:publisher}, \rdfproperty{sioc:has\_creator}, \rdfproperty{sioc:has\_modifier}, \rdfproperty{sioc:owner\_of}, or \rdfproperty{foaf:maker}; and also for the specification of the data origin using \rdfproperty{dcterms:provenance}, \rdfproperty{dcterms:source}, \rdfproperty{sioc:earlier\_version}, or \rdfproperty{sioc:previous\_version}. With SHACL core, we defined shape \shape{\ref{shape:security_authenticity_void}} that checks for the usage of at least one of the properties \rdfproperty{dcterms:contributor}, \rdfproperty{dcterms:creator}, \rdfproperty{dcterms:publisher}, and also for the usage of the properties \rdfproperty{dcterms:source} and \rdfproperty{dcterms:provenance}, while targeting the class \rdfproperty{void:Dataset}. We limit the shape to only these properties because they are the ones that are used in the VoID vocabulary. 

Note that the class \texttt{dcat:Dataset} could also be targeted in this case, since this vocabulary also allows the usage of these properties.

\begin{minipage}{\linewidth}
\begin{lstlisting}[basicstyle=\ttfamily\scriptsize, caption={Security - Authenticity of the dataset}, label={shape:security_authenticity_void}]
ex:AuthenticityOfDatasetShape a sh:NodeShape ;
    sh:targetClass void:Dataset;
    sh:or(
        [ sh:path dcterms:contributor; sh:minCount 1; ]
        [ sh:path dcterms:creator; sh:minCount 1; ]
        [ sh:path dcterms:publisher; sh:minCount 1; ]
    );
    sh:or(
        [ sh:path dcterms:source; sh:minCount 1; ]
        [ sh:path dcterms:provenance; sh:minCount 1;]
    ).
\end{lstlisting}
\end{minipage}

The validation report for \shape{\ref{shape:security_authenticity_void}} will indicate a violation if an instance of \rdfproperty{void:Dataset} lacks any of the properties used to specify an author or the source of the dataset. The DQ measure calculated from the validation result is a \BinaryMeasure.

\subsubsection{Performance}\label{section:performance_appendix}

For this dimension, \BaseSurvey~proposed four metrics, where the metric P1 was already discussed in \autoref{section:accessibility}. The remaining metrics (P2-P4) are focused on the efficiency of a system that either stores the dataset or provides access to it (e.g., SPARQL endpoint). Therefore, these metrics describe system-level behaviors, such as low-latency or high throughput, rather than constraints on nodes and properties. Hence, SHACL core can't be used to cover these metrics. 

\subsection{Intrinsic}\label{section:intrinsic_appendix}

This section presents the shape definitions for the dimensions \dq{Syntactic Validity}, \dq{Semantic Accuracy}, \dq{Completeness}, and \dq{Conciseness}. It also includes the remaining shape definitions for the dimension \dq{Consistency}, which were not included in \autoref{section:instrinsic}.

\subsubsection{Syntactic Validity}

\dq{Syntactic validity} refers to the extent to which an RDF document complies with the rules of its serialization format. For this dimension, \BaseSurvey~identified three metrics. The first metric (SV1) checks for syntax errors in RDF documents, either using validators (e.g., W3C RDF Validation Service\footnote{\url{http://w3.org/RDF/Validator/documentation}}) or crowd-sourcing. The idea behind this metric is to verify that an RDF document is structurally valid and can be parsed without errors. SHACL core cannot be used for this metric, since SHACL works after parsing the RDF document. 

The second metric (SV2) for this dimension verifies that values are syntactically accurate. For this metric, they define multiple aspects to check (SV2A1-SV2A4). The first aspect (SVA1) verifies the usage of allowed values for a certain property, for which we defined the shapes 
\shape{\ref{shape:accuracy_allowed_values_range}}, \shape{\ref{shape:accuracy_allowed_values_list}} and \shape{\ref{shape:accuracy_allowed_values_has_value}}. 

\shape{\ref{shape:accuracy_allowed_values_range}} uses \textit{Value Range Constraints}, which let you define a range of allowed values for a specific property. This is useful for values that can be compared with operators such as $<$, $>$, $>=$ and $<=$. \shape{\ref{shape:accuracy_allowed_values_has_value}} states that a property has at least one value that is the specified RDF term. Note that the property can have other values. The last shape, \shape{\ref{shape:accuracy_allowed_values_list}} also checks for allowed values, but in this case against a list, which can contain literals or IRIs.

\begin{minipage}[t]{0.48\linewidth}
\begin{lstlisting}[basicstyle=\ttfamily\scriptsize, caption={Syntactic Validity - Range of \newline allowed values}, label={shape:accuracy_allowed_values_range}]
ex:RangeAllowedValuesShape a sh:NodeShape ;
    sh:targetSubjectsOf PROPERTY_URI; 
    sh:property [
        sh:path PROPERTY_URI;
        sh:minInclusive MIN_VALUE ; % or sh:minExclusive
        sh:maxInclusive MAX_VALUE; % or sh:maxExclusive
    ].
\end{lstlisting}
\end{minipage}
\hfill
\begin{minipage}[t]{0.48\linewidth}
\begin{lstlisting}[basicstyle=\ttfamily\scriptsize, caption={Syntactic Validity - Allowed values \newline (at least one value)}, label={shape:accuracy_allowed_values_has_value}]
ex:AllowedValuesAtLeastOneShape a sh:NodeShape ;
    sh:targetSubjectsOf PROPERTY_URI; 
    sh:property [
        sh:path PROPERTY_URI;
        sh:hasValue RDF_TERM ;
    ].
\end{lstlisting}
\end{minipage}

\begin{minipage}[t]{0.48\linewidth}
\begin{lstlisting}[basicstyle=\ttfamily\scriptsize, caption={List of allowed values}, label={shape:accuracy_allowed_values_list}]
ex:AllowedValuesShape a sh:NodeShape ;
    sh:targetSubjectsOf PROPERTY_URI; 
    sh:property [
        sh:path PROPERTY_URI;
        sh:in ( LIST_ALLOWED_VALUES ) ;
    ].
\end{lstlisting}
\end{minipage}
\hfill
\begin{minipage}[t]{0.48\linewidth}
\begin{lstlisting}[basicstyle=\ttfamily\scriptsize, caption={Syntactic Validity - Syntactic rules}, label={shape:accuracy_syntactic_rules}]
ex:SyntacticRulesShape a sh:NodeShape ;
    sh:targetSubjectsOf PROPERTY_URI; 
    sh:property [
        sh:path PROPERTY_URI;
        sh:pattern "PATTERN";
    ].
\end{lstlisting}
\end{minipage}

The second aspect SV2A2 checks if values follow syntactic rules, which refer to the type of
characters and/or the pattern of literal values \cite{SWIQAFrber2011}. In this case, we defined shape \shape{\ref{shape:accuracy_syntactic_rules}}, which uses the constraint \rdfproperty{sh:pattern} to state the pattern that values of a property should follow.

Each defined shape for the aspects SV2A1 and SV2A2 must be instantiated with a specific property. Moreover, each shape targets subjects of the specified property; hence, the validation report will flag a violation for each subject that uses these properties with a value that is not allowed (\shape{\ref{shape:accuracy_allowed_values_range}}, \shape{\ref{shape:accuracy_allowed_values_list}}), doesn't follow the specified pattern (\shape{\ref{shape:accuracy_syntactic_rules}}), or lacks the specified value (\shape{\ref{shape:accuracy_allowed_values_has_value}}). Note that for all these shapes, we need to consider assumption \assumption{3}, as we need domain knowledge to instantiate the shapes with the allowed values (\shape{\ref{shape:accuracy_allowed_values_range}}, \shape{\ref{shape:accuracy_allowed_values_list}}, \shape{\ref{shape:accuracy_allowed_values_has_value}}) or pattern (\shape{\ref{shape:accuracy_syntactic_rules}}).

The third aspect (SV2A3) verifies if values conform to the specific ``RDF pattern'' and that the ``types'' are defined for specific resources. 
Upon exchange with the authors of \cite{Zaveri2012}, we were able to capture the meaning of ``RDF pattern'' as ``the expected structural and semantic patterns in the data that arise from community practice''. For example, while the VoID vocabulary does not define a specific property to specify the homepage of a dataset, it is common practice to use \rdfproperty{foaf:homepage} for this purpose. This is something that arises from community practice, but it's not stated in the formal definition of the vocabulary. In this case, SHACL core can cover this aspect, because shapes can specify the expected structure for class instances. We illustrate a shape for this aspect in \shape{\ref{shape:accuracy_rdf_pattern}}; however, note that this depends on the vocabulary, and it's restricted to stating the class and expected (or commonly used) properties. Moreover, for this shape, it is necessary to consider assumption \assumption{3}, since to instantiate this shape we need knowledge about community practices. 

\begin{minipage}[t]{0.48\linewidth}
\centering
\begin{lstlisting}[basicstyle=\ttfamily\scriptsize, caption={Syntactic Validity - VoID RDF \newline pattern}, label={shape:accuracy_rdf_pattern}]
ex:RDFPatternVOIDShape a sh:NodeShape ;
    sh:targetClass void:Dataset; 
    sh:property [
        sh:path foaf:homepage;
        sh:minCount 1; 
        sh:datatype xsd:string;
    ].
\end{lstlisting}
\end{minipage}
\hfill
\begin{minipage}[t]{0.48\linewidth}
\begin{lstlisting}[basicstyle=\ttfamily\scriptsize, caption={Syntactic Validity - Malformed \newline literal}, label={shape:accuracy_malformed_literal}]
ex:MalformedLiteralShape a sh:NodeShape ;
    sh:targetSubjectsOf PROPERTY_URI; 
    sh:property [
        sh:path PROPERTY_URI ;
        sh:datatype DATATYPE_URI;
    ].
\end{lstlisting}
\end{minipage}


The fourth aspect (SV2A4) states the usage of outlier techniques and clustering for detecting wrong values; however, SHACL core doesn't allow the usage of advanced techniques. Hence, this last aspect can't be covered with SHACL core.

Overall, we argue that SHACL core can partially cover the metric SV2, as we were unable to define shapes to check all aspects of this metric.

The last metric (SV3) identifies malformed literals, which can be literals that are members of another datatype or that don't conform to the lexical syntax of their datatype. In this case, we defined \shape{\ref{shape:accuracy_malformed_literal}}, which enforces the constraint \rdfproperty{sh:datatype}, that specifies the expected datatype for the values of the property. This constraint also raises a violation if the value is an ill-typed literal, for the datatypes supported by SPARQL 1.1 \cite{w3c_shacl}. For datatypes not supported by SPARQL 1.1, shape \shape{\ref{shape:accuracy_malformed_literal}} should be modified to use \rdfproperty{sh:pattern} to define the expected value format (datatype).


To instantiate \shape{\ref{shape:accuracy_malformed_literal}}, we need to obtain the properties that have a datatype range defined in the ontology or vocabularies used by the dataset. Moreover, the validation report for this shape outputs a result for each entity that uses the property with a malformed literal (either an incorrect datatype or incorrect syntax for the used datatype). The DQ measure calculated from the validation result is a \CompositeMeasure.

\subsubsection{Semantic Accuracy}
\dq{Semantic accuracy} is the extent to which data values accurately reflect real-world facts. \BaseSurvey~identified five metrics for this dimension. Metric SA1 focuses on identifying outliers by using advanced techniques for outlier detection or by leveraging statistical distributions. However, SHACL core does not permit the use of either approach.

The second metric (SA2) verifies that there are no inaccurate values by checking three aspects: whether functional dependencies hold in the data (SA2A1), by comparing literal values of a resource (SA2A2), or via crowd-sourcing (SA2A3). The only way to compare values of different properties with SHACL core is by leveraging \textit{Property Pair Constraint} components. However, these constraints allow you to compare values between different properties when the triples share the same subject. To verify the aspect SA2A1, the comparison needs to be done across multiple entities; hence, triples won't share the same subject. Additionally, the aspect SA2A3 can't be covered with SHACL core, since SHACL is a declarative language, and crowd-sourcing involves human input. 
Therefore, for this metric, we were only able to define shape \shape{\ref{shape:semantic_accuracy_inaccurate_values}} for the aspect SA2A2, which defines a constraint between literal values of a resource, using the constraint \rdfproperty{sh:equals} as an example. Note that this shape can be modified to consider the other \textit{Property Pair} constraints \rdfproperty{sh:lessThan}, \rdfproperty{sh:lessThanOrEquals}, or \rdfproperty{sh:disjoint}.

\begin{minipage}{\linewidth}
\begin{lstlisting}[basicstyle=\ttfamily\scriptsize, caption={Semantic Accuracy - Inaccurate values}, label={shape:semantic_accuracy_inaccurate_values}]
ex:InaccurateValuesShape a sh:NodeShape ;
    sh:targetSubjectsOf PROPERTY_URI_1; 
    sh:property [
        sh:path PROPERTY_URI_1 ;
        sh:equals PROPERTY_URI_2;
    ].
\end{lstlisting}
\end{minipage}

For \shape{\ref{shape:semantic_accuracy_inaccurate_values}}, the validation report will output a violation for each subject that uses \texttt{PROPERTY\_URI\_1}, where the set of values for \rdfproperty{PROPERTY\_URI\_1} and \rdfproperty{PROPERTY\_URI\_2} don't conform to the specified constraint in the shape. For this shape, we must consider assumption \assumption{3}, since domain knowledge is required to determine the necessary constraints and properties for its instantiation.

The third metric (SA3) verifies there are no inaccurate labels and classifications by calculating the ratio \( 1 - (\frac{\text{inaccurate instances}}{\text{total no. of instances}} \times \frac{\text{balanced distance metric}}{\text{total no. of instances}} ) \). Inaccurate labeling happens when an object of the real-world has been correctly represented in the dataset, but has been incorrectly labeled (e.g. the value of \rdfproperty{rdfs:label} is incorrect). Moreover, inaccurate classification refers to instances in the dataset that aren't classified with the correct type. For example, a dataset that contains information about a University includes instances of their professors; however, some have been classified as instances of the class \textit{Person}, instead of \textit{Professor}, which is the most precise one~\cite{Lei2007}.

With SHACL core, we defined shapes \shape{\ref{shape:semantic_accuracy_inaccurate_annotations}} and \shape{\ref{shape:semantic_accuracy_inaccurate_classifications}}, which target each node and enforce that they have a specific label or type, respectively. Note that to instantiate both shapes, we need to consider assumption \assumption{3}, since we need domain knowledge, such as a gold standard, to know the correct labels and types for each entity.

\begin{minipage}[t]{0.48\linewidth}
\begin{lstlisting}[basicstyle=\ttfamily\scriptsize, caption={Semantic Accuracy - No inaccurate \newline annotations}, label={shape:semantic_accuracy_inaccurate_annotations}]
ex:NoInaccurateAnnotationsShape a sh:NodeShape ;
    sh:targetNode ex:Person1; 
    sh:property [
        sh:path rdfs:label;
        sh:in ( "Person 1"@en "Persona 1"@es );
    ].
\end{lstlisting}
\end{minipage}
\hfill
\begin{minipage}[t]{0.48\linewidth}
\begin{lstlisting}[basicstyle=\ttfamily\scriptsize, caption={Semantic Accuracy - No inaccurate \newline classifications}, label={shape:semantic_accuracy_inaccurate_classifications}]
ex:NoInaccurateClassificationsShape a sh:NodeShape ;
    sh:targetNode ex:Person1; 
    sh:property [
        sh:path rdf:type;
        sh:in ( ex:Person ex:Human );
    ].
\end{lstlisting}
\end{minipage}

In this case, the validation report will output a violation if the value for \rdfproperty{rdfs:label} or \rdfproperty{rdf:type} is not in the list of ``allowed values'', for each node used to instantiate the shapes. From the result of the validation report, one can calculate a \RatioMeasure~like \( 1 - (\frac{\text{inaccurate instances}}{\text{total no. of instances}} ) \). However, with SHACL core, we cannot calculate the formula presented for this metric, as it requires applying a similarity measure (such as the Balanced Distance Metric from the original work \cite{Lei2007}) between the annotation and a reference classification. SHACL core only supports exact value matching, not similarity-based comparisons. Therefore, we state that SHACL core can partially address this metric.

The fourth metric (SA4) verifies there's no misuse of properties by leveraging profiling statistics to identify discordant values. In this case, it's obvious SHACL core cannot be used, since it doesn't provide a way to profile the data. 

The fifth metric (SA5) states the detection of valid rules by calculating the ratio of the number of semantically valid rules to the number of non-trivial rules. The authors who proposed this metric use a semantic network to represent knowledge, where vertices represent dataset attributes or domain concepts, and association edges represent association rules between vertices \cite{Chen2010}. Additionally, domain experts can label these association edges with "KNOWN" and "BASIC" tags to indicate the support of that association rule.
Since constructing this network can be time-consuming for users, the authors introduce methods to generate new rules by using induction and analogy methods, based on existing rules within the network. 
To assess the quality of the generated rules, they calculate the ratio \(\frac{\text{semantically valid rules}}{\text{number of non-trivial rules}}\), where the non-trivial rules are the ones that provide new associations. 

Overall, it's clear that SHACL core can't cover the metric SA5, since generating new hypotheses from association rules using induction or analogy is not a capability of this language. Moreover, SHACL cannot classify generated rules as being semantically valid or invalid.

\subsubsection{Consistency}\label{section:consistency_appendix}

In this section, we describe the coverage with SHACL core for the remaining metrics defined for \dq{Consistency}. The first metric (CN1) verifies that there are no entities that are members of disjoint classes. For this metric, we defined \shape{\ref{shape:consistency_entities_in_disjoint_classes}}, which targets one class (\rdfproperty{CLASS\_URI}) and enforces the constraint that its entities shouldn't be instances of another class (\rdfproperty{DISJOINT\_CLASS\_URI}).

\begin{minipage}[t]{\linewidth}
\begin{lstlisting}[basicstyle=\ttfamily\scriptsize, caption={Consistency - Entities in disjoint classes}, label={shape:consistency_entities_in_disjoint_classes}]
ex:EntitiesDisjointClassesShape a sh:NodeShape ;
    sh:targetClass CLASS_URI; 
    sh:not [ sh:class DISJOINT_CLASS_URI].
\end{lstlisting}
\end{minipage}

\shape{\ref{shape:consistency_entities_in_disjoint_classes}} needs to be instantiated with URIs of classes that are defined as disjoint in the ontology and used in the dataset. Therefore, we need to consider assumption \assumption{2}.
The validation result for \shape{\ref{shape:consistency_entities_in_disjoint_classes}} will output a violation for each instance of the class \rdfproperty{CLASS\_URI} that is also an instance of \rdfproperty{DISJOINT\_CLASS\_URI}. The DQ measure calculated from the validation result is a \CompositeMeasure.

The second metric (CN2) verifies there are no misplaced classes or properties, by checking that there are no classes used as properties (as predicates in triples), and no properties used as classes (as objects of triples with the predicate \rdfproperty{rdf:type}) \cite{HoganHPDP10}. This metric is covered using SHACL core with shapes \shape{\ref{shape:consistency_misplaced_properties}} and \shape{\ref{shape:consistency_misplaced_classes}}. \shape{\ref{shape:consistency_misplaced_properties}} targets a property and restricts incoming \rdfproperty{rdf:type} links, ensuring properties aren’t misused as classes.

\noindent
\begin{minipage}[t]{0.48\linewidth}
\begin{lstlisting}[basicstyle=\ttfamily\scriptsize, caption={Consistency - No misplaced \newline properties}, label={shape:consistency_misplaced_properties}]
ex:MisplacedPropertiesShape a sh:NodeShape ;
    sh:targetNode PROPERTY_URI; 
    sh:property [
        sh:path [sh:inversePath rdf:type];
        sh:maxCount 0; 
    ].
\end{lstlisting}
\end{minipage}
\hfill
\begin{minipage}[t]{0.48\linewidth}
\begin{lstlisting}[basicstyle=\ttfamily\scriptsize, caption={Consistency - No misplaced classes}, label={shape:consistency_misplaced_classes}]
ex:MisplacedClassesShape a sh:NodeShape ;
  sh:targetSubjectsOf rdf:type;
  sh:or (
    [sh:path rdf:type; sh:hasValue rdfs:Class;]
    [sh:path rdf:type; sh:hasValue rdf:Property;]
    [sh:path rdf:type; sh:hasValue owl:NamedIndividual;]
    [
        sh:path CLASS_URI;
        sh:maxCount 0;
    ]
  ).
\end{lstlisting}
\end{minipage}

\shape{\ref{shape:consistency_misplaced_properties}} needs to be instantiated with the properties' URIs that are defined in the ontology. Therefore, we need to consider assumption \assumption{2}. The validation result will output a violation for each property that is used as a class. The DQ measure calculated from the validation result is a \CompositeMeasure.

\shape{\ref{shape:consistency_misplaced_classes}} targets entities and constraints that there should be no values for a property with the URI of the class. For this shape, we need to consider assumptions \assumption{1} and \assumption{2}, given that we are targeting entities and we need to instantiate the shape with classes defined in the ontology. For \shape{\ref{shape:consistency_misplaced_classes}}, the validation report will output a violation for each of the entities that use the class as a property. In this case, we also defined the DQ measure as a \CompositeMeasure.

The third metric (CN3) checks whether \rdfproperty{owl:DatatypeProperty} and \rdfproperty{owl:ObjectProperty} are misused. For this metric, the authors in \cite{Zaveri2012} state that this should be verified through the ontology maintainer or engineer. Although this task is specified as manual, we successfully defined two shapes to verify the misuse of these property types. While shape \shape{\ref{shape:consistency_misuse_datatype_properties}} targets subjects of an \rdfproperty{owl:DatatypeProperty} and constraints that the values are literals, shape \shape{\ref{shape:consistency_misuse_object_properties}} targets subjects of an \rdfproperty{owl:ObjectProperty} and constraints that the values are IRIs or blank nodes.
We target the subjects of the property so that the validation results outputs the subjects that misuse these properties.

\noindent
\begin{minipage}[t]{0.48\linewidth}
\begin{lstlisting}[basicstyle=\ttfamily\scriptsize, caption={Consistency - No misuse of Datatype \newline properties}, label={shape:consistency_misuse_datatype_properties}]
ex:MisuseDatatypePropertiesShape a sh:NodeShape ;
    sh:targetSubjectsOf PROPERTY_URI;
    sh:property [
        sh:path PROPERTY_URI;
        sh:nodeKind sh:Literal;
    ].
\end{lstlisting}
\end{minipage}
\hfill
\begin{minipage}[t]{0.48\linewidth}
\begin{lstlisting}[basicstyle=\ttfamily\scriptsize, caption={Consistency - No misuse of Object \newline properties}, label={shape:consistency_misuse_object_properties}]
ex:MisuseObjectPropertiesShape a sh:NodeShape ;
    sh:targetSubjectsOf PROPERTY_URI;
    sh:property [
        sh:path PROPERTY_URI;
        sh:nodeKind sh:BlankNodeOrIRI;
    ].
\end{lstlisting}
\end{minipage}

In this case, both shapes need to be instantiated with properties defined in the ontology, and used in the dataset, as \rdfproperty{owl:DatatypeProperty} or \rdfproperty{owl:ObjectProperty}, respectively. Therefore, we need to consider assumption \assumption{2}. Moreover, for this metric, we calculate two different measures, one for the misuse of \rdfproperty{owl:DatatypeProperty} and the other for the misuse of \rdfproperty{owl:ObjectProperty}. In both cases, the DQ measure is a \CompositeMeasure. 

Metric CN4 verifies that deprecated classes (\rdfproperty{owl:DeprecatedClass}) and properties (\rdfproperty{owl:DeprecatedProperty}) are not used. \BaseSurvey~states that the ontology maintainer should detect the usage of these classes and properties. However, with SHACL core, we defined shape \shape{\ref{shape:consistency_deprecated_class}} to check for the usage of deprecated classes. This shape targets entities and ensures that they are not typed as instances of any of the deprecated classes. Hence, the validation report will output a violation for each entity that is an instance of any of the deprecated classes. For this shape, we need to consider assumptions \assumption{1} and \assumption{2} because we are targeting entities, and we need the list of deprecated classes from the ontology.

\noindent
\begin{minipage}[t]{0.48\linewidth}
\begin{lstlisting}[basicstyle=\ttfamily\scriptsize, caption={Consistency - Usage of deprecated \newline classes}, label={shape:consistency_deprecated_class}]
ex:DeprecatedClassesShape a sh:NodeShape ;
    sh:targetSubjectsOf rdf:type;
    sh:or (
        [sh:path rdf:type; sh:hasValue rdfs:Class;]
        [sh:path rdf:type; sh:hasValue rdf:Property;]
        [sh:path rdf:type; sh:hasValue owl:NamedIndividual;]
        [
            sh:path rdf:type;
            sh:not [ sh:in (CLASSES_LIST); ];
        ]
    ).
\end{lstlisting}
\end{minipage}
\hfill
\begin{minipage}[t]{0.48\linewidth}
\begin{lstlisting}[basicstyle=\ttfamily\scriptsize, caption={Consistency - Usage of deprecated \newline properties}, label={shape:consistency_deprecated_property}]
ex:DeprecatedPropertiesUsageShape a sh:NodeShape ;
    sh:targetSubjectsOf rdf:type;
    sh:or (
        [sh:path rdf:type; sh:hasValue rdfs:Class;]
        [sh:path rdf:type; sh:hasValue rdf:Property;]
        [sh:path rdf:type; sh:hasValue owl:NamedIndividual;]
        [ sh:path PROPERTY_URI; sh:maxCount 0; ]
    ).
\end{lstlisting}
\end{minipage}

Moreover, to detect the usage of deprecated properties, we target entities and check that none of them use the property (See \shape{\ref{shape:consistency_deprecated_property}}). The validation report for this shape will output the entities that use the deprecated property at least once. Once again, for this shape we also need to consider assumptions \assumption{1} and \assumption{2}.

For \shape{\ref{shape:consistency_deprecated_class}}, the DQ measure calculated from the validation result is a \BinaryMeasure, while for  \shape{\ref{shape:consistency_deprecated_property}}~it's a \CompositeMeasure.

Metric CN6 checks the absence of ontology hijacking, which occurs when terms from the ontology of the dataset are redefined by a third party, which ends up affecting inference and reasoning \cite{HoganHPDP10}. This metric entails two aspects: identifying the redefinition of terms in third-party ontologies and checking if inference is affected by this new definition. The first aspect can't be covered with SHACL core, as it lacks a mechanism to determine whether a property or class originates from the original vocabulary or is a redefinition. The second aspect entails detecting if reasoning is affected, which implies checking whether two ontologies' inferences are different due to local redefinitions of properties or classes, which goes beyond the capabilities of SHACL core. 

Metric CN7 checks for no negative dependencies or correlation among properties by using association rules. This is defined in \cite{Bohm2010}, where they first cluster the data into semantically correlated subsets and then they apply advanced techniques to identify correlations between properties, inverse relations, among others. The authors state that their tool is able to detect positive and negative association rules, which show occurrence dependencies among properties in a cluster. In particular, negative association rules state that properties on both sides of the association rule never appear together in instances of the cluster. An example of negative association rules is $\mathrm{title}\rightarrow \neg\,\mathrm{name}$, which specifies that no instance has both properties, $\mathrm{title}$ and $\mathrm{name}$, at the same time. We consider that this metric can be partially addressed using SHACL core by translating association rules into expressions that use only logical operators, such as $\land$, $\lor$, and $\neg$, and then representing these expressions as SHACL constraints to validate whether the rules hold on the data. It is important to note that, while association rules are typically mined based on support and confidence from data, the translation presented here treats the rule as a strict logical constraint.

Formally, association rules are defined as an implication of the form $X \rightarrow Y$ where $X$ and $Y$ are a set of items, called itemsets, and $X \cap Y = \emptyset$ \cite{Antonie2004}. In \cite{Bohm2010}, itemsets contain properties, and each itemset represents the existence of properties for entities in a cluster. 

To illustrate how the shape is defined, we consider a \textit{confined negative association rule}, which can take the form $X \rightarrow \neg Y$, $\neg X \rightarrow Y$, or $\neg X \rightarrow \neg Y$, where the antecedent or consequent must be a conjunction of negated or non-negated attributes \cite{Antonie2004}. While we use a rule of the form $X \rightarrow \neg Y$ to explain the shape construction, any of the other forms can also be translated.

Therefore, given the following negative association rule 
\[
 (A_1 \land \ldots \land A_n ) \rightarrow ( \neg B_1 \land \ldots \land \neg B_m ),
\]
we can use the logical equivalence $p \rightarrow q$ \texttt{eq} $\neg p \lor q$, to express the operator $\rightarrow$ with the operators $\neg$ and $\lor$, obtaining the following rule:
\[
\neg ( A_1 \land \ldots \land A_n ) \lor ( \neg B_1 \land \ldots \land \neg B_m ).
\]

By applying De Morgan, we can obtain the final formula: 
\begin{equation}
\label{eq:association_rule}
( \neg A_1 \lor \ldots \lor \neg A_n ) \lor ( \neg B_1 \land \ldots \land \neg B_m ).
\end{equation}

Hence, for the rewritten association rule in (\ref{eq:association_rule}), we defined shape \shape{\ref{shape:consistency_negative_dependencies}}, which expresses the negative association rule. Note that from the formula (\ref{eq:association_rule}), $\neg A_{i}$ and $\neg B_{i}$ state that the attribute $A_{i}$ or $B_{i}$ should not be present. Therefore, instead of using the constraint \rdfproperty{sh:minCount 1} with \rdfproperty{sh:not}, we use the constraint \rdfproperty{sh:minCount 0}.
Moreover, in this shape, we target nodes of a certain class that should comply with the association rule. A validation result for this shape would output all instances of the class that don't conform to this association rule.

\begin{minipage}[t]{\linewidth}
\begin{lstlisting}[basicstyle=\ttfamily\scriptsize, caption={Consistency - Negative dependencies}, label={shape:consistency_negative_dependencies}]
ex:NegativeDependenciesShape a sh:NodeShape ;
    sh:targetClass CLASS_URI;
    sh:or (
        [   % First group: OR of negated A's
            sh:or (
                [sh:path ex:A_1; sh:minCount 0;]
                ...
                [sh:path ex:A_n; sh:minCount 0;]
            )
        ]
        [   % Second group: AND of negated B's
            sh:and (
                [sh:path ex:B_1; sh:minCount 0;]
                ...
                [sh:path ex:B_m; sh:minCount 0;]
            )
        ]
    ).\end{lstlisting}
\end{minipage}

For this shape, we assume that association rules are defined for instances of a specific class. Therefore, this shape needs to be instantiated with the corresponding class. Lastly, this shape relies on assumption \assumption{3}, since the association rules must come from domain knowledge.

Metric CN8 checks that there are no inconsistencies in spatial data. \BaseSurvey~cited \cite{Mostafavi2004} where the definition of constraints for checking inconsistencies in spatial data is implemented using Prolog and the GeoMedia software. Moreover, the formulation of these constraints depends on how the spatial data is represented, particularly concerning specific ontologies. Therefore, we won't provide a shape for this metric, and we mark it as not covered in \autoref{table:intrinsic_metrics}.

Metric CN9 verifies that properties are used with the correct domain and range, as defined in the ontology or vocabulary. The original authors suggest using SPARQL queries as constraints to detect correct property usage. In our case, we used SHACL core to define these constraints and were able to fully cover this metric with two shapes. \shape{\ref{shape:consistency_correct_domain}} targets subjects of a property. It uses a class constraint to restrict the types of entities that use the property, and \shape{\ref{shape:consistency_correct_range}} targets objects of the property and constraints its values, with either a specific datatype or class, depending on the range definition of the property.

\noindent
\begin{minipage}[t]{0.48\linewidth}
\begin{lstlisting}[basicstyle=\ttfamily\scriptsize, caption={Consistency - Correct domain \newline (specific class)}, label={shape:consistency_correct_domain}]
ex:CorrectDomainShape a sh:NodeShape ;
    sh:targetSubjectsOf PROPERTY_URI;
    sh:class CLASS.
\end{lstlisting}
\end{minipage}
\hfill
\begin{minipage}[t]{0.48\linewidth}
\begin{lstlisting}[basicstyle=\ttfamily\scriptsize, caption={Consistency - Correct range \newline (specific datatype or class)}, label={shape:consistency_correct_range}]
ex:CorrectRangeShape a sh:NodeShape ;
    sh:targetSubjectsOf PROPERTY_URI;
    sh:property [
        sh:path PROPERTY_URI;
        sh:datatype DATATYPE;  % or sh:class CLASS
    ].
\end{lstlisting}
\end{minipage}

Note that other variants of these shapes can be defined depending on the domain or range of the property. For example, for \shape{\ref{shape:consistency_correct_range}}, if the range is \rdfproperty{rdfs:Literal}, \rdfproperty{owl:Thing}, or \rdfproperty{rdfs:Resource}, it does not specify a concrete datatype or class. In such cases, the correct usage can be checked by applying the constraint \rdfproperty{sh:nodeKind sh:Literal} (See \shape{\ref{shape:consistency_correct_range_literal}}), \rdfproperty{sh:nodeKind sh:BlankNodeOrIRI} (See \shape{\ref{shape:consistency_correct_range_owl_thing}}), or a combination of both constraints using \rdfproperty{sh:or} (See \shape{\ref{shape:consistency_correct_range_rdfs_resource}}). Similarly, for \shape{\ref{shape:consistency_correct_domain}}, if the domain is defined as \rdfproperty{owl:Thing}, correct usage can be enforced by validating the node kind with \rdfproperty{sh:BlankNodeOrIRI} (See \shape{\ref{shape:consistency_correct_domain_owl_thing}}). 
\noindent

\begin{minipage}[t]{0.48\linewidth}
\begin{lstlisting}[basicstyle=\ttfamily\scriptsize, caption={Consistency - Correct domain \newline (owl:Thing)}, label={shape:consistency_correct_domain_owl_thing}]
ex:CorrectDomainShape a sh:NodeShape;
    sh:targetSubjectsOf PROPERTY_URI;
    sh:nodeKind sh:BlankNodeOrIRI.
\end{lstlisting}
\end{minipage}
\hfill
\begin{minipage}[t]{0.48\linewidth}
\begin{lstlisting}[basicstyle=\ttfamily\scriptsize, caption={Consistency - Correct range \newline (owl:Thing)}, label={shape:consistency_correct_range_owl_thing}]
ex:CorrectRangeShape a sh:NodeShape;
    sh:targetSubjectsOf PROPERTY_URI;
    sh:property [
        sh:path PROPERTY_URI;
        sh:nodeKind sh:BlankNodeOrIRI;  
    ].
\end{lstlisting}
\end{minipage}

\noindent
\begin{minipage}[t]{0.48\linewidth}
\begin{lstlisting}[basicstyle=\ttfamily\scriptsize, caption={Consistency - Correct range \newline (rdfs:Literal)}, label={shape:consistency_correct_range_literal}]
ex:CorrectRangeShape a sh:NodeShape ;
    sh:targetSubjectsOf PROPERTY_URI;
    sh:property [
        sh:path PROPERTY_URI;
        sh:nodeKind sh:Literal;  
    ].
\end{lstlisting}
\end{minipage}
\hfill
\begin{minipage}[t]{0.48\linewidth}
\begin{lstlisting}[basicstyle=\ttfamily\scriptsize, caption={Consistency - Correct range \newline (rdfs:Resource)}, label={shape:consistency_correct_range_rdfs_resource}]
ex:CorrectRangeShape a sh:NodeShape ;
    sh:targetSubjectsOf PROPERTY_URI;
    sh:property [
        sh:path PROPERTY_URI;
        sh:or (
            [ sh:nodeKind sh:BlankNodeOrIri; ]
            [ sh:nodeKind sh:Literal; ]
        );
    ].
\end{lstlisting}
\end{minipage}

All shapes defined for this metric must be instantiated with properties from the ontology that have a defined range or domain. Therefore, we consider assumption \assumption{2}. The validation report will output a violation for each entity that uses the property incorrectly (either a wrong domain or range, depending on the shape). The DQ measure for all shapes is a \CompositeMeasure.

Lastly, metric CN10 verifies that there are no inconsistent values by generating a set of axioms for all the properties in the dataset, and then manually checking the correctness of the generated axioms. This metric is proposed in \cite{Zaveri2013} where the authors generate (inverse) functional, irreflexive, and asymmetric axioms for the different properties using DL-Learner\footnote{\url{https://aksw.org/Projects/DLLearner.html}} and then manually check the correctness of these axioms.


In this case, SHACL core can't be used, as this metric entails generating axioms from the data, which exceeds SHACL core's capabilities. However, we argue that SHACL core can partially cover this metric because it can check the correctness of irreflexive, (inverse) functional, and asymmetric properties. 
To check the correctness of irreflexive properties, we defined shape \shape{\ref{shape:consistency_irreflexive_property}}, which ensures that no entity is related to itself via the property \rdfproperty{PROPERTY\_URI}. A violation is reported for any entity that appears as both subject and object of \rdfproperty{PROPERTY\_URI} in a triple. The DQ measure is a \CompositeMeasure.

\noindent
\begin{minipage}[t]{0.48\linewidth}
\begin{lstlisting}[basicstyle=\ttfamily\scriptsize, caption={Consistency - No inconsistent \newline values (Irreflexive property)}, label={shape:consistency_irreflexive_property}]
ex:IrreflexivePropertyShape a sh:NodeShape ;
    sh:targetSubjectsOf PROPERTY_URI;
    sh:disjoint PROPERTY\_URI.
\end{lstlisting}
\end{minipage}
\hfill
\begin{minipage}[t]{0.48\linewidth}
\begin{lstlisting}[basicstyle=\ttfamily\scriptsize, caption={Consistency - No inconsistent \newline values (Functional property)}, label={shape:consistency_functional_property}]
ex:FunctionalPropertyShape a sh:NodeShape ;
    sh:targetSubjectsOf PROPERTY_URI;
    sh:property [
        sh:path PROPERTY_URI;
        sh:maxCount 1;
    ].
\end{lstlisting}
\end{minipage}


For functional properties, we defined shape \shape{\ref{shape:consistency_functional_property}}, which ensures that each subject has at most one value for the functional property. A violation is reported for any subject with multiple values for \rdfproperty{PROPERTY\_URI}. The DQ measure is a \CompositeMeasure.

In the case of inverse functional properties, to check for their correctness, the shape \shape{\ref{shape:consistency_uniqueness_inverse_functional_property}} defined for the metric CN5 can be used.

Finally, with regards to asymmetric properties, we defined \shape{\ref{shape:consistency_asymmetric_property}} that checks that the set of nodes that point to the focus node (\rdfproperty{[sh:inversePath PROPERTY\_URI]}) is disjoint with the set of nodes pointed by the focus node (\rdfproperty{sh:disjoint PROPERTY\_URI}). If these sets overlap, it means that there exists one triple \rdftriple{ex:A}{PROPERTY\_URI}{ex:B}, for which the triple \rdftriple{ex:B}{PROPERTY\_URI}{ex:A} exists, violating the asymmetry constraint.

\begin{minipage}[t]{\linewidth}
\begin{lstlisting}[basicstyle=\ttfamily\scriptsize, caption={Consistency - No inconsistent values (Asymmetric property)}, label={shape:consistency_asymmetric_property}]
ex:AsymmetricPropertyShape a sh:NodeShape ;
    sh:targetSubjectsOf PROPERTY_URI;
    sh:property [
        sh:path [sh:inversePath PROPERTY_URI];
        sh:disjoint PROPERTY_URI;
    ].
\end{lstlisting}
\end{minipage}



\subsubsection{Conciseness}\label{section:conciseness}

\dq{Conciseness} refers to minimizing redundancy at both the schema and data levels.
For this dimension, \BaseSurvey~identified three metrics (CS1-CS3). The first metric (CS1) refers to high intensional conciseness, which can be measured by calculating the ratio 
\(\frac{\text{\# unique properties/classes of a dataset}}{\text{\# of properties/classes in a target schema}}\)  (for the case of linked data, the schema is represented by the ontology or vocabularies used by the dataset) \cite{SieveMendes2012}. In this case, to identify non-unique properties and classes, we would need to recognize equivalent properties and classes with different names or identifiers (e.g., a dataset can use the properties \textit{name} and \textit{firstName}. While they have different names, they refer to the same concept) \cite{SieveMendes2012}. However, this entails understanding the semantics of properties and classes to decide if they are equivalent in meaning, which is something SHACL core cannot do. 

The second metric (CS2) refers to high extensional conciseness, where they identified two ways of measuring this: 
\begin{enumerate}
    \item \( \displaystyle \frac{\text{\# unique objects of a dataset}}{\text{\# objects representations in the dataset}}\)
    \item \( \displaystyle 1 - \frac{\text{\# instances that violate the uniqueness rule}}{\text{\# relevant instances}} \)  
\end{enumerate}


In the first case, to obtain the number of unique objects, we need to verify that there are no two instances which are equal but use different identifiers (in linked data, this refers to the URI). We would need to check that there are no two entities with different URIs with the same (or similar) values for all their properties. However, this would require comparing property values across different entities, which SHACL core doesn’t support. \textit{Property Pair} constraints only allow comparisons between properties of the same subject. Moreover, to determine the similarity between the values of properties, we would need to use some similarity measure, which is not something that can be done with SHACL core.

For the second ratio (defined in \cite{SWIQAFrber2011}, which states that uniqueness rules define properties where each literal value occurs only once), we can leverage SHACL core to identify properties that are defined as unique but are not used correctly. Hence, we defined the shape \shape{\ref{shape:conciseness_uniqueness_rule}} which targets objects of a specific property and checks that each object value is only referenced by one subject. Hence, if there's more than one subject pointing to the value of the property, these entities violate the uniqueness rule defined by the property. In this case, we need to consider assumption \assumption{3}, since we need domain knowledge that tells us which properties define a uniqueness rule.

\begin{minipage}[t]{\linewidth}
\begin{lstlisting}[basicstyle=\ttfamily\scriptsize, caption={Extensional conciseness - Uniqueness rule}, label={shape:conciseness_uniqueness_rule}]
ex:UniquenessRuleShape a sh:NodeShape ;
  sh:targetObjectsOf PROPERTY_URI ; 
  sh:property [ 
    sh:path [ sh:inversePath PROPERTY_URI ] ;
    sh:maxCount 1 ;
  ] .
\end{lstlisting}
\end{minipage}

The validation report for \shape{\ref{shape:conciseness_uniqueness_rule}} outputs a violation for each of the values of the property that are used more than once. 

As some aspects of CS2 cannot be covered, we consider that SHACL core can partially cover it. Finally, metric CS3 verifies that there's no usage of unambiguous annotations, such as labels and descriptions. In this case, SHACL core can't be used, as recognizing ambiguous annotations requires understanding the semantics of annotations and determining if they are precise enough to map each entity in your dataset to a single real-world object~\cite{Lei2007}.

\subsubsection{Completeness}\label{section:completeness}
\dq{Completeness} measures how much required information is present in a dataset. 
For this dimension, \BaseSurvey~identified four metrics. Metric CP1 evaluates schema completeness, which involves verifying whether all the properties and classes defined in the schema (vocabularies and ontologies) are actually used in the dataset. With SHACL core, we were only able to define a single shape for this metric, which checks if classes defined in a schema are used in the dataset. For this, we target the specific class and check that there's at least one triple with the class as the object and \rdfproperty{rdf:type} as the predicate (See \shape{\ref{shape:completeness_schema_completeness_class_usage}}). To ensure that this triple represents a true instance declaration, the shape also requires that the subject of this triple is not an \rdfproperty{owl:NamedIndividual}. This is achieved through the use of \rdfproperty{sh:qualifiedValueShape} and \rdfproperty{sh:qualifiedMinCount} constraints. This distinction is important because some vocabularies and ontologies include predefined individuals, which should not be counted as instances created in the dataset.

\noindent
\begin{minipage}[t]{0.48\linewidth}
\begin{lstlisting}[basicstyle=\ttfamily\scriptsize, caption={Completeness - Schema \newline completeness (Class usage)}, label={shape:completeness_schema_completeness_class_usage}]
ex:NotNamedIndividualShape a sh:NodeShape;
    sh:property [
        sh:path rdf:type ;
        sh:not [ sh:hasValue owl:NamedIndividual ] ;
    ].

ex:SchemaCompletenessClassUsageShape a sh:NodeShape ;
    sh:targetNode CLASS_URI ;
    sh:property [
        sh:path [ sh:inversePath rdf:type ] ;   
        sh:minCount 1 ;   
        sh:qualifiedValueShape [
            sh:node ex:NotNamedIndividualShape ;
        ];                     
        sh:qualifiedMinCount 1 ;
    ].
\end{lstlisting}
\end{minipage}
\hfill
\begin{minipage}[t]{0.48\linewidth}
\begin{lstlisting}[basicstyle=\ttfamily\scriptsize, caption={Completeness - Property \newline completeness}, label={shape:completeness_property_completeness_general}]
ex:PropertyCompletenessShape a sh:NodeShape ;
  sh:targetSubjectsOf PROPERTY_URI; 
  sh:property [ 
    sh:path PROPERTY_URI;
    sh:minCount COUNT;
  ] .
\end{lstlisting}
\end{minipage}



The validation report for \shape{\ref{shape:completeness_schema_completeness_class_usage}} outputs a violation for each class that isn't used in the dataset. The DQ measure in this case is a \CompositeMeasure.

We were unable to define a shape to check for the usage of properties, as SHACL core can only target subjects or objects in triples. However, to verify property usage, we must check if at least a triple uses that property; therefore, we need to target the predicate in triples. Therefore, with SHACL core, we are only able to partially cover this metric.

Metric CP2 verifies property completeness, for which two aspects were identified. The first aspect calculates the ratio \(\frac{\texttt{\# values present for a property}}{\texttt{total amount of values that the property should have}}\). In this case, we can use SHACL core's cardinality constraints to state the expected number of values that the property should have (See \shape{\ref{shape:completeness_property_completeness_general}}). Note that this shape could be modified to target a specific class and check if instances of the class have the expected number of values for the property. The validation report for \shape{\ref{shape:completeness_property_completeness_general}} outputs all the entities that use the property \rdfproperty{PROPERTY\_URI} with fewer values than the minimum stated in the constraint \rdfproperty{sh:minCount}. Note that from the validation result, we are not able to calculate the exact ratio stated in the metric. Still, we can calculate the ratio \(\frac{\text{\# entities with correct number of values}}{\text{\# entities that use the property}}\). Additionally, for \shape{\ref{shape:completeness_property_completeness_general}} we need to consider assumption \assumption{3}, as we need domain knowledge to instantiate the expected number of values for the property. The second aspect identified for metric CP2 leverages statistical distributions of properties to characterize them and then detect completeness. However, SHACL core doesn't provide a way of calculating statistical distributions. Overall, metric CP2 can be partially covered with SHACL core.

The third metric (CP3) checks for population completeness, which verifies if all real-world objects are present in the dataset, and it's defined with the ratio \(\frac{\texttt{\# real-world entities in dataset}}{\texttt{total \# real-world entities}}\). 

To illustrate why SHACL core can partially cover this metric, we use the following example: We want to check population completeness for States in Germany. We consider two RDF modeling options, but others are possible. The first one represents ``Germany'' as an entity with states as property values (e.g., \rdftriple{ex:Germany}{ex:hasState}{ex:Berlin}), and the second one models ``States of Germany'' as a class with states as instances (e.g., \rdftriple{ex:Berlin}{rdf:type}{ex:GermanState}).
In both cases, we would need to check that there are 16 unique values/entities. However, the same entity could be duplicated, just by having a different URI (e.g., \rdfproperty{ex:Berlin} and \rdfproperty{ex:BerlinState}). Therefore, to verify this with SHACL core, we need to enforce 2 constraints (See shapes \shape{\ref{shape:completeness_population_completeness_property}} and \shape{\ref{shape:completeness_population_completeness_class}}), one with the exact number of values that should be present (using \rdfproperty{sh:minCount} and \rdfproperty{sh:maxCount}), and the other with the actual values that should be present (using \rdfproperty{sh:in}). Note that SHACL core treats triples as sets, so exact duplicates are ignored. However, different IRIs or blank nodes representing the same conceptual entity will trigger a validation error unless explicitly included in the expected value list. 

These shapes may fall short in cases where equivalent entities are declared (e.g., \rdftriple{ex:Berlin}{owl:sameAs}{ex:BerlinState}). SHACL core does not take into account this equivalence, so if both values appear as a state of Germany (in any of the two approaches), unless they are in the expected value list and the number of values is updated accordingly, there will be a validation result. Therefore, we state that this metric can be partially covered.

Lastly, instantiating both shapes requires domain knowledge to specify the expected values and their count, as well as to select the appropriate shape based on how these cases are modeled in the graph (assumption \assumption{3}).

\noindent
\begin{minipage}[t]{0.48\linewidth}
\begin{lstlisting}[basicstyle=\ttfamily\scriptsize, caption={Completeness - Population completeness (Property approach)}, label={shape:completeness_population_completeness_property}]
ex:PopulationCompletenessShape_Property
a sh:NodeShape;
  sh:targetNode ex:Germany;
  sh:property [
    sh:path ex:hasState ;
    sh:minCount 16;
    sh:maxCount 16;
    sh:in (ex:Berlin ... ex:Bavaria);
  ].
\end{lstlisting}
\end{minipage}
\hfill
\begin{minipage}[t]{0.48\linewidth}
\begin{lstlisting}[basicstyle=\ttfamily\scriptsize, caption={Completeness - Population completeness (Class approach)}, label={shape:completeness_population_completeness_class}]
ex:PopulationCompletenessShape_Class
a sh:NodeShape;
  sh:targetNode ex:GermanState;
  sh:property [
    sh:path [ sh:inversePath rdf:type ] ;
    sh:minCount 16;
    sh:maxCount 16;
    sh:in (ex:Berlin ... ex:Bavaria);
  ].
\end{lstlisting}
\end{minipage}


The last metric (CP4) checks for interlinking completeness and defines two aspects. The first aspect measures interlinking completeness by calculating the ratio of instances that are interlinked in the dataset to the total number of instances. For this aspect, we were able to define a shape that checks if entities are interlinked, by checking if subjects of \rdfproperty{rdf:type} have an interlinking property, such as \rdfproperty{owl:sameAs} (See \shape{\ref{shape:completeness_interlinking_completeness}}). For this shape, we need to consider assumptions \assumption{1} and \assumption{2}, because we are targeting entities.

\begin{minipage}[t]{\linewidth}
\begin{lstlisting}[basicstyle=\ttfamily\scriptsize, caption={Completeness - Interlinking completeness}, label={shape:completeness_interlinking_completeness}]
ex:InterlinkingCompletenessShape a sh:NodeShape ;
  sh:targetSubjectsOf rdf:type; 
  sh:or (
    [ sh:path rdf:type; sh:hasValue rdfs:Class; ]
    [ sh:path rdf:type; sh:hasValue rdf:Property; ]
    [ sh:path rdf:type; sh:hasValue owl:NamedIndividual; ]
    [ sh:path owl:sameAs; sh:minCount 1; ]
  ).
\end{lstlisting}
\end{minipage}

The validation report for \shape{\ref{shape:completeness_interlinking_completeness}} outputs all the entities that don't have a value for the property \rdfproperty{owl:sameAs}. In this case, the DQ measure is a \RatioMeasure.

The second aspect checks for linkset type completeness. A linkset is a collection of triples where all subjects are in one dataset, all objects are in another dataset, and the predicate of all the triples uses an interlinking property (e.g. \rdfproperty{owl:sameAs}). This metric measures how complete a linkset is by checking if there are links between entities of equivalent classes. Hence, the linkset is incomplete if there are equivalent classes between the datasets, but the linkset does not contain any links between entities of these types~\cite{Albertoni2013}. For this metric, SHACL cannot be used, since this metric entails checking whether any link exists between arbitrary instances of two classes. However, SHACL requires constraints to be evaluated against a specific target node, whereas this metric checks for the existence of a triple involving instances of two different classes, a condition that is not tied to any single node.

\subsection{Contextual}\label{section:contextual_appendix}
This section presents the shape definitions for the dimensions \dq{Relevancy}, \dq{Trustworthiness}, and \dq{Understandability} from the \textit{Contextual} group. It also includes the remaining shape definitions for the dimension \dq{Timeliness}, which were not included in \autoref{section:contextual}.

\subsubsection{Relevancy} 

\dq{Relevancy} refers to the ability to obtain information that aligns with the task at hand and is important to the user’s query.
For this dimension, \BaseSurvey~identified two metrics. Metric R1 looks to identify relevant data by assigning a numerical value (rank) to RDF documents or statements, or via crowd-sourcing. However, SHACL is a constraint language, which means it does not support assigning values to triples, such as ranks, or enabling processes like crowd-sourcing that require human input.

Moreover, metric R2 measures if the data retrieved is appropriate for the task at hand, by looking at the number of entities and their level of detail (e.g., number of properties). We argue that SHACL core can partially cover this metric, since we can define a shape that checks that entities of a certain class have values for specific properties. The nodes that don’t conform to the shape won’t have the required properties. To obtain a coverage value, we would need to process the validation result to count the nodes that don't conform to the shape and subtract them from the number of instances of the class. We illustrate an example shape for this metric in \shape{\ref{shape:relevancy_coverage}}, where we consider a generic class \rdfproperty{ex:Person} representing people, and define that every person should have exactly one first name and one last name. Note that this shape relies on domain knowledge (assumption \assumption{3}) that specifies the class and expected properties.

\begin{minipage}[t]{\linewidth}
\begin{lstlisting}[basicstyle=\ttfamily\scriptsize, caption={Relevancy - Coverage}, label={shape:relevancy_coverage}]
ex:CoverageOfEntitiesShape a sh:NodeShape ;
  sh:targetClass ex:Person; 
  sh:property [
    sh:path ex:firstName;
    sh:minCount 1;
    sh:maxCount 1;
  ]
  sh:property [
    sh:path ex:lastName;
    sh:minCount 1;
    sh:maxCount 1;
  ].
\end{lstlisting}
\end{minipage}

The validation report for \shape{\ref{shape:relevancy_coverage}} lists all \rdfproperty{ex:Person} entities missing any of the required properties (\rdfproperty{ex:firstName} or \rdfproperty{ex:lastName}).

\subsubsection{Understandability}\label{section:understandability}
\dq{Understandability} refers to how easily data can be comprehended without ambiguity and effectively used by a human user. For this dimension, \BaseSurvey~identified six metrics. 
Metric U1 checks for the usage of human-readable labels in entities, classes, and properties, as well as the indication of metadata of the dataset (title, description, and homepage). SHACL core can partially cover this metric, since we can define a shape that checks for the existence of labels in entities, classes, and properties. Still, we can't check if the labels are human-readable. 

The shapes for entities, classes, and properties are defined in \shape{\ref{shape:understandability_human_readable_labels}}, \shape{\ref{shape:understandability_human_readable_labels_classes}}, and \shape{\ref{shape:understandability_human_readable_labels_properties}}, respectively.
Note that \shape{\ref{shape:understandability_human_readable_labels}}, which targets entities, relies on assumptions \assumption{1} and \assumption{2}. Additionally, \shape{\ref{shape:understandability_human_readable_labels_classes}}, and \shape{\ref{shape:understandability_human_readable_labels_properties}} rely on assumption \assumption{2}, given that they target \rdfproperty{rdfs:Class} and \rdfproperty{rdf:Property}.

\begin{minipage}[t]{\linewidth}
\begin{lstlisting}[basicstyle=\ttfamily\scriptsize, caption={Understandability - Human-readable labels in entities}, label={shape:understandability_human_readable_labels}]
ex:LabelForEntitiesShape a sh:NodeShape;
  sh:targetSubjectsOf rdf:type; 
  sh:or (
    [ sh:path rdf:type; sh:hasValue rdfs:Class; ]
    [ sh:path rdf:type; sh:hasValue rdf:Property; ]
    [ sh:path rdf:type; sh:hasValue owl:NamedIndividual; ]
    [ sh:path rdfs:label; sh:minCount 1; ]
  ).
\end{lstlisting}
\end{minipage}

\begin{minipage}[t]{0.48\linewidth}
\begin{lstlisting}[basicstyle=\ttfamily\scriptsize, caption={Understandability - Human-readable \newline labels in Classes}, label={shape:understandability_human_readable_labels_classes}]
ex:LabelForClassesShape a sh:NodeShape;
  sh:targetClass rdfs:Class; 
  sh:property [
    sh:path rdfs:label; sh:minCount 1;
  ].
\end{lstlisting}
\end{minipage}
\hfill
\begin{minipage}[t]{0.48\linewidth}
\begin{lstlisting}[basicstyle=\ttfamily\scriptsize, caption={Understandability - Human-readable \newline labels in Properties}, label={shape:understandability_human_readable_labels_properties}]
ex:LabelForPropertiesShape a sh:NodeShape ;
  sh:targetClass rdf:Property; 
  sh:property [
    sh:path rdfs:label; sh:minCount 1;
  ].
\end{lstlisting}
\end{minipage}

The validation reports for \shape{\ref{shape:understandability_human_readable_labels}}, \shape{\ref{shape:understandability_human_readable_labels_classes}}, and \shape{\ref{shape:understandability_human_readable_labels_properties}} output a violation for each entity, class, or property missing an \rdfproperty{rdfs:label}. The DQ measures calculated from the validation report are all \textit{ratio measures}.

In addition, we also defined shape \shape{\ref{shape:understandability_dataset_metadata}}, which checks for the indication of metadata, such as the name, description, and website of the dataset, in the VoID description. In this shape, we target the class \rdfproperty{void:Dataset}, and check for the existence of the properties \rdfproperty{dcterms:title}, \rdfproperty{dcterms:description}, and \rdfproperty{foaf:homepage}. Although these properties aren't from the VoID vocabulary, \cite{void2011} mentions their use to convey this information. \shape{\ref{shape:understandability_dataset_metadata}} targets \rdfproperty{void:Dataset}, but it can also be adapted for DCAT by targeting \rdfproperty{dcat:Dataset} and using \rdfproperty{dcterms:title},  \rdfproperty{dcterms:description}, and \rdfproperty{dcat:landingPage}.

\begin{minipage}[t]{\linewidth}
\begin{lstlisting}[basicstyle=\ttfamily\scriptsize, caption={Understandability - Dataset metadata}, label={shape:understandability_dataset_metadata}]
ex:UnderstandabilityDatasetMetadataShape a sh:NodeShape ;
  sh:targetClass void:Dataset; 
  sh:property [
    sh:path dcterms:title;
    sh:minCount 1;
    sh:nodeKind sh:Literal;
  ];
  sh:property [
    sh:path dcterms:description;
    sh:minCount 1;
    sh:nodeKind sh:Literal;
  ];
  sh:property [
    sh:path foaf:homepage;
    sh:minCount 1;
    sh:class foaf:Document;
  ].
\end{lstlisting}
\end{minipage}

The validation result for \shape{\ref{shape:understandability_dataset_metadata}} outputs a violation if any of the properties are either missing or do not conform with the expected range. In this case, the DQ measure is a \BinaryMeasure.

For the metrics U2, U3, and U5, we defined a single shape to cover them with SHACL core (See \shape{\ref{shape:understandability_extra_metadata}}), since we need to target the class \rdfproperty{void:Dataset} for all of them. Additionally, to fully cover the metric U3, we also had to define a separate shape, which is described later (See \shape{  \ref{shape:understandability_uri_regex_namespace_compliance}}).
To cover the metric U2, we should check for the indication of one or more exemplary URIs via \rdfproperty{void:exampleResource}. To cover U3, we should check for the indication of a regular expression for the URI of the entities via \rdfproperty{void:uriRegexPattern}, or for the indication of the base URI for the entities via \rdfproperty{void:uriSpace}. Lastly, to cover U5 we should check if there's an indication of the vocabularies used in the dataset via \rdfproperty{void:vocabulary}.

\begin{minipage}[t]{\linewidth}
\begin{lstlisting}[basicstyle=\ttfamily\scriptsize, caption={Understandability - Dataset metadata}, label={shape:understandability_extra_metadata}]
ex:UnderstandabilityExtraMetadataShape a sh:NodeShape ;
  sh:targetClass void:Dataset; 
  sh:property [
    sh:path void:exampleResource;
    sh:minCount 1;
  ];
  sh:property [
    sh:path void:vocabulary;
    sh:minCount 1;
  ];
  sh:or ([
        sh:path void:uriRegexPattern;
        sh:minCount 1;
    ]
    [
        sh:path void:uriSpace;
        sh:minCount 1;
        sh:nodeKind sh:Literal;
    ]).
\end{lstlisting}
\end{minipage}

The validation report for \shape{\ref{shape:understandability_extra_metadata}} outputs a violation for instances of \rdfproperty{void:Dataset} that are missing the properties \rdfproperty{void:exampleResource}, \rdfproperty{void:vocabulary}, and \rdfproperty{void:uriRegexPattern} or \rdfproperty{void:uriSpace}. For the property \rdfproperty{void:uriSpace}, it will also output a violation if it's present but the values don't conform to the specified range.
From \shape{\ref{shape:understandability_extra_metadata}}, we have to output DQ measures for the different metrics U2, U3, and U5, and in all cases, the DQ measure is a \BinaryMeasure.

Although shapes \shape{\ref{shape:understandability_dataset_metadata}} and \shape{\ref{shape:understandability_extra_metadata}} target the same class and could be merged, we keep them separate to highlight that \shape{\ref{shape:understandability_dataset_metadata}} can be adapted for the DCAT vocabulary (by switching the target class and properties).

Additionally, for metric U3, we check that entities use URIs matching the pattern or base namespace defined by \rdfproperty{void:uriRegexPattern} or \rdfproperty{void:uriSpace}. For this, we defined shape \shape{\ref{shape:understandability_uri_regex_namespace_compliance}}, which targets entities and validates URI compliance. Assumptions \assumption{1} and \assumption{2} are required, as we focus on entities. The validation report flags each entity with a non-conforming URI, and the DQ measure is a \RatioMeasure. Note that shape \shape{\ref{shape:understandability_uri_regex_namespace_compliance}} depends on the existence of values for \rdfproperty{void:uriRegexPattern} or \rdfproperty{void:uriSpace}.

\begin{minipage}[t]{\linewidth}
\begin{lstlisting}[basicstyle=\ttfamily\scriptsize, caption={Understandability - URI regex or namespace compliance for entities}, label={shape:understandability_uri_regex_namespace_compliance}]
ex:URIRegexComplianceShape a sh:NodeShape ;
  sh:targetSubjectsOf rdf:type; 
  sh:or (
    [sh:path rdf:type; sh:hasValue rdfs:Class;]
    [sh:path rdf:type; sh:hasValue rdf:Property;]
    [sh:path rdf:type; sh:hasValue owl:NamedIndividual;]
    [sh:pattern "^URI_REGEX_PATTERN";]
    [sh:pattern "^URI_SPACE";]
  ).
\end{lstlisting}
\end{minipage}

For the metrics U4 and U6, we can't use SHACL core. On the one hand, U4 checks for the indication of an exemplary SPARQL query, which, according to \cite{Flemming2011}, can be in the documentation of the dataset or its source. However, SHACL core works over RDF graphs, and the documentation of the dataset or source is usually an HTML document \cite{Flemming2011}. Moreover, none of the standard vocabularies for annotating metadata on RDF datasets (VoID and DCAT) contains a property to state an exemplary SPARQL query. On the other hand, U6 checks for the efficiency and effectiveness of message boards and mailing lists, but these aspects can't be checked with SHACL core, since it's a constraint language that works over RDF graphs. 

\subsubsection{Trustworthiness}

\dq{Trustworthiness} is defined as the extent to which information is accepted as correct, reliable, and credible.
For this dimension, \BaseSurvey~ identified seven metrics (TW1 - TW7), where only one of them can be fully covered with SHACL core. Metrics TW2 and TW3 mention annotating the data with trust values, for example, using a trust ontology. In this case, even though SHACL core can't annotate the data, it can check if entities have trust annotations. For example, we can consider one of the proposed approaches in \cite{Jacobi2011} where they state the use of a trust ontology which defines the property \rdfproperty{ex:trustvalue} to assign trust values to resources. Therefore, we defined shape \shape{\ref{shape:trustworthiness_trust_values_entities}} that checks if entities have a value for this property. It's important to note that currently, there's no standard vocabulary or ontology for assigning trust values to data. In the future, if a new vocabulary arises, the shape should be updated to accommodate it. For \shape{\ref{shape:trustworthiness_trust_values_entities}}, we need to consider assumptions \assumption{1} and \assumption{2}, given that we are targeting entities, and we also need the existence of a vocabulary or ontology that states the property used to indicate a trust value.

\begin{minipage}[t]{0.48\linewidth}
\begin{lstlisting}[basicstyle=\ttfamily\scriptsize, caption={Trustworthiness - Trust values \newline in entities}, label={shape:trustworthiness_trust_values_entities}]
ex:TrustValuesEntitiesShape a sh:NodeShape ;
  sh:targetSubjectsOf rdf:type; 
  sh:or (
    [sh:path rdf:type; sh:hasValue rdfs:Class;]
    [sh:path rdf:type; sh:hasValue rdf:Property;]
    [sh:path rdf:type; sh:hasValue owl:NamedIndividual;]
    [
        sh:path ex:trustvalue;
        sh:minCount 1;
    ]
  ).
\end{lstlisting}
\end{minipage}
\hfill
\begin{minipage}[t]{0.48\linewidth}
\begin{lstlisting}[basicstyle=\ttfamily\scriptsize, caption={Trustworthiness - Trusted \newline contributors and providers}, label={shape:trustworthiness_trusted_provider}]
ex:TrustedContributorProviderShape a sh:NodeShape ;
  sh:targetClass void:Dataset; 
  sh:property [
    sh:path dcterms:provider;
    sh:in ( LIST_TRUSTED_PROVIDERS );
  ];
  sh:property [
    sh:path dcterms:contributor;
    sh:in ( LIST_TRUSTED_CONTRIBUTORS );
  ].
\end{lstlisting}
\end{minipage}

The validation report for \shape{\ref{shape:trustworthiness_trust_values_entities}} outputs a violation for each entity that doesn't have a value for the property \rdfproperty{ex:trustvalue}.

Furthermore, the metric TW5 verifies the trustworthiness of the information provider by checking three aspects. 
The first aspect (TW5M1) computes the trustworthiness of the information provider by creating decision networks, which cannot be accomplished with SHACL core, given that it's a constraint language. The second aspect (TW5M2) verifies whether the dataset’s contributor or provider is in a list of trusted sources. For this, we defined shape \shape{\ref{shape:trustworthiness_trusted_provider}}, which targets the class \rdfproperty{void:Dataset} and checks that the values of \rdfproperty{dcterms:contributor} or \rdfproperty{dcterms:provider} are included in a predefined list (as literals or IRIs). Specifying this list requires domain knowledge (assumption \assumption{3}). Alternatively, the shape could be adapted to exclude untrusted sources using the constraint \rdfproperty{sh:not} with \rdfproperty{sh:in}. The validation report for \shape{\ref{shape:trustworthiness_trusted_provider}} outputs a violation for each instance of \rdfproperty{void:Dataset} with a contributor or provider not in the trusted list. This shape could also be reused for \rdfproperty{dcat:Dataset} instances.

The third aspect (TW5M3) assesses the trust level of the publisher, represented by a value from 1 (lowest) to 9 (highest). This idea originates from \cite{Golbeck2003}, where the FOAF vocabulary is extended with nine properties to let people rate others they know. However, those properties apply specifically to \rdfproperty{foaf:Person} instances. Since our use case involves datasets rather than individuals, we adopt the approach from \cite{Jacobi2011}, which uses a generic \rdfproperty{ex:trustvalue} property to annotate trust levels. Based on this, we defined \shape{\ref{shape:trustworthiness_level_trust_publisher}} that verifies whether a publisher has an associated trust value in the range [1,9]. Moreover, for this last aspect, we need to consider assumption \assumption{2}, given that we need the existence of a vocabulary or ontology, where the property used to annotate trust values of entities is defined.

\begin{minipage}[t]{0.48\linewidth}
\begin{lstlisting}[basicstyle=\ttfamily\scriptsize, caption={Trustworthiness - Level of trust \newline of the publisher}, label={shape:trustworthiness_level_trust_publisher}]
ex:LevelOfTrustPublisherShape a sh:NodeShape ;
  sh:targetObjectsOf dcterms:publisher; 
  sh:property [
    sh:path ex:trustvalue;
    sh:minCount 1;
    sh:minInclusive 1;
    sh:maxInclusive 9;
    sh:datatype xsd:integer;
  ].
\end{lstlisting}
\end{minipage}
\hfill
\begin{minipage}[t]{0.48\linewidth}
\begin{lstlisting}[basicstyle=\ttfamily\scriptsize, caption={Trustworthiness - Trust through \newline association}, label={shape:trustworthiness_trust_through_association}]
ex:TrustThroughAssociationShape a sh:NodeShape ;
  sh:targetSubjectsOf rdf:type; 
  sh:or (
    [sh:path rdf:type; sh:hasValue rdfs:Class;]
    [sh:path rdf:type; sh:hasValue rdf:Property;]
    [sh:path rdf:type; sh:hasValue owl:NamedIndividual;]
    [
        sh:path prov:wasAttributedTo;
        sh:in ( LIST_TRUSTED_AUTHORS );
        sh:minCount 1;
    ]
  ).
\end{lstlisting}
\end{minipage}


The validation report for \shape{\ref{shape:trustworthiness_level_trust_publisher}} outputs a violation for each of the objects of \rdfproperty{dcterms:publisher} that either doesn't have a trust value (via the property \rdfproperty{ex:trustvalue}), the value is outside the range [1,9], or the value is not an integer. 

Nevertheless, we are not able to cover all aspects defined by this metric, so we argue that TW5 can be partially covered by SHACL core.

The metric TW6 evaluates whether trust can be inferred through associations (e.g., resources linked to a trusted dataset author are considered more trustworthy). This can be supported using provenance annotations via ontologies like PROV-O\footnote{\url{https://www.w3.org/TR/2013/REC-prov-o-20130430/}}, which allows linking entities to authors using \rdfproperty{prov:wasAttributedTo}. Based on this, we defined \shape{\ref{shape:trustworthiness_trust_through_association}} to check if entities have such links. The shape can also be adapted to flag entities linked to untrusted authors using \rdfproperty{sh:not} with \rdfproperty{sh:in}. Validation results include entities with no author or linked to a non-trusted author. This shape requires assumptions \assumption{1} and \assumption{2}, as it targets entities and assumes knowledge of the provenance property used. Moreover, to be able to instantiate this shape, we need domain knowledge to specify the list of trustworthy authors (assumption \assumption{3}).

To conclude, the metrics TW1 and TW4 state the computation of trust values using different methods. However, SHACL is a constraint language and it doesn't allow the computation of values that can then be assigned to entities, triples or the dataset. Furthermore, the definition for the metric TW7 states the assignment of trust ratings to datasets by humans or by analyzing external pages. As we've stated before, SHACL core doesn't provide a way of assigning trust values to the dataset, nor allows human input.

\subsubsection{Timeliness} 
In the following, we present the remaining shapes' definitions for the dimension \dq{Timeliness} that weren't described in \autoref{section:contextual}.

Metric T2 checks for the freshness of the dataset based on the data source, by measuring the distance between the last modified date of the dataset and the data source. In this case, we can't use SHACL core, since there's no constraint that let's us calculate the distance between two values. However, what we can do with SHACL core is verify the freshness of the dataset, by targeting the class \rdfproperty{void:Dataset} or \rdfproperty{dcat:Dataset} and checking if the value for \rdfproperty{dcterms:modified} is after some point in time (see \shape{\ref{shape:timeliness_dataset}}), meaning the dataset is up to date. For this shape, we need to consider assumption \assumption{3}, since we require domain knowledge indicating when the dataset is considered up-to-date

\begin{minipage}[t]{\linewidth}
\begin{lstlisting}[basicstyle=\ttfamily\scriptsize, caption={Timeliness - Outdated dataset}, label={shape:timeliness_dataset}]
ex:TimelinessDatasetShape a sh:NodeShape ;
    sh:targetClass void:Dataset; 
    sh:property [
        sh:path dcterms:modified;
        sh:minInclusive "DATE_RANGE_MIN_BOUND";
    ].
\end{lstlisting}
\end{minipage}

The validation report for this shape outputs a violation for the instances of \rdfproperty{void:Dataset} whose value for the property \rdfproperty{dcterms:date} is older than the date specified in \rdfproperty{DATE\_RANGE\_MIN\_BOUND}.

\subsection{Representational}\label{section:representational_appendix}
This section presents the shape definitions for the dimensions \dq{Representational Conciseness}, \dq{Interoperability}, and \dq{Interpretability} from the \textit{Representational} group. It also includes the definitions of the remaining shapes for the dimension \dq{Versatility}, which were not included in \autoref{section:representational}.

\subsubsection{Representational conciseness}
\dq{Representational Conciseness} refers to how data is represented in a way that is both compact and well-formatted, while also being clear and complete.
For this dimension, \BaseSurvey~identified two metrics. The first metric (RC1) checks for the usage of long URIs or those that use query parameters. For this metric, we defined shape \shape{\ref{shape:representational_conciseness_short_uris}} that restricts the lengths of URIs using \rdfproperty{sh:maxLength}, and \shape{\ref{shape:representational_conciseness_parameters}} that uses a regex pattern to identify query parameters by checking for strings of the form \rdfproperty{?param\_name=param\_value} in URIs. For both shapes, we need to consider assumptions \assumption{1} and \assumption{2}, because we are targeting entities.

\noindent
\begin{minipage}[t]{0.48\linewidth}
\begin{lstlisting}[basicstyle=\ttfamily\scriptsize, caption={Representational conciseness - \newline Short URIs}, label={shape:representational_conciseness_short_uris}]
ex:URIsLengthShape a sh:NodeShape ;
  sh:targetSubjectsOf rdf:type; 
  sh:or (
    [sh:path rdf:type; sh:hasValue rdfs:Class;]
    [sh:path rdf:type; sh:hasValue rdf:Property;]
    [sh:path rdf:type; sh:hasValue owl:NamedIndividual;]
    [sh:maxLength LENGTH_VALUE;]
  ).
\end{lstlisting}
\end{minipage}
\hfill
\begin{minipage}[t]{0.48\linewidth}
\begin{lstlisting}[basicstyle=\ttfamily\scriptsize, caption={Representational conciseness - \newline Parameters in URIs}, label={shape:representational_conciseness_parameters}]
ex:URIsParametersShape a sh:NodeShape ;
  sh:targetSubjectsOf rdf:type; 
  sh:or (
    [sh:path rdf:type; sh:hasValue rdfs:Class;]
    [sh:path rdf:type; sh:hasValue rdf:Property;]
    [sh:path rdf:type; sh:hasValue owl:NamedIndividual;]
    [sh:not [ sh:pattern "\\?.+=.*" ; ] ;]
  ).
\end{lstlisting}
\end{minipage}

The validation report for \shape{\ref{shape:representational_conciseness_short_uris}} outputs a violation for every entity that has a URI with a length greater than \rdfproperty{LENGTH\_VALUE}, while the report for \shape{\ref{shape:representational_conciseness_parameters}} outputs a violation for every entity that has a URI with parameters. In both cases, the DQ measure is a \RatioMeasure.

The second metric (RC2) verifies the absence of RDF prolix features, such as RDF reification, containers and collections. 
To address this, we defined \shape{\ref{shape:representational_conciseness_prolix_rdf}}, 
which targets all entities and verifies they are not instances of \rdfproperty{rdf:Statement}, \rdfproperty{rdf:List}, \rdfproperty{rdf:Seq}, \rdfproperty{rdf:Bag}, or \rdfproperty{rdf:Alt}. This shape requires assumptions \assumption{1} and \assumption{2}, given that we target entities. The validation report outputs a violation for each entity of these types, and the DQ measure is a \RatioMeasure.

\begin{minipage}[t]{\linewidth}
\begin{lstlisting}[basicstyle=\ttfamily\scriptsize, caption={Representational conciseness - Use of prolix RDF features}, label={shape:representational_conciseness_prolix_rdf}]
ex:ProlixRDFFeaturesShape a sh:NodeShape ;
  sh:targetSubjectsOf rdf:type; 
  sh:or (
    [sh:path rdf:type; sh:hasValue rdfs:Class;]
    [sh:path rdf:type; sh:hasValue rdf:Property;]
    [sh:path rdf:type; sh:hasValue owl:NamedIndividual;]
    [  sh:not [ 
        sh:or (
            [ sh:class rdf:Statement ]
            [ sh:class rdf:List ]
            [ sh:class rdf:Seq ]
            [ sh:class rdf:Bag ]
            [ sh:class rdf:Alt ]
        )];
    ]
  ).
\end{lstlisting}
\end{minipage}

\subsubsection{Interoperability}\label{section:interoperability_appendix}

\dq{Interoperability} refers to the extent to which data is represented using standardized vocabularies and formats, enabling uniform interpretation and seamless integration with other datasets and systems.
For this dimension, \BaseSurvey~identified two metrics. Metric ITO1 checks whether existing terms from relevant vocabularies are reused. With SHACL core, one can define a shape that targets classes from known vocabularies and verifies the use of properties defined for those classes. As an example, we present shape \shape{\ref{shape:interop_reuse_existing_vocab_terms}} based on the FOAF vocabulary, which targets the class \rdfproperty{foaf:Person} and enforces the presence of at least one value for \rdfproperty{foaf:givenName} and \rdfproperty{foaf:familyName}.

\begin{minipage}[t]{\linewidth}
\begin{lstlisting}[basicstyle=\ttfamily\scriptsize, caption={Interoperability - Re-use of existing terms}, label={shape:interop_reuse_existing_vocab_terms}]
ex:ReUseExistingVocabularyTerms a sh:NodeShape ;
  sh:targetClass foaf:Person ; 
  sh:property [
    sh:path foaf:givenName;
    sh:minCount 1;
  ];
  sh:property [
    sh:path foaf:familyName;
    sh:minCount 1;
  ].
\end{lstlisting}
\end{minipage}

The validation result for \shape{\ref{shape:interop_reuse_existing_vocab_terms}} outputs a violation for each instance of \rdfproperty{foaf:Person} that doesn't use the specified properties (\rdfproperty{foaf:firstName} and \rdfproperty{foaf:lastName}).

Metric ITO2 verifies the usage of relevant vocabularies for the domain. To verify this, one would need to examine the usage of a vocabulary, which involves checking whether a class or property defined in the vocabulary is utilized in the dataset. 
As mentioned in Appendix \ref{section:completeness} for the metric CP1, with SHACL core we can check the usage of classes with the shape  \shape{\ref{shape:completeness_schema_completeness_class_usage}}. However, we cannot check the usage of properties, since SHACL core does not provide a mechanism to verify whether a given property is used as a predicate in any triple. Therefore, this metric can be partially covered with SHACL core.
Additionally, for both shapes defined for the metrics of this dimension, we need to know in advance which are the relevant vocabularies for the domain to be able to instantiate the shapes; therefore, we need to consider assumption \assumption{3}.

\subsubsection{Versatility}\label{section:versatility_appendix}
In the following, we present the remaining shapes' definitions for the dimension \dq{Versatility} that weren't described in \autoref{section:representational}.

Metric V1, defined in \cite{Flemming2011}, verifies if the data is available in different serialization formats. \shape{\ref{shape:versatility_ser_form_void}} shows the shape defined for this metric, where we check for the existence of the property \rdfproperty{void:feature} and the correctness of its values, as defined in the VoID vocabulary. 

\begin{minipage}[t]{\linewidth}
\begin{lstlisting}[basicstyle=\ttfamily\scriptsize, caption={Versatility - Serialization formats VoID}, label={shape:versatility_ser_form_void}]
ex:SerializationFormatsShape a sh:NodeShape ;
  sh:targetClass void:Dataset; 
  sh:property [
    sh:path void:feature;
    sh:minCount 1;
    sh:maxCount 5;
    sh:in (
        <http://www.w3.org/ns/formats/N3>
        <http://www.w3.org/ns/formats/N-Triples>
        <http://www.w3.org/ns/formats/RDF_XML>
        <http://www.w3.org/ns/formats/RDFa>
        <http://www.w3.org/ns/formats/Turtle>
    );
  ].
\end{lstlisting}
\end{minipage}

The validation report for this shape outputs a violation for each instance of \rdfproperty{void:Dataset} that doesn't have a value for the property \rdfproperty{void:feature}, or that the value for this property is not in the list of allowed values. The DQ measure in this case is a \BinaryMeasure.

Lastly, \shape{\ref{shape:versatility_languages_labels_extension}} presents an extension for \shape{\ref{shape:versatility_languages_labels}}, presented in \autoref{section:representational}, that checks for specific languages (\rdfproperty{sh:languageIn}) and that there's a single label for each language (\rdfproperty{sh:uniqueLang}). Note that for this extension, we also need to consider assumption \assumption{3}, apart from \assumption{1} and \assumption{2}.

\begin{minipage}[t]{\linewidth}
\begin{lstlisting}[basicstyle=\ttfamily\scriptsize, caption={Versatility - Languages in labels of entities (Extension)}, label={shape:versatility_languages_labels_extension}]
ex:DifferentLanguagesLabelsExtensionShape a sh:NodeShape ;
  sh:targetSubjectsOf rdfs:label; 
  sh:or (
    [ sh:path rdf:type; sh:hasValue rdfs:Class; ]
    [ sh:path rdf:type; sh:hasValue rdf:Property; ]
    [ sh:path rdf:type; sh:hasValue owl:NamedIndividual; ]
    [
        sh:path rdfs:label;
        sh:datatype rdf:langString;
        sh:languageIn ( REQUIRED_LANGUAGES );
        sh:uniqueLang true;
    ]
  ).
\end{lstlisting}
\end{minipage}

\subsubsection{Interpretability}\label{section:interpretability}
\dq{Interpretability} refers to whether information is represented using suitable notations and can be effectively processed by machines.
For this dimension, \BaseSurvey~defined four metrics. Metric ITP1 verifies the use of self-descriptive formats, which checks for the usage of URIs to identify entities and terms used to describe them. We argue SHACL core can be used to check this metric, since we can define a shape that checks that entities are identified with IRIs (see \shape{\ref{shape:interpretability_self_descriptive_formats}}), and we can also check if objects of triples are IRIs (See \shape{\ref{shape:interpretability_self_descriptive_formats_properties}}). Note that in the last case, we need to target every property used in the dataset and constrain its value to be an IRI.
Additionally, for \shape{\ref{shape:interpretability_self_descriptive_formats}} we have to consider the assumptions \assumption{1} and \assumption{2}, given that we are targeting entities.

\begin{minipage}[t]{0.48\linewidth}
\begin{lstlisting}[basicstyle=\ttfamily\scriptsize, caption={Interpretability - Use of self-descriptive \newline formats}, label={shape:interpretability_self_descriptive_formats}]
ex:SelfDescriptiveFormatEntitiesShape 
  a sh:NodeShape ;
  sh:targetSubjectsOf rdf:type; 
  sh:or (
    [sh:path rdf:type; sh:hasValue rdfs:Class;]
    [sh:path rdf:type; sh:hasValue rdf:Property;]
    [sh:path rdf:type; sh:hasValue owl:NamedIndividual;]
    [ sh:nodeKind sh:IRI; ]
  ).
\end{lstlisting}
\end{minipage}
\hfill
\begin{minipage}[t]{0.48\linewidth}
\begin{lstlisting}[basicstyle=\ttfamily\scriptsize, caption={Interpretability - Use of self-descriptive \newline formats (properties)}, label={shape:interpretability_self_descriptive_formats_properties}]
ex:SelfDescriptiveFormatPropertiesShape 
  a sh:NodeShape ;
  sh:targetObjectsOf PROPERTY_URI; 
  sh:nodeKind sh:IRI.
\end{lstlisting}
\end{minipage}

The validation report for \shape{\ref{shape:interpretability_self_descriptive_formats}} outputs a violation for each entity that is not identified with an IRI. In this case, the DQ measure is a \RatioMeasure. For \shape{\ref{shape:interpretability_self_descriptive_formats_properties}}, the validation report outputs a violation for every object of \rdfproperty{PROPERTY\_URI} that it’s not an IRI. In this case, the DQ measure is a \CompositeMeasure.

For the metric ITP2, we need to detect the use of appropriate language, symbols, units, datatypes, and clear definitions. In this case, we argue SHACL core can partially cover this metric, since we can define syntax checks to verify the appropriate usage of datatypes (using the constraint \rdfproperty{sh:datatype}), symbols (using the constraint \rdfproperty{sh:pattern}), and language (using the constraints \rdfproperty{sh:datatype rdfs:langString} and \rdfproperty{sh:languageIn}). SHACL core is capable of verifying the syntax of literal values, but it does not evaluate their semantics. For instance, it can check if literals have language tags and whether these tags correspond to specified languages. However, it cannot determine if the literal values are appropriate for those languages. Furthermore, for this metric, we consider assumption \assumption{3}, as we require domain knowledge to identify the correct languages, symbols, datatypes, and so on.

Metric ITP3 checks for the invalid usage of undefined classes and properties, which means detecting the usage of properties and classes that don't have a formal definition, either in an ontology or vocabulary. To cover this metric, we defined \shape{\ref{shape:interpretability_undefined_classes}}, which checks that classes are typed as \rdfproperty{rdfs:Class}, and \shape{\ref{shape:interpretability_undefined_properties}}, which checks that properties are typed as \rdfproperty{rdf:Property}. For the shape \shape{\ref{shape:interpretability_undefined_classes}}, we considered named classes. Additionally, for these two shapes, we also have to consider assumption \assumption{2}, since these shapes have to be validated against the ontology. 
Moreover, due to \assumption{2}, the ontology has the triple \rdftriple{c}{rdf:type}{rdfs:Class} for each class, and the triple \rdftriple{p}{rdf:type}{rdf:Property} for each property. Therefore, by using the constraint \rdfproperty{sh:hasValue}, the class or property can have other types, such as \rdfproperty{owl:Class} for classes or \rdfproperty{owl:OntologyProperty} for properties. 

\begin{minipage}[t]{0.48\linewidth}
\begin{lstlisting}[basicstyle=\ttfamily\scriptsize, caption={Interpretability - Undefined classes}, label={shape:interpretability_undefined_classes}]
ex:UndefinedClassShape a sh:NodeShape ;
  sh:targetNode CLASS_URI; 
  sh:property [
    sh:path rdf:type;
    sh:hasValue rdfs:Class;
    sh:minCount 1;
  ].
\end{lstlisting}
\end{minipage}
\hfill
\begin{minipage}[t]{0.48\linewidth}
\begin{lstlisting}[basicstyle=\ttfamily\scriptsize, caption={Interpretability - Undefined \newline properties}, label={shape:interpretability_undefined_properties}]
ex:UndefinedPropertyShape a sh:NodeShape ;
  sh:targetNode PROPERTY_URI; 
  sh:property [
    sh:path rdf:type;
    sh:hasValue rdf:Property;
    sh:minCount 1;
  ].
\end{lstlisting}
\end{minipage}


The validation reports for shapes~\shape{\ref{shape:interpretability_undefined_classes}} and~\shape{\ref{shape:interpretability_undefined_properties}} output a violation for each undefined class (i.e., not typed or not of type \rdfproperty{rdfs:Class}) and undefined property (i.e., not typed or not of type \rdfproperty{rdf:Property}), respectively. In both cases, the DQ measure is a \CompositeMeasure. 

Finally, the last metric (ITP4) checks for the usage of blank nodes. We defined \shape{\ref{shape:interpretability_blank_nodes}}, which verifies that entities are not blank nodes, since the Best Practices for Publishing Linked Data \cite{BestPractices_2014} state the usage of HTTP URIs for identifying resources. Note that blank nodes can be used in the dataset, as long as they are not used to identify entities. Moreover, for this shape, we need to consider assumptions \assumption{1} and \assumption{2}, given that we are targeting entities.

\begin{minipage}[t]{\linewidth}
\begin{lstlisting}[basicstyle=\ttfamily\scriptsize, caption={Interpretability - Usage of blank nodes}, label={shape:interpretability_blank_nodes}]
ex:BlankNodesUsageEntitiesShape a sh:NodeShape ;
  sh:targetSubjectsOf rdf:type; 
  sh:or (
    [sh:path rdf:type; sh:hasValue rdfs:Class;]
    [sh:path rdf:type; sh:hasValue rdf:Property;]
    [sh:path rdf:type; sh:hasValue owl:NamedIndividual;]
    [ sh:not [ sh:nodeKind sh:BlankNode ]; ]
  ).
\end{lstlisting}
\end{minipage}


The validation report for \shape{\ref{shape:interpretability_blank_nodes}} outputs a violation for each entity that is a blank node, and the DQ measure is a \RatioMeasure. 

\section{Prototype}\label{section:evaluation_appendix}
This section provides further details about the developed prototype.\\

\subsection{Shape instantiation}
To improve runtime efficiency, we merged shapes with the same target, such as those for metrics U1, U2, U3, and U5, all targeting \rdfproperty{void:Dataset}. Their property shapes were combined into a single shape, and we use \rdfproperty{sh:resultPath} to distinguish the results in the validation report.

Additionally, many shapes apply constraints over all entities (e.g. \shape{\ref{shape:performance_hash_uris_entities}}). However, we couldn’t merge such shapes due to how the library PySHACL handles validation results for logical constraint components (i.e., \rdfproperty{sh:or}, \rdfproperty{sh:and}, \rdfproperty{sh:not}, and \rdfproperty{sh:xone}). For example, to merge shapes \shape{\ref{shape:performance_hash_uris_entities}} and \shape{\ref{shape:understandability_human_readable_labels}}, we would need to use the structure presented in \shape{\ref{example_structure_merged_rdf_type}}, due to the ``filter'' for classes, properties, and named individuals.

\begin{minipage}[t]{\linewidth}
\begin{lstlisting}[basicstyle=\ttfamily\scriptsize, caption={Structure to merge shapes that target entities}, label={example_structure_merged_rdf_type}]
ex:EntitiesShape a sh:NodeShape;
  sh:targetSubjectsOf rdf:type; 
  sh:or (
    [ sh:path rdf:type; sh:hasValue rdfs:Class; ]
    [ sh:path rdf:type; sh:hasValue rdf:Property; ]
    [ sh:path rdf:type; sh:hasValue owl:NamedIndividual; ]
    [ sh:and (
            [ sh:pattern "^[^#]*$"; ] % Usage of slash URIs
            [ sh:path rdfs:label; sh:minCount 1;  ] % Labesl in entities
        )
    ]
  ).
\end{lstlisting}
\end{minipage}

However, for these types of constraints, the validation report generated by PySHACL does not indicate which specific property shape caused the violation (e.g., when the URI contains a \# or lacks a label), only that the focus node does not conform to one or more shapes within the \rdfproperty{sh:or} constraint. As a result, we wouldn’t be able to get the specific violation to then compute the corresponding DQ measure.

\subsection{DQ measures for SHACL shapes}

\autoref{tab:shapes_metrics} shows an overview of the defined shapes with their respective DQ measure used during the implementation of the proof of concept. We encode the type of each DQ measure in the table using B, R, or C to indicate \BinaryMeasure, \RatioMeasure, and \CompositeMeasure~types, respectively. Moreover, for DQ measures of type \CompositeMeasure, when the \textit{Aggregation} ratio refers to “property (or class) correctly used,” we define a property (or class) as correctly used if the value of its corresponding \textit{Individual score} is 1. 

\begin{table}[h!]
\centering
\tiny
\begin{tabularx}{\textwidth} { 
  | >{\centering\arraybackslash}p{1.5cm} 
  | >{\centering\arraybackslash}p{1.4cm}  
  | >{\centering\arraybackslash}p{0.5cm}
  | >{\centering\arraybackslash}X
  | >{\centering\arraybackslash}X | }
    \hline
    \textbf{Dimension} & \textbf{Shape} & \textbf{Type} & \textbf{Individual score} & \textbf{Aggregation} \\ \hline
    Availability & \shape{\ref{shape:availability_void_shape}} & B & 1 if there's an indication of a data dump, 0 otherwise & ~ \\ \hline
    
    Licensing & \shape{\ref{shape:licensing_void_shape}} & B & 1 if there's an indication of a machine-readable license, 0 otherwise & ~ \\ \hline
    
    Interlinking 
    & \shape{\ref{shape:interlinking_external_uris}} & R & \(\frac{\texttt{\# violations}}{ \texttt{\# entities  with interlinking property}}\) & ~ \\ \hline

    Security & \shape{\ref{shape:security_authenticity_void}} & B & 1 if there's an indication of the source and author of the dataset, 0 otherwise & ~ \\ \hline

    Performance & \shape{\ref{shape:performance_hash_uris_entities}} & R & \( \frac{\texttt{\# violations}}{ \texttt{\# entities}}\) & ~ \\ \hline

    Syntactic Validity &
    \shape{\ref{shape:accuracy_malformed_literal}} & C & \( \frac{\texttt{\# violations} }{ \texttt{\# subjects that use the property} }\) & \( \frac{\texttt{\# properties correctly used}}{ \texttt{\# properties with a datatype range}}\) \\ \hline
    
    \multirow{13}{*}{Consistency}
    & \shape{\ref{shape:consistency_entities_in_disjoint_classes}} & C & \(\frac{\texttt{\# violations}}{\texttt{\# entities of the target class}} \) & \(\frac{\texttt{\# disjoint classes correctly used}}{\texttt{\# disjoint classes}}\) \\ \cline{2-5}
    
    & \shape{\ref{shape:consistency_misplaced_properties}} & C & 1 if the property isn't used as a class, otherwise 0 & \(\frac{\texttt{\# properties correctly used}}{\texttt{\# properties defined in ontology}} \) \\ \cline{2-5}
    
    & \shape{\ref{shape:consistency_misplaced_classes}} & C & 1 if the class isn't used as a property, otherwise 0 & \(\frac{\texttt{\# classes correctly used}}{\texttt{\# classes defined in ontology}} \) \\ \cline{2-5}
    
    & \shape{\ref{shape:consistency_misuse_datatype_properties}} & C & \(\frac{\texttt{\# violations}}{ \texttt{\# subjects that use owl:DatatypeProperty}}\) & \(\frac{\texttt{\# owl:DatatypeProperty correctly used}}{\texttt{\# owl:DatatypeProperty used}}\) \\ \cline{2-5}

    & \shape{\ref{shape:consistency_misuse_object_properties}} & C & \(\frac{\texttt{\# violations}}{ \texttt{\# subjects that use owl:ObjectProperty}}\) & \(\frac{\texttt{\# owl:ObjectProperty correctly used}}{\texttt{\# owl:ObjectProperty used}}\) \\ \cline{2-5}
    
    & \shape{\ref{shape:consistency_deprecated_property}} & C & \(\frac{\texttt{\# violations}}{\texttt{\# entities that use the deprecated property}}\) & \(\frac{\texttt{\# deprecated properties unused}}{\texttt{\# deprecated properties}}\) \\ \cline{2-5}
    
    & \shape{\ref{shape:consistency_deprecated_class}} & B & 1 if none of the deprecated classes are used, otherwise 0 & ~ \\ \cline{2-5}
    
    & \shape{\ref{shape:consistency_uniqueness_inverse_functional_property}} & C & 1 if the inverse functional property is used correctly, otherwise 0 & \(\frac{\texttt{\# owl:InverseFunctionalProperty correctly used}}{\texttt{\# owl:InverseFunctionalProperty used}}\) \\ \cline{2-5}
    
    & \shape{\ref{shape:consistency_correct_domain}}, \shape{\ref{shape:consistency_correct_domain_owl_thing}} & C & \(\frac{\texttt{\# violations}}{\texttt{\# subjects that use the property}}\) & \(\frac{\texttt{\# properties with a defined domain correctly used}}{\texttt{\# properties with a defined domain}}\) \\ \cline{2-5}
    
    & \shape{\ref{shape:consistency_correct_range}}, \shape{\ref{shape:consistency_correct_range_literal}}, \shape{\ref{shape:consistency_correct_range_owl_thing}}, \shape{\ref{shape:consistency_correct_range_rdfs_resource}} & C & \(\frac{\texttt{\# violations}}{\texttt{\# subjects that use the property}}\) & \(\frac{\texttt{\# properties with a defined range correctly used}}{\texttt{\# properties with a defined range}}\) \\ \cline{2-5}
    
    & \shape{\ref{shape:consistency_irreflexive_property}} & C & \(\frac{\texttt{\# violations}}{\texttt{\# subjects that use the property}}\) & \(\frac{\texttt{\# owl:IrreflexiveProperty correctly used}}{\texttt{\# owl:IrreflexiveProperty used}}\) \\ \cline{2-5}
    
    & \shape{\ref{shape:consistency_functional_property}} & C & \(\frac{\texttt{\# violations}}{\texttt{\# subjects that use the property}}\) & \(\frac{\texttt{\# owl:FunctionalProperty correctly used}}{\texttt{\# owl:FunctionalProperty used}}\) \\ \cline{2-5}

    & \shape{\ref{shape:consistency_asymmetric_property}} & C & \(\frac{\texttt{\# violations}}{\texttt{\# subjects that use the property}}\) & \(\frac{\texttt{\# owl:AsymmetricProperty correctly used}}{\texttt{\# owl:AsymmetricProperty used}}\) \\ \hline

    \multirow{2}{*}{Complete\-ness}
    & \shape{\ref{shape:completeness_schema_completeness_class_usage}} & C & 1 if the class is used, 0 otherwise & \(\frac{\texttt{\# used classes}}{\texttt{\# classes defined in the ontology}}\) \\ \cline{2-5}
    
    & \shape{\ref{shape:completeness_interlinking_completeness}} & R & \(\frac{\texttt{\# violations}}{\texttt{\# entities}}\) & ~ \\ \hline

    \multirow{8}{*}{Understandability}
     & \shape{\ref{shape:understandability_human_readable_labels}} & R & \(\frac{\texttt{\#  violations}}{\texttt{\# entities}}\) & ~ \\ \cline{2-5}
    
    & \shape{\ref{shape:understandability_human_readable_labels_classes}} & R & \(\frac{\texttt{\# violations}}{\texttt{\# classes in ontology}}\) & ~ \\ \cline{2-5}
    
    & \shape{\ref{shape:understandability_human_readable_labels_properties}} & R & \(\frac{\texttt{\# violations}}{\texttt{\# properties in ontology}}\) & ~ \\ \cline{2-5}
    
    & \shape{\ref{shape:understandability_dataset_metadata}} & B & 1 if there's an indication of the title, description and homepage of the dataset, 0 otherwise & ~ \\ \cline{2-5}
    
    & \shape{\ref{shape:understandability_extra_metadata}} (exemplary resources) & B & 1 if there's an indication of exemplary resources, 0 otherwise & ~ \\ \cline{2-5}
    
    & \shape{\ref{shape:understandability_extra_metadata}} (regex pattern and URI namespace) &  B & 1 if there's an indication of a regex pattern or namespace for the entites, 0 otherwise & ~ \\ \cline{2-5}
    
    & \shape{\ref{shape:understandability_uri_regex_namespace_compliance}} & R & \(\frac{\texttt{\# violations}}{\texttt{\# entities}}\) & ~ \\ \cline{2-5}
    
    & \shape{\ref{shape:understandability_extra_metadata}} (vocabularies used) & B & 1 if there's an indication of the vocabularies used in the dataset, 0 otherwise & ~ \\ \hline

    \multirow{3}{*}{\parbox{1.5cm}{Representa\-tional concise\-ness}}
    & \shape{\ref{shape:representational_conciseness_short_uris}} & R & \(\frac{\texttt{\# violations}}{\texttt{\# entities}}\) & ~ \\ \cline{2-5}
    & \shape{\ref{shape:representational_conciseness_parameters}} & R & \(\frac{\texttt{\# violations}}{\texttt{\# entities}}\) & ~ \\ \cline{2-5}
    & \shape{\ref{shape:representational_conciseness_prolix_rdf}} & R & \(\frac{\texttt{\# violations}}{\texttt{\# entities}}\) & ~ \\ \hline
    
    \multirow{3}{*}{Versatility}
    & \shape{\ref{shape:versatility_ser_form_void}} & B & 1 if there's an indication of the serialization formats for the dataset, 0 otherwise & ~ \\ \cline{2-5}
    & \shape{\ref{shape:versatility_languages_labels}} & R & \(\frac{\texttt{\# violations}}{\texttt{\# entities with labels}}\) & ~ \\ \cline{2-5}
    & \shape{\ref{shape:versatility_languages_descriptions}} & R & \(\frac{\texttt{\# violations}}{\texttt{\# entities with descriptions}}\) & ~ \\ \hline
    
    \multirow{5}{*}{\parbox{1.5cm}{Interpretabi\-lity}}
    & \shape{\ref{shape:interpretability_self_descriptive_formats}}  & R & \(\frac{\texttt{\# violations}}{\texttt{\# entities}}\) & ~ \\ \cline{2-5}
    
    & \shape{\ref{shape:interpretability_self_descriptive_formats_properties}} & C & 1 if the property is used with an IRI, 0 otherwise & \(\frac{\texttt{\# properties correctly used}}{\texttt{\# properties used in the dataset}}\) \\ \cline{2-5}
    
    & \shape{\ref{shape:interpretability_undefined_classes}} & C & 1 if the class is defined, 0 otherwise & \(\frac{\texttt{\# classes defined}}{\texttt{ \# classes used in the dataset}}\) \\ \cline{2-5}
    
    & \shape{\ref{shape:interpretability_undefined_properties}} & C & 1 if the property is defined, 0 otherwise & \(\frac{\texttt{\# properties defined}}{\texttt{\# properties used in the dataset}}\) \\ \cline{2-5}
    
    & \shape{\ref{shape:interpretability_blank_nodes}} & R & \(\frac{\texttt{\# violations}}{\texttt{\# entities}}\) & ~ \\ \hline

\end{tabularx}
\caption{Overview of instantiated shapes and associated DQ measures}
\label{tab:shapes_metrics}
\end{table}

\section{SHACL extensions}\label{section:extensions_appendix}
This section provides further details about existing SHACL extensions and their applicability to DQA.\\

\paragraph{SHACL-SPARQL.} 
SHACL-SPARQL extends SHACL core by enabling features like flexible targets, node bindings, cross-entity comparisons, and basic arithmetic operations. However, even SHACL-SPARQL is not enough to cover all aspects for comprehensive DQA\@. Specifically, it falls short in verifying semantic aspects of the data, computing network-based measures over subgraphs (e.g., interlinking degree), accessing resources on the web, or verifying system-level aspects.

\paragraph{SHACL advanced features.} SHACL Advanced Features were published as a working group note of the \textit{RDF Data Shapes Working Group} in 2017. These advanced features present functionalities such as defining custom targets, functions, node expressions, and rules to derive new triples, which mostly rely on SPARQL. 
These additions offer more flexibility (e.g., selecting target nodes with SPARQL queries or performing arithmetic checks) but still rely heavily on SPARQL. As a result, they share the same limitations as SHACL-SPARQL and remain insufficient for comprehensive DQA.

\paragraph{DASH Data Shapes Vocabulary.} The \textit{DASH Data Shapes Vocabulary} extends SHACL with new target types and constraint components aimed at specific validation scenarios, such as string formatting, property value comparisons, and structural checks. While it introduces some useful constraints (e.g., \texttt{dash:stem}, \texttt{dash:subSetOf}), many are new constraints for the existing constraint components rather than new validation capabilities. As such, DASH adds specificity to the constraints but does not significantly expand the coverage of DQ metrics.

\paragraph{SHACL with inference.}
SHACL validators may optionally support inference through entailment regimes~\cite{w3c_shacl}. 
However, inference does not extend the expressiveness of SHACL core but expands the evaluated RDF graph. 
We do not explicitly evaluate SHACL with inference since our focus is on the DQ assessment capabilities of SHACL rather than on graph enrichment techniques.
}{

}

\end{document}